\RequirePackage{fix-cm}
\documentclass{svjour3}                     
\smartqed  

\usepackage{graphicx}
\usepackage{bm}
\usepackage{subcaption}
\usepackage{empheq}
\usepackage{enumitem}
\usepackage{amssymb}
\usepackage{url}
\newtheorem{prop}[theorem]{Proposition}
\numberwithin{equation}{section}
\newcommand{\bv}[1]{\mathbf{#1}}

\begin{document}

\title{Resonances in a Chaotic Attractor Crisis of the Lorenz Flow}

\author{Alexis Tantet \and Valerio Lucarini \and Henk A. Dijkstra}
\institute{A. Tantet \and V. Lucarini \at
Universit\"at Hamburg, Center for Earth System Research and Sustainability,
Meteorologisches Institut, Hamburg, Germany\\
\email{alexis.tantet@uni-hamburg.de}
\and
V. Lucarini \at
Department of Mathematics and Statistics, University of Reading, Reading, UK
\and
V. Lucarini \at
Centre for the Mathematics of Planet Earth, University of Reading, Reading, UK
\and
H. A. Dijkstra \at
Institute for Marine and Atmospheric research Utrecht, Department of Physics and 
Astronomy,  University of Utrecht, Utrecht, The Netherlands}

\date{\today}

\maketitle

\begin{abstract}

Local bifurcations of stationary points and limit cycles have successfully been
characterized in terms of the critical exponents of these solutions.
Lyapunov exponents and their associated covariant Lyapunov vectors
have been proposed as tools for supporting the understanding of critical transitions in chaotic dynamical systems.
However, it is in general not clear how the statistical properties of dynamical systems change across
a boundary crisis during which a chaotic attractor collides with a saddle.

This behavior is investigated here for a boundary crisis in the Lorenz flow,
for which neither the Lyapunov exponents nor the covariant Lyapunov vectors
provide a criterion for the crisis.
Instead, the convergence of the time evolution of probability densities to the invariant measure,
governed by the semigroup of transfer operators,
is expected  to slow down at the approach of the crisis.
Such convergence is described by the eigenvalues of the generator of this semigroup,
which can be divided into two families, referred to as the \emph{stable} and \emph{unstable}
Ruelle-Pollicott resonances, respectively.
The former describes the convergence of densities to the attractor (or escape from a repeller) and
is estimated from many short time series sampling the phase space.
The latter is responsible for the decay of correlations, or mixing,
and can be estimated from a long times series, invoking ergodicity.

It is found numerically for the Lorenz flow that the stable resonances do approach the imaginary axis
during the crisis, as is indicative of the loss of global stability of the attractor.
On the other hand, the unstable resonances, and a fortiori the decay of correlations,
do not flag the proximity of the crisis,
thus questioning the usual design of early warning indicators of boundary crises of chaotic attractors
and the applicability of response theory close to such crises.


\end{abstract}

\section{Introduction}
\label{sec:introduction}

It is a problem of fundamental relevance in mathematical, natural, and applied sciences to understand under which conditions a system may undergo abrupt changes under perturbation and, if so,
predict when these changes will occur.
Much of our understanding of such transitions comes from the
bifurcation theory of autonomous dynamical systems
\cite{Guckenheimer1983,Ruelle1989a,Kuznetsov1998},
with extensions to nonautonomous \cite{Kloeden2011} and random \cite{Arnold2003} dynamical systems.
In particular, local bifurcations, taking place for example when a stationary point or a limit cycle loses stability,
are characterized by the critical exponents of these invariant sets.
They yield a local measure of the relaxation rate of trajectories to these sets.
As the latter become less stable, these exponents approach zero,
resulting in the slowing down of the convergence of trajectories to the attractor.
This \emph{critical slowing down} has allowed to design early-warning signals of
critical transitions by monitoring the rate of decay of correlations \cite{Held2004a},
peaks in power spectra \cite{Kleinen2003a}, or of recovery from perturbations \cite{Nes2007}.
See \cite{Scheffer2009}, for a review,
and \cite{Lenton2011a} for applications to climate science.

Most physical systems of interest are, however,
chaotic, in the sense that they support an invariant measure
with positive Lyapunov exponents \cite{Eckmann1985,Young2002}.
Such sets can be involved in global bifurcations during which they may lose their
attracting character or undergo topological transformations \cite[Chap.~17]{Ruelle1989a}.
A particularly important type of bifurcation is the \emph{boundary crisis} \cite{Grebogi1983},
during which a chaotic invariant set ceases to be attracting
due to, loosely speaking, its collision with a saddle
\footnote{A saddle is defined here as a possibly chaotic invariant set,
which is attracting neither in forward nor in backward time
(see~\cite{Ashwin1996} for a more rigorous definition).}, the so-called \textit{edge state}. 
Most trajectories then undergo a transient before they converge to another attractor, if any.
Such crises have been identified in the H\'enon map and the Lorenz system \cite{Grebogi1983},
and are also found in high-dimensional turbulent flows \cite{Skufca2006,Schneider2007,Eckhardt2008}
and climate models \cite{Bodai2015,Lucarini2017}.

Much less is known regarding changes in the statistical
properties of these systems, and, in particular, if critical slowing down may be observed.
While for chaotic attractors the notion of critical exponents
can be generalized to that of Lyapunov exponents characterizing
the stability of chaotic trajectories \cite{Oseledets1968,Eckmann1985},
the latter do not in general allow to infer to what extent a chaotic set is attracting.
The Lyapunov spectrum is indeed calculated from a linearization about
orbits on the invariant set and thus only provides local information.
On the other hand, the size of the basin of attraction
leads to the notion of \emph{global stability} of an attractor (see e.g.~\cite{LaSalle1976}).
For example, during the boundary crisis of an attractor,
the size of the basin of attraction shrinks as the criticality is neared and vanishes
when the set becomes unstable.
It has been proposed in \cite{Faranda2014e} to approach the problem
studying large fluctuations of the systems using extreme value theory and relating the change in the qualitative properties of the extremes to the approach to the critical transition (see also~\cite{Faranda2014b}).

From a statistical physics point of view,
the divergence of nearby trajectories characterized by the positive Lyapunov exponents
is a manifestation of chaos at a microscopic level.
Macroscopically, the decay with time of correlations
associated with the loss of memory to initial conditions
of ensembles (as they mix) is a clear manifestation of chaos.
The evolution in time of the correlation function between any appropriate observables
is fully determined by the semigroup of transfer operators
$\mathcal{P}_t^\mu, t \ge 0$ \cite{Lasota1994},
governing the evolution of densities with respect to an invariant measure $\mu$
(e.g.~supported by an attractor).

It is a classical result from ergodic theory \cite{Halmos1956,Arnold1968}
that correlations between observables vanish for time lags going to infinity only if there are
no  eigenvalues of the transfer operator in the unit disk other than the eigenvalue 1.
A more difficult problem, which is still a matter of investigation
\cite{Baladi2001,Young2013}, is to characterize the rate of mixing.
This rate depends on the position in the complex plane of the poles of the correlation spectrum of a pair of observables, which correspond to the \emph{Ruelle-Pollicott resonances} \cite{Pollicott1985,Ruelle1986}.
The latter are given by the eigenvalues with nonzero real part
of the generator of the semigroup of transfer operators
acting on anisotropic Banach spaces adapted to the dynamics of
contraction and expansion of chaotic systems
\cite{Liverani1995a,Blank2001a,Gouezel2006,Butterley2007,Faure2014,Baladi2017}.
Note that, while both the presence of positive Lyapunov exponents and mixing are a manifestation of chaos, 
their relationship is nontrivial \cite{Collet2004,Alves2004,Pires2011,Slipantschuk2013}.
In the case of uniformly hyperbolic systems,
correlations are expected to undergo an initial fast decay  associated with large Lyapunov exponents,
but have an asymptotic decay bounded above
by the smallest positive Lyapunov exponent \cite{Collet2004}.
On the other hand, nonuniformly hyperbolic systems
may have arbitrarily slow mixing rates.

While transfer operators acting on densities
with respect to the invariant measure
allow to study the ergodic and mixing properties,
global information about the invariant set supporting this measure (e.g.~an attractor)
should be studied from transfer operators $\mathcal{P}_t^m, t\ge 0$, acting on densities
with respect to the Lebesgue measure (i.e.~in phase space).
This has recently led to new developments in the theory and applications
of the stability of dynamical systems \cite{Vaidya2008,Mauroy2016}.
In this case, the spectrum of the generator of the transfer semigroup,
does not only capture the rate of mixing, but also
the rate of convergence (escape) of densities
to (from) an invariant measure supported by an attractor (repeller).
It follows that, as an attractor becomes less attracting at the approach of a crisis,
densities are expected to take more time to converge to the invariant measure,
resulting in the slowing down of the decay of correlations, i.e.~in the mixing.
This slowing down should thus be associated with a decrease of the spectral gap
in the spectrum of the transfer semigroup.
This characterization of the stability of chaotic attractors has been used in \cite{Tantet2015a}	
to give numerical evidences that, in a high-dimensional climate model undergoing a boundary crisis,
the spectral gap in some eigenvalues of the transfer operators
in heavily projected spaces indeed shrinks,
explaining the slowing down of the decay of correlations observed from simulations.

It is, however, still unclear whether this change in the spectral properties of the
transfer semigroup is a generic property of dynamical systems undergoing an attractor crisis.
Moreover, the question emerges on whether such changes can be detected from
time series on the attractor alone - which can be investigated looking at the properties of $P_t^\mu$ - or if perturbations of the system away from the attractor are needed, so that the eigenvalues of $P_t^m$ are the key objects of interest. We refer to the resonances that belong to the spectrum of $P_t^m$ but not of $P_t^\mu$ as the  \emph{stable} resonances, which are associated to the nearing of the orbits towards the attractor, while we use the expression \emph{unstable} resonances for the eigenvalues of $P_t^\mu$. 

This issue is directly related to the generalization of the Fluctuation-Dissipation
Theorem (FDT) to nonequilibrium systems by \cite{Ruelle2009} (see also~\cite{Cessac2007}).
In this regard, the problem is that in the case of deterministic systems the natural fluctuations explore only the unstable manifold of the attractor, whereas generally an external perturbation will impact both the unstable and stable directions. A specific example in the context of a geophysical system is considered in \cite{Gritsun2017}.
On the other hand, for systems with an invariant measure
which is absolutely continuous with respect to the Lebesgue measure,
such as stochastically perturbed dynamical systems leading to hypoelliptic diffusions \cite{Hairer2010},
both transfer semigroups can be identified and the FDT is expected to hold. 
Note that a theory of the linear response to noise
of the statistics of the system has been developed by \cite{Lucarini2012c},
while the changes in the spectrum of the transfer semigroup due to noise
have been studied by \cite{Gaspard2002a} and by \cite{Tantet2016},
for the particular case of the Hopf bifurcation.

In this study, a chaotic boundary crisis in the Lorenz flow
is analyzed in terms of eigenvalues of the transfer semigroup.
Although this appears are a case study, we think that it can be of more general interest
by revealing possible statistical properties of more general chaotic dynamical systems.
The Lorenz flow is system of three ordinary differential equations
derived in the seminal paper \cite{Lorenz1963a}
via a spectral truncation of the fluid equations for Rayleigh-B\'enard convection.
As one of the first examples of low-dimensional flows with robust chaos and because
it exhibits a wide range of dynamical phenomena, it has been extensively studied
(see~\cite{Sparrow1982} for a review).
Most of the dynamical properties of Lorenz-like attractors have been obtained for the
geometric Lorenz system introduced in \cite{Guckenheimer1979}.
However, these properties were later found to persist
for general singular-hyperbolic attractors,
yielding a paradigm for robust low-dimensional chaotic systems \cite{Araujo}.
In particular, the standard Lorenz attractor was proved
to be a singular-hyperbolic chaotic attractor in \cite{Tucker1999}.
As a consequence, the Lorenz attractor supports a unique physical measure
with non-zero Lyapunov exponents \cite[Chap.~7]{Araujo}.
Numerical evidences suggest that the standard Lorenz flow
obeys linear \cite{Reick2002a} and nonlinear response theory \cite{Lucarini2009b}.
However, whether the Lorenz flow is mixing and has exponential decay of correlations is
still an open problem \cite[Chap.~10.2]{Araujo}.
In addition, the Lorenz flow is known \cite{Grebogi1983}
to undergo a boundary crisis for some parameter values,
while its low-dimensionality renders numerical applications tractable.
For these reasons, the Lorenz flow constitutes an interesting test bed
to better understand the relationship
between the eigenvalues of the transfer semigroup and the dynamics of the boundary crisis.

The analysis presented here is numerical and descriptive in nature,
but should allow to build intuition for further studies.
The Lorenz flow and the bifurcations of interest for this study
are briefly introduced in section 2.
The results presented in this section are mainly a reproduction of the ones presented in \cite{Sparrow1982}.
However, we also discuss the fact that that neither the Lyapunov exponents
nor the alignment of the covariant Lyapunov vectors,
can give an indication of the approach of the crisis.
This provides evidence that, contrary to what was found for other types of crises
\cite{Rollins1984,Pompe1988,Mehra1996,Beims2016},
these important dynamical quantities are in general not useful to flag vicinity to a boundary crises.
In section 3, two different methods are presented to approximate the transfer semigroups $P_t^m$
and $P_t^\mu$ to yield the stable and unstable resonances, respectively.
The main results are presented in section 4, for different parameter values about the crisis,
allowing to monitor the changes in the resonances during the crisis and to analyze
whether these changes can be observed from the dynamics on the attractor alone.
A summary and discussion is given on the implications of these results
regarding the possibility of designing early-warning indicators of chaotic attractor crises,
the effect of the addition of noisy perturbations as well as the viability of response theory
close to the crisis.
In Appendix \ref{sec:transfer}, we recapitulate some concepts
of ergodic theory of dynamical systems that we deem relevant for the interpretation
of our results ; The expert reader might want to skip this material.
Appendix \ref{sec:robust} gives evidence that the numerical results of section
\ref{sec:results} are robust to the resolution, sampling and transition time.

\section{Attractor crisis in the Lorenz system}
\label{sec:crisis}

Here, we summarize some general properties of the Lorenz flow, as well as the bifurcations
that will be important for the rest of this study.

\subsection{Dissipativity and boundedness}
\label{sec:dissip}

The Lorenz flow $\Phi_t$, with time $t$ in $\mathbb{R}$,
is generated by the following set of Ordinary Differential Equations (ODEs),
\begin{align}
	\begin{cases}
		\dot{x} &= \sigma (y - x) \\
		\dot{y} &= x (\rho - z) - y \\
		\dot{z} &= xy - \beta z
	\end{cases}
	\quad (x, y, z) \in \mathbb{R}^3,
	\label{eq:Lorenz}
\end{align}
where $\rho, \sigma$ and $\beta$ are positive parameters
and the dot indicates differentiation with respect to time.
In this study, the parameters $\sigma$ and $\beta$ are set to classical values
of $10$ and $8 / 3$, respectively.
On the other hand, the Rayleigh number $\rho$ will be varied
from $0$, for which all trajectories converge to the stationary point at the origin,
to the classical value of $28$, at which Lorenz obtained the celebrated "butterfly" attractor.

The ODE \eqref{eq:Lorenz} is invariant with respect to the change of variable
$(x, y, z) \to (-x, -y, z)$, so that to each solution corresponds another symmetrically related one.
The Jacobian of the vector field $F$ of the right-hand side of \eqref{eq:Lorenz} is given by,
\begin{align}
	DF(x, y, z) = 
	\begin{pmatrix}
		-\sigma	& \sigma	& 0 \\
		\rho - z	& -1		& -x \\
		y		& x		& -\beta
	\end{pmatrix},
	\quad (x, y, z) \in \mathbb{R}^3.
	\label{eq:Jacobian}
\end{align}
The contraction rate of volumes under $F$ is given 
by its divergence,
\begin{align}
	\mathrm{div}~F = \mathrm{trace}~DF = -1 - \sigma -\beta,
	\label{eq:volContraction}
\end{align}
It is constant and, for the values of $\sigma$ and $\beta$ considered here,
negative so that volumes contract uniformly in phase space.
The vector field $F$ is therefore said to be dissipative.

From the quadratic nature of the equations \eqref{eq:Lorenz},
it was shown in \cite{Lorenz1963a} (see also~\cite{Lorenz1979b})
that solutions initiated in a specific ellipsoid remain there forever.
In particular, for $\sigma = 10$ and $\beta = 8 / 3$,
trajectories remain in the ball $R_o$ bounded by the sphere $S_o$
of radius $\rho + \sigma$ and center $(0, 0, \sigma + \rho)$.
This probability together with that of positive volume contraction
ensures that trajectories initialized inside $R_o$ remain in $R_o$
and must converge to a non-wandering set of Lebesgue measure 0.
The sphere $S_o$ is used in sections \ref{sec:approxSpectrum} and \ref{sec:results}
to bound the domain on which the transfer operators are approximated.

\subsection{Route to chaos in the Lorenz flow}

In this section, we briefly describe the series of local and global bifurcations
leading the chaotic attractor crisis of interest (e.g.~\cite{Guckenheimer1983,Kuznetsov1998}
for references on bifurcation theory).
These results are summarized in the bifurcation diagram in figure \ref{fig:diagram}.
They are not new and should rather be considered as a reproduction
of those discussed in \cite{Sparrow1982}.
However, they will be useful to interpret later results discussed
in section \ref{sec:results}.
\begin{figure}[ht]
	\centering
	\includegraphics[width=\textwidth]{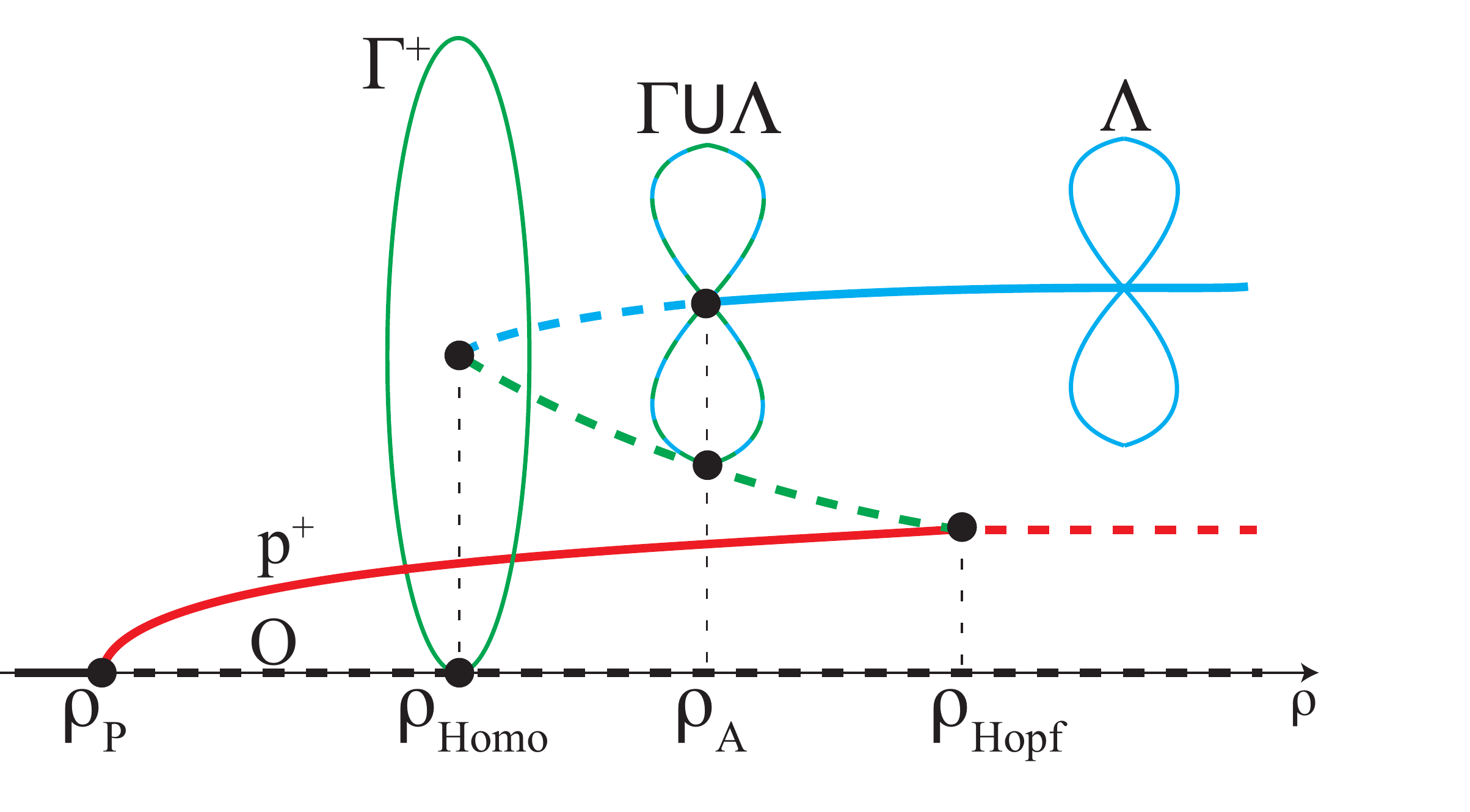}
	\caption{Schematic bifurcation diagram of the Lorenz flow,
	 for $0 \le \rho \le 28$ and with fixed $\sigma = 10$ and $\beta = 8 / 3$.
	 The branch of the stationary points $O$ and $p^+$, the periodic orbit $\Gamma^+$
	 and the chaotic set $\Lambda$ are marked in black, red, green and cyan, respectively.
	 Stable (unstable) branches are marked by a plain (dashed) line.
	 The various dots at $\rho_P = 1 , \rho_\mathrm{Homo} \approx  13.93$,
	 $\rho_A \approx 24.06$ and $\rho_\mathrm{Hopf} \approx 24.74$
	indicate the position of the pitchfork bifurcation, the homoclinic biurcation,
	the boundary attractor crisis and the Hopf bifurcation, respectively.}
	\label{fig:diagram}
\end{figure}

We start by increasing the control parameter $\rho$ from zero.
For $\rho = 0$, the stationary point $O$ at the origin is stable.
Solving the algebraic equation $F(x) = 0$ analytically
reveals that two symmetric stationary points $p^+$ and $p^-$
other than $O$ exist for $\rho > \rho_P = 1$.
The eigenvalues of the Jacobian $DF$ evaluated at each point reveals
that $O$ loses stability at $\rho_P$ while the points $p^\pm$ are stable.
A pitchfork bifurcation thus occurs at $\rho_P$,
as is illustrated in the left of the diagram in figure \ref{fig:diagram},
on which the branch of the stationary point $O$ is represented in black,
while that for $p^+$ is represented in red.

It is then numerically found that the Jacobian evaluated at $p^\pm$
quickly acquires a complex conjugate pair of eigenvalues.
The latter eventually cross the imaginary axis at $\rho_{\mathrm{Hopf}} \approx 24.74$.
This loss of stability of the two stationary points is associated with two symmetric subcritical Hopf bifurcations.
That is, two symmetric unstable periodic orbits, denoted $\Gamma^{+}$ and $\Gamma^{-}$,
merge with the stationary points $p^+$ and $p^-$, respectively.
The branch of the orbit $\Gamma^+$ is represented in green in the diagram in figure \ref{fig:diagram}.
\begin{figure}[ht]
	\centering
	\begin{subfigure}{0.43\textwidth}
		\includegraphics[width=\textwidth]{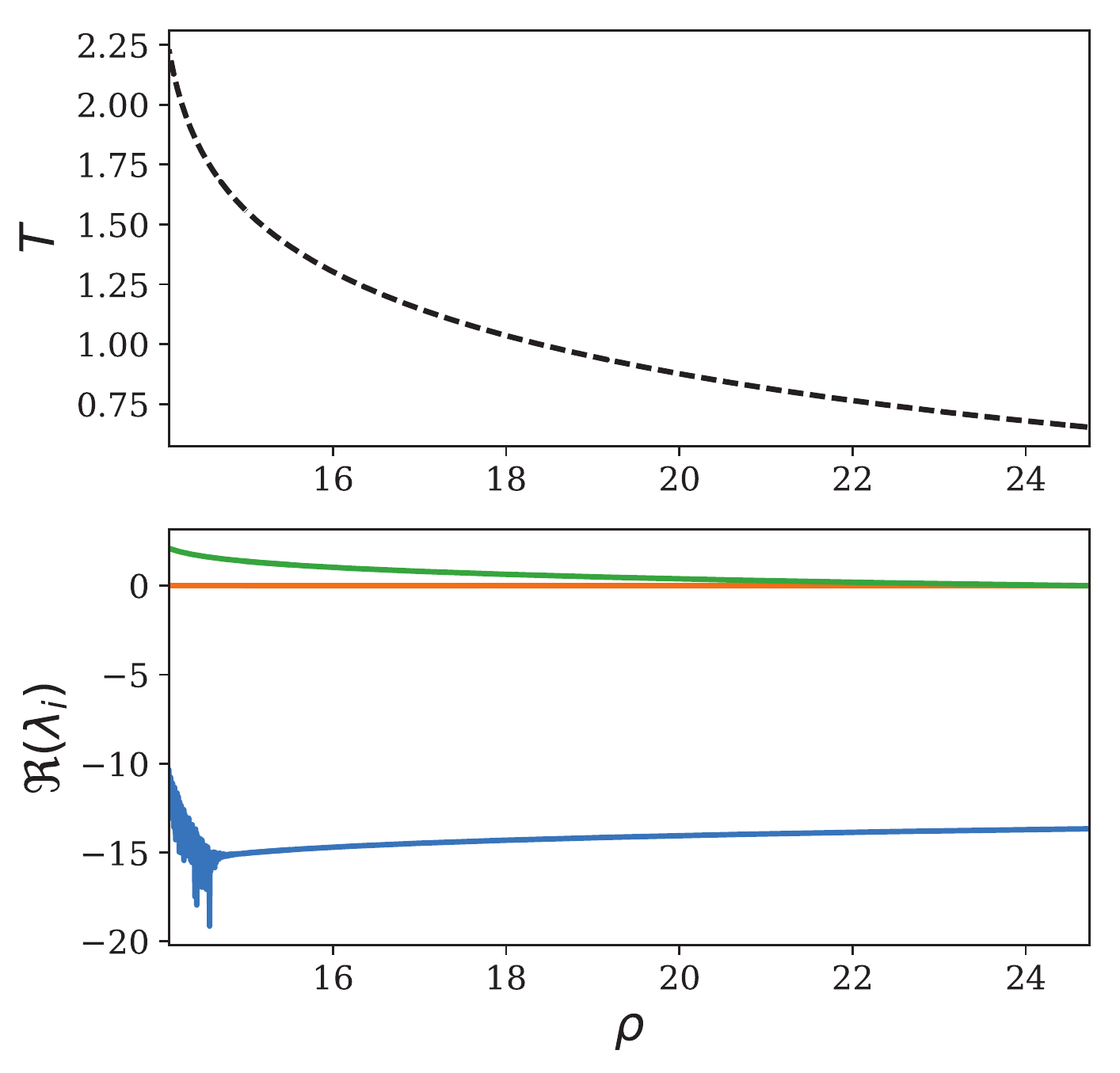}
	\end{subfigure}
	\begin{subfigure}{0.55\textwidth}
		\includegraphics[width=\textwidth]{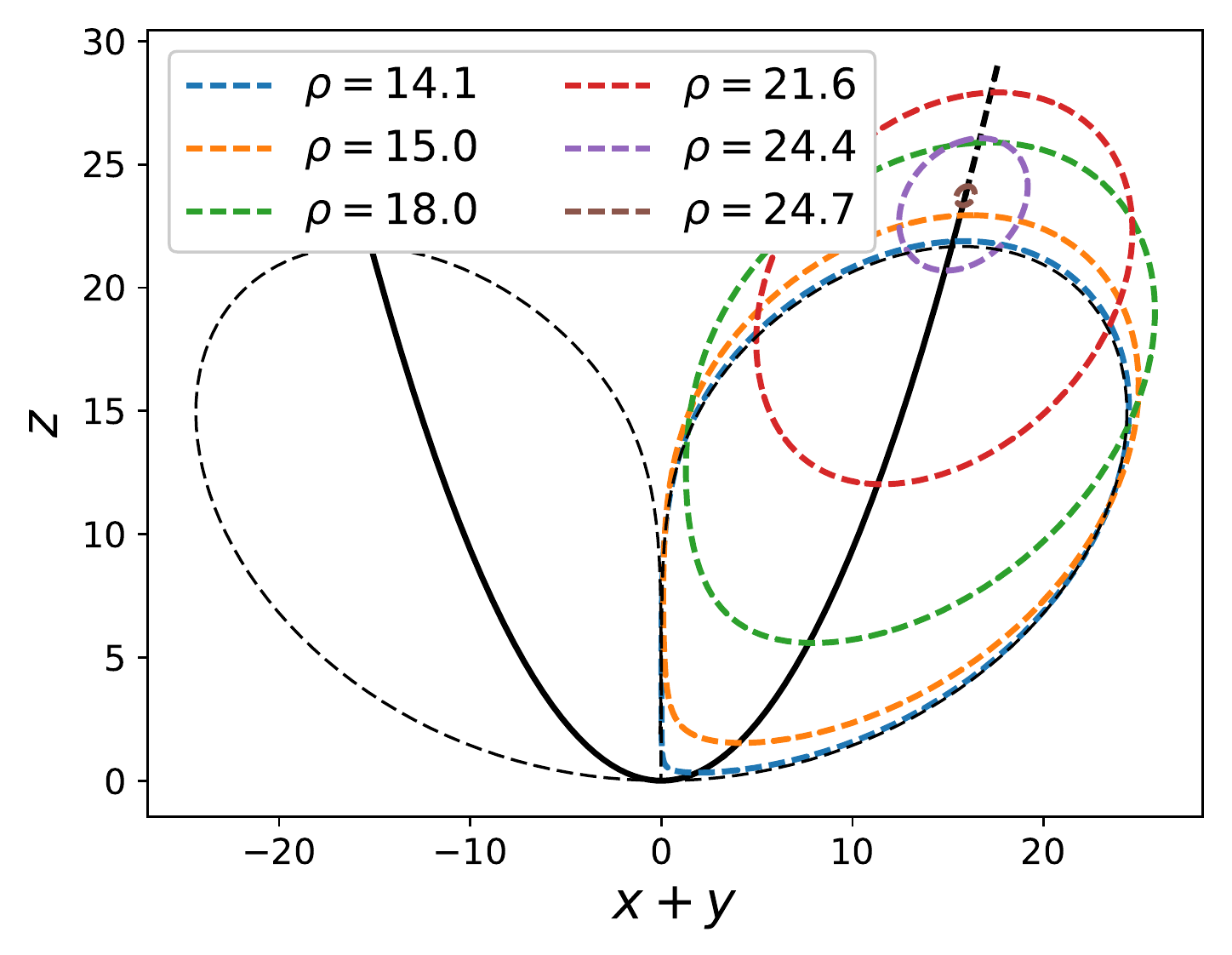}
	\end{subfigure}
	\caption{Continuation of the unstable periodic orbit $\Gamma^+$.
	The upper and lower left panels  respectively represent the period and
	the real parts of the Floquet exponents of $\Gamma^+$ versus $\rho$.
	The right panel represents, in the $(x + y, z)$ plane , the orbit $\Gamma^+$
	in thick dashed colored lines and for different values of $\rho$.
	The thick black line represents the two asymmetric stationary points $p^+$ and $p^-$
	versus $\rho$.
	Finally, the thin black dashed line represents a numerical integration of the
	homoclinic orbit at $\rho_{\mathrm{Homo}} \approx 13.9265$.}
\label{fig:poCont}
\end{figure}

Starting from $\rho = \rho_\mathrm{Hopf}$,
we now \emph{decrease} $\rho$ to continue the unstable periodic orbits $\Gamma^\pm$
as they emerge from the stationary points $p^\pm$.
The two orbits $\Gamma^\pm$
are continued numerically via pseudo-arclength continuation
(see e.g.~\cite[Chap.~10]{Kuznetsov1998}), yielding a numerical integration of the orbit,
its period $T$, as well as the fundamental matrix $D \Phi_T$ at time $T$.
The latter allows to calculate the Floquet exponents characterizing the stability
of the limit cycle \cite{Guckenheimer1983}
as the complex logarithm of the eigenvalues of $D \Phi_T$ divided by $T$ \cite{Hartman1964}.
The results of this continuation are represented in figure \ref{fig:poCont}.
On the upper left panel, one can see that the period of the limit cycles is increasing rapidly
as $\rho$ is decreased below the Hopf bifurcation.
The real parts of the Floquet exponents are represented in the lower left panel.
The Floquet exponent zero is associated with the direction of the flow,
while one of the exponents is positive and the last is negative,
indicating that the periodic orbits $\Gamma^\pm$ are unstable and of saddle type.
As expected from normal form theory \cite{Guckenheimer1983,Kuznetsov1998},
the instability of the periodic orbits grows away from the Hopf bifurcation,
i.e. as $\rho$ is decreased from $\rho_{\mathrm{Hopf}}$ and their radius increases.
This can be seen on the right panel of the same figure \ref{fig:poCont},
where the periodic orbits for different values of $\rho$ are represented in the $(x + y, z)$ plane.

The increase in the period of the periodic orbits $\Gamma^\pm$
and the decrease of their distance to the stationary point $O$
at the origin give numerical evidence that the periodic orbits eventually collide with $O$.
It is thus expected that, at a particular value $\rho = \rho_{\mathrm{Homo}}$,
the union of each periodic orbit with $O$ yields a homoclinic orbit on which
$O$ is an asymptotic point for both forward and backward trajectories.
The numerical method used here does not permit the continuation of the periodic orbits
all the way to the homoclinic bifurcation.
However, direct numerical integration of a trajectory starting in the neighborhood of the stationary
point allows to get an approximation of the homoclinic orbit.
The latter is obtained at $\rho = 13.93 \approx \rho_{\mathrm{Homo}}$,
in agreement with \cite{Sparrow1982}.
It is plotted by a thin dashed black line on the right panel of figure \ref{fig:poCont}
and is also represented by the green ellipse in the diagram in figure \ref{fig:diagram}.
This homoclinic bifurcation has been studied in the geometric Lorenz system
\cite{Guckenheimer1979,Guckenheimer1983} where a nontrivial invariant set $\Lambda$,
with the structure of a singular horseshoe \cite[Chap.~3]{Araujo},
is known to emerge from its unfolding, i.e. for $\rho > \rho_{\mathrm{Homo}}$.
An infinity of unstable periodic orbits are embedded in this horseshoe,
including the two simple periodic orbits $\Gamma^\pm$.
This set is thus responsible for transient chaos \cite{Kaplan1979a}, but is not yet attracting
for $\rho$ close to $\rho_{\mathrm{Homo}}$.

\subsection{Attractor crisis}

The bifurcation of interest in this study occurs as $\rho$ is \emph{increased} from $\rho_\mathrm{Homo}$,
when the chaotic invariant set $\Lambda$ becomes an attractor
\footnote{This attractor must be only partially hyperbolic, since it contains the singularity
at the origin, which prevents the stable and unstable manifolds to be continuous.}.
The study of the one-dimensional Lorenz map \cite{Lorenz1963a,Sparrow1982,Guckenheimer1983},
gives evidence that this crisis occurs for $\rho = \rho_A \approx 24.06$.
That is, for $\rho_{\mathrm{Homo}} < \rho < \rho_A$,
the chaotic set $\Lambda$ is of saddle type,
while for $\rho > \rho_A$ and $\rho$ not too large, $\Lambda$ is an attractor.
The branch of the chaotic set $\Lambda$ is represented in cyan in the diagram in figure \ref{fig:diagram},
with $\Lambda$ represented by a lemniscate.
Importantly, for $\rho_\mathrm{Homo} < \rho < \rho_\mathrm{Hopf}$,
the strange attractor $\Lambda$ coexists with the two unstable periodic orbits $\Gamma^\pm$
as well as with two other attractors, the stable stationary points $p^\pm$.
Following \cite{Sparrow1982}, an informal phenomenological description
of this crisis is supported here by numerical integration.

\begin{figure}[ht]
	\centering
	\begin{subfigure}{0.48\textwidth}
		\includegraphics[width=\textwidth]{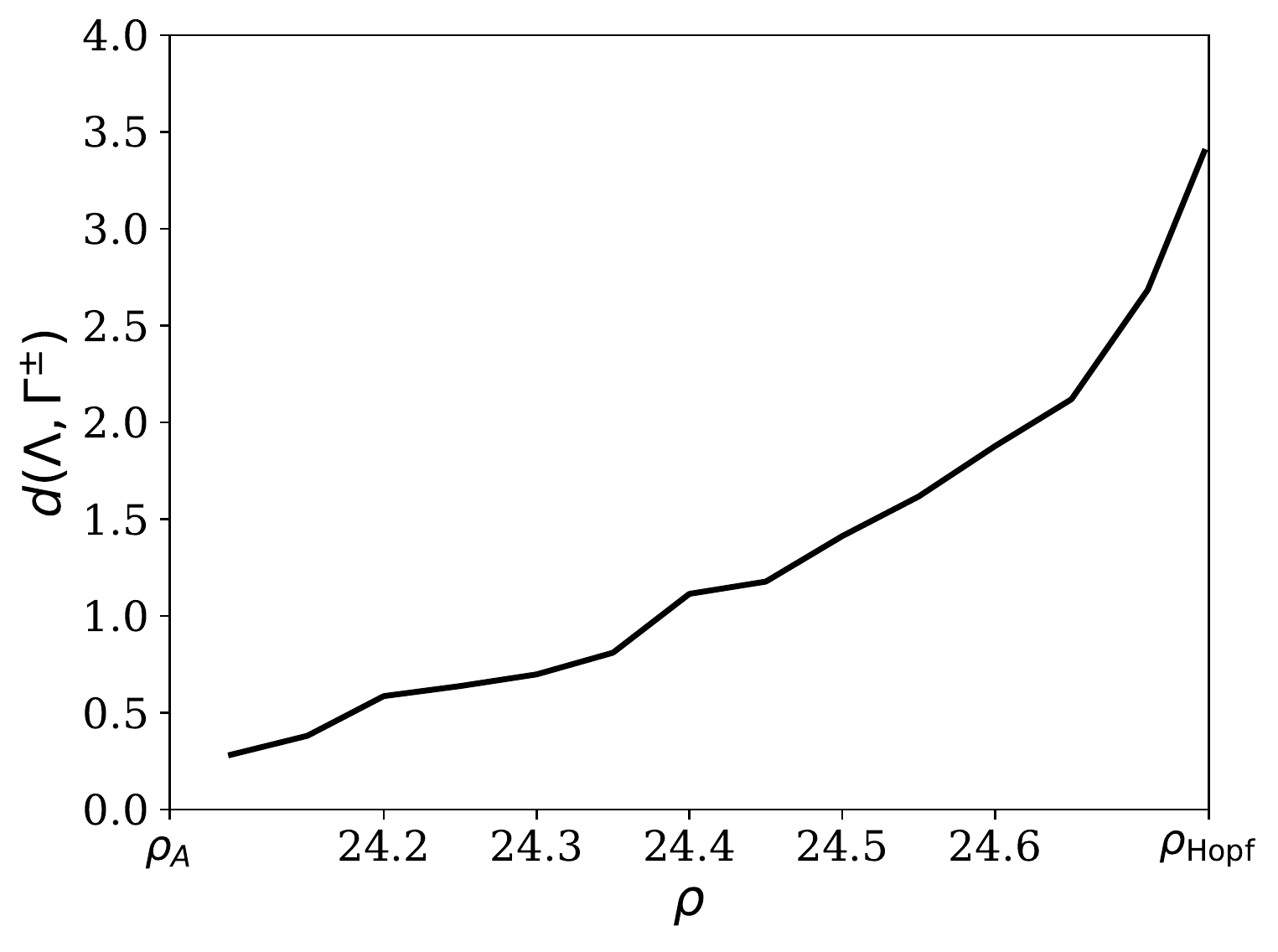}
	\end{subfigure}
	\begin{subfigure}{0.48\textwidth}
		\includegraphics[width=\textwidth]{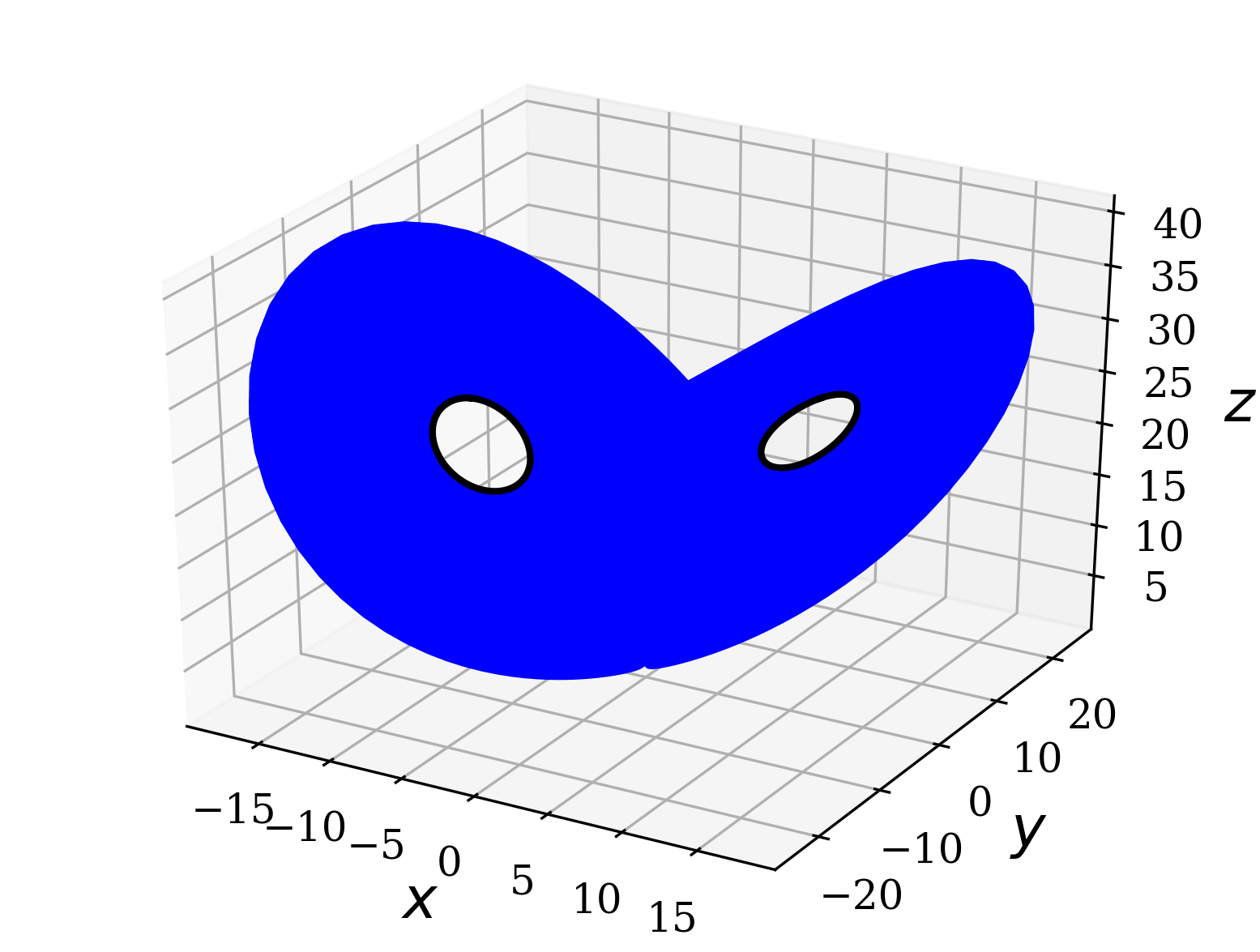}
	\end{subfigure}
	\caption{a) Minimal distance between a long aperiodic orbit on $\Lambda$
	and the periodic orbits $\Gamma^\pm$ versus $\rho$.\\
	b) Long aperiodic orbit on $\Lambda$ (blue line)
	and periodic orbits $\Gamma^\pm$ (thick black line) for (a) $\rho = 24.1$.
	The orbits have been integrated with a Runge-Kutta scheme of order 4
	with a time step of $10^{-3}$ time units, for $10^5$ time units.}
	\label{fig:poChaotic}
\end{figure}

Figure \ref{fig:poChaotic}.(a) represents the minimal Euclidean distance between
an aperiodic orbit of the chaotic set $\Lambda$
and the periodic orbits $\Gamma^{\pm}$, versus $\rho$.
One can see that, close to the crisis, this numerical calculation of the distance is close to zero.
This is more obvious from figure \ref{fig:poChaotic}.(b),
in which a long integration of an aperiodic orbit is represented in blue
together with the periodic orbits $\Gamma^\pm$, in black, for $\rho \approx \rho_A$.
In fact, below the crisis, for $\rho < \rho_A$, the periodic orbits $\Gamma^\pm$
are embedded in the chaotic saddle $\Lambda$ \cite{Sparrow1982}.
On the other hand, this distance increases as $\rho$ is increased from $\rho_A$
due to the shrinkage of $\Gamma^\pm$ as the Hopf bifurcation is approached.
The chaotic set $\Lambda$ thus appears to become on attractor
only once the periodic orbits $\Gamma^\pm$ have left it.
Since all the orbits in $\Lambda$ for $\rho < \rho_A$ cannot be mapped 
to those of $\Lambda$ for $\rho > \rho_A$, the flow on this set is not
structurally stable at the boundary crisis; there is a genuine global bifurcation at $\rho = \rho_A$.

\begin{figure}[ht]
	\centering
	\begin{subfigure}{0.48\textwidth}
		\includegraphics[width=\textwidth]{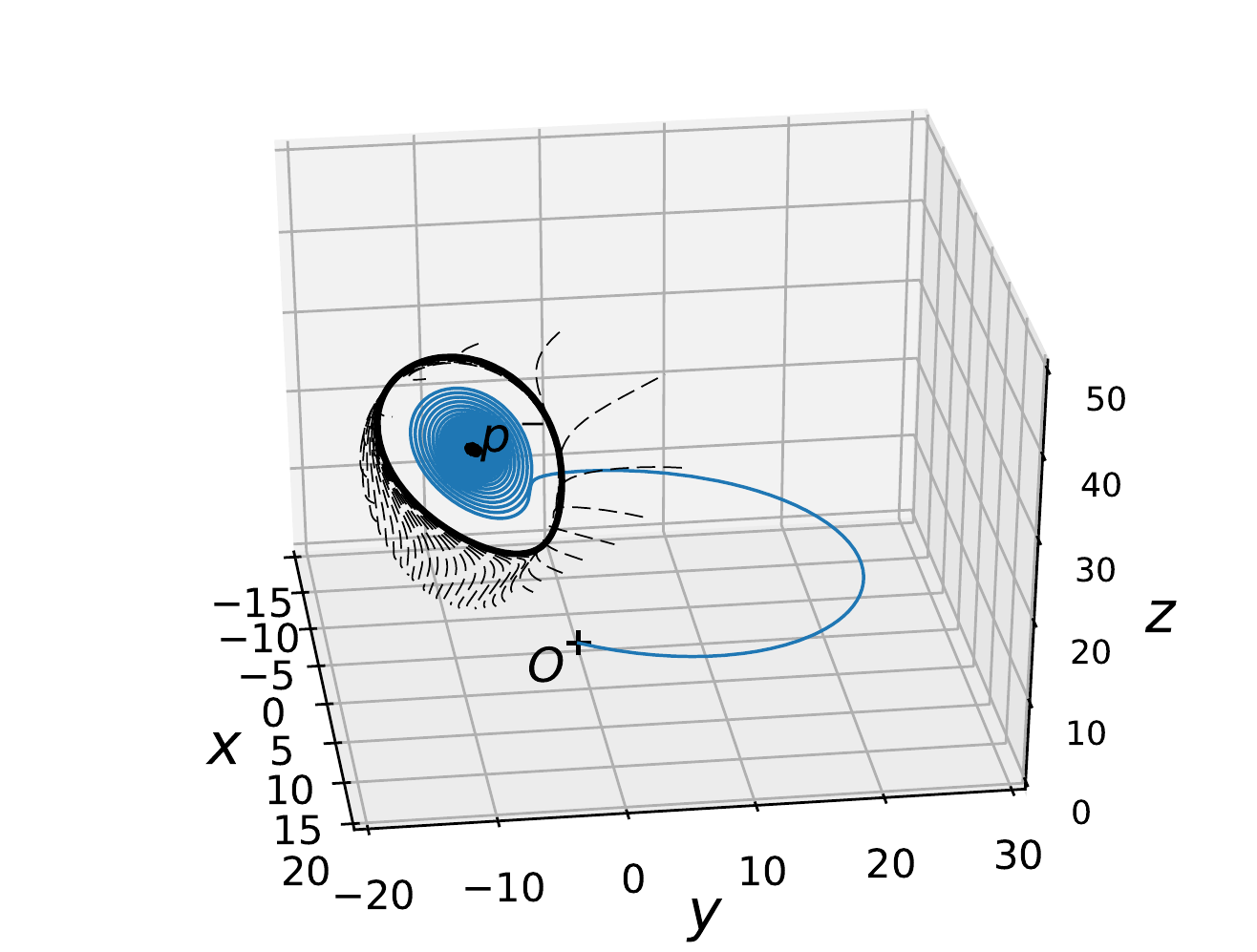}
		\caption{$\rho = 20.$}
	\end{subfigure}
	\begin{subfigure}{0.48\textwidth}
		\includegraphics[width=\textwidth]{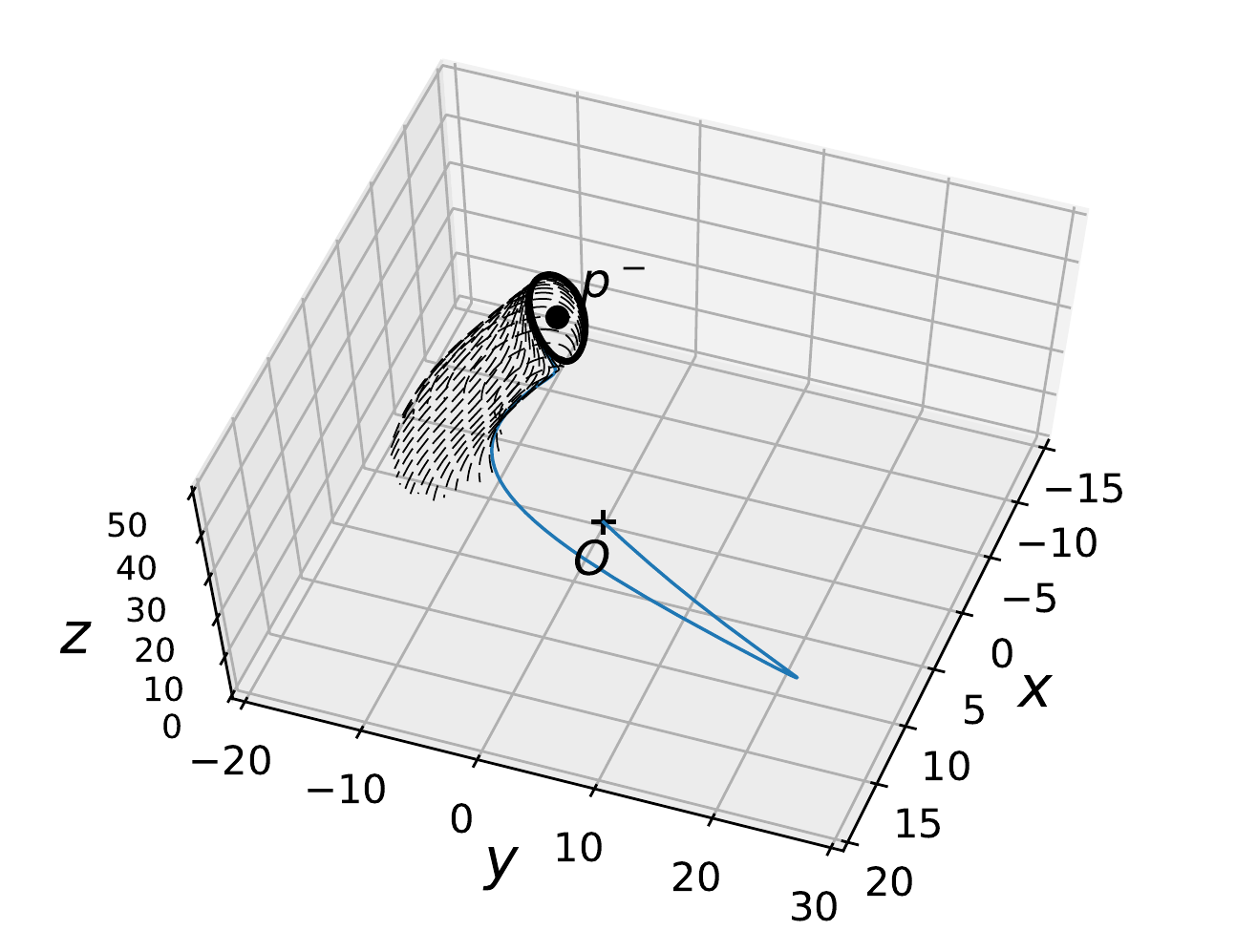}
		\caption{$\rho = 24.06$}
	\end{subfigure}
	\begin{subfigure}{0.48\textwidth}
		\includegraphics[width=\textwidth]{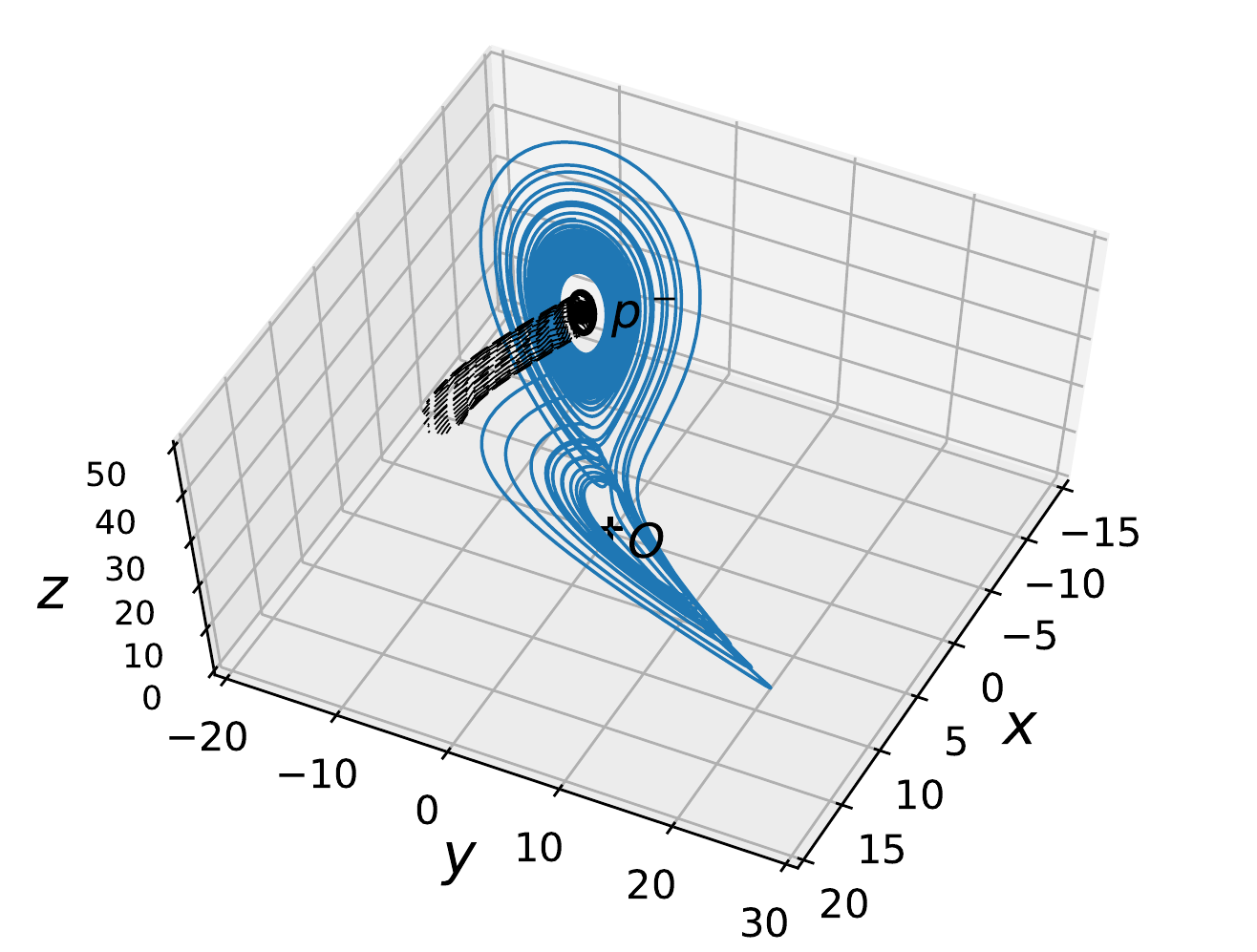}
		\caption{$\rho = 24.6$}
	\end{subfigure}
	\caption{Periodic orbit (thick black line),
	backward trajectories initialized in the neighbourhood of the periodic orbit
	along its stable manifold (dashed black lines)
	and unstable manifold of the stationary point at the origin (blue line) for
	(a) $\rho = 20.$, (b) $\rho = 24.06$ and (c) $\rho = 24.6$.}
	\label{fig:heteroCritic}
\end{figure}

This bifurcation is better understood from the geometry of the stable manifold
of the stationary point $O$.
Numerical integrations of one branch of the unstable manifold of $O$
are represented in blue in figure \ref{fig:heteroCritic},
for different values of $\rho$ about the crisis.
The periodic orbit $\Gamma^-$ is also represented by a thick black line,
together with orbits belonging to the stable manifold of $\Gamma^-$, represented by thin black dashed lines.
These numerical results suggest that:
\begin{enumerate}
\item Between the homoclinic bifurcation and the attractor crisis
(for $\rho_{\mathrm{Homo}} < \rho < \rho_A$, Fig.~\ref{fig:heteroCritic}.(a)),
the unstable manifold of $O$ is connected to the stable manifold
of the stable stationary point $p^-$ emanating from the pitchfork bifurcation,
i.e. orbits repelled by $O$ converge to $p^-$,
\item At the crisis ($\rho = \rho_A$, Fig.~\ref{fig:heteroCritic}.(b)),
as a consequence of the shrinkage of the periodic orbit $\Gamma^-$ with increasing $\rho$, ,
the trajectories along the unstable manifold of $O$ no longer convergences to $p^-$,
but instead connect to the stable manifold of the periodic orbit $\Gamma^-$,
\item After the crisis, ($\rho > \rho_A$ Fig.~\ref{fig:heteroCritic}.(c)),
the small size of $\Gamma^-$ prevents the unstable manifold of $O$
to dive into the stable manifold of $\Gamma^-$. These trajectories having
to converge to an attractor (see Sect.~ section \ref{sec:dissip}),
must wander along the chaotic set $\Lambda$, which is then attracting.
\end{enumerate}
%


\subsection{Lyapunov exponents and covariant Lyapunov vectors}
\label{sec:Lyap}

Local bifurcations of stationary points or limit cycles
are fully characterized by the crossing of the imaginary axis
by the characteristic exponents of these invariant sets,
i.e. the eigenvalues of the Jacobian for the stationary points, the Floquet
exponents for the limit cycles \cite{Guckenheimer1983,Kuznetsov1998}.
In these cases a bifurcation point corresponds to a loss of hyperbolicity
due to the crossing of the imaginary axis by one or several critical exponents.

The notion of hyperbolicity can be generalized to nontrivial sets.
A very strong form of hyperbolicity is that of \emph{uniform hyperbolicity} for which
the tangent space can be split into a one-dimensional space in the direction of the flow
and contracting and expanding spaces,
the stable and unstable spaces, respectively, with uniform decay and growth bounds,
respectively \cite[Part~4]{Katok1996}.

A weaker form of hyperbolicity can be introduced making reference to the Lyapunov exponents.
Lyapunov exponents describing the growth rate of perturbations applied to a trajectory
provide a generalization of the characteristic exponents to chaotic invariant sets
\cite{Oseledets1968,Eckmann1985}.
Nonuniform hyperbolic systems do not obey uniform hyperbolicity but admit
an invariant measure with nonzero Lyapunov exponents almost everywhere,
except for the exponent associated with the direction of the flow \cite{Barreira2002}.
In this case, the covariant vectors in the direction of which
the growth of perturbations is given by the respective Lyapunov exponents,
namely the Covariant Lyapunov Vectors (CLVs, e.g.~\cite{Kuptsov2012}).
The CLVs associated with positive (neutral, negative) Lyapunov exponents are tangent to the
unstable (center, stable) manifold along an orbit and span the
unstable (center, stable) tangent space.
In general, the angles between the manifolds can be arbitrarily small,
as opposed to the uniform hyperbolic case.

The set $\Lambda$ in the Lorenz flow is chaotic, so that it has
one positive Lyapunov exponent $\lambda_+$, one zero exponent $\lambda_0$
corresponding to the direction of the flow and one negative exponent $\lambda_-$,
which is larger in absolute value than $\lambda_+$, in order for volumes to contract.
Moreover, the stationary point at the origin belongs to $\Lambda$.
These results were first proved for the geometric Lorenz attractor \cite{Guckenheimer1979}
and later for the Lorenz flow for the classical parameter values
$\rho = 28, \sigma = 10$ and $\beta = 8 / 3$ \cite{Tucker1999}.
The Lorenz attractor is thus at most \emph{partially hyperbolic} \cite{Pesin2004}.
In fact, this type of partially hyperbolic flow with a singularity in the attractor
happens to be the prototype of robust chaotic flow in three dimensions \cite{Araujo}
and has been called \emph{singular hyperbolic}.
In the case of the Lorenz flow, the tangent space splits into a stable space, with contraction,
and a center-unstable space where, loosely speaking, weak contraction is allowed in addition to expansion.
This implies that the angle between the center-unstable space and the stable space
is uniformly bounded away from zero \cite{Pesin2004}, and so are the angles between
the stable CLV and each of the unstable and central CLVs.
However, due to the stationary point at the origin, where the center manifold
does not exist, the angle between the unstable and the central CLVs may be arbitrarily small
as this singularity is approached.

\begin{figure}[ht]
\centering
\includegraphics[width=0.47\textwidth]{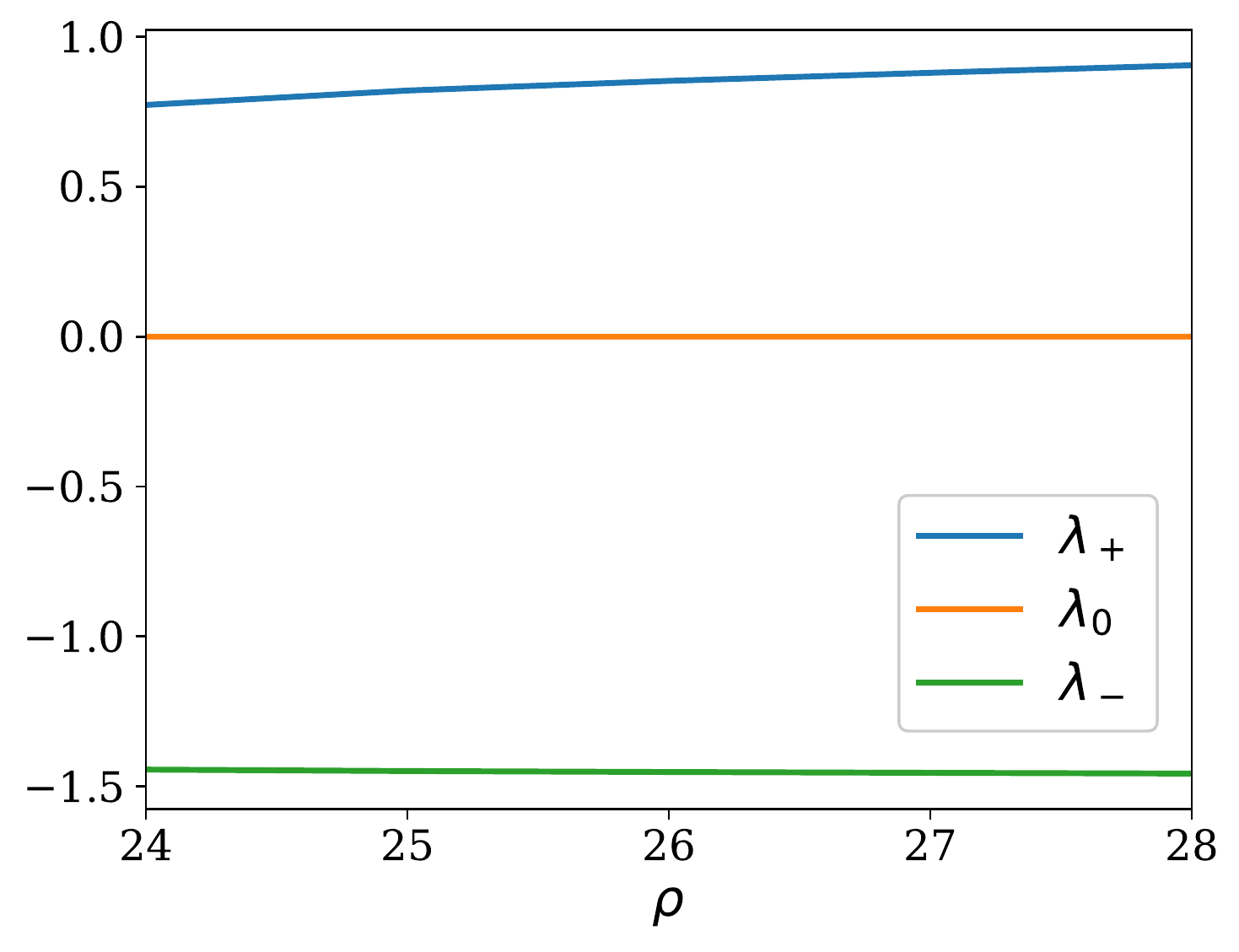}
\caption{Lyapunov exponents of $\Lambda$ versus $\rho$.
The negative Lyapunov exponent $\lambda_{-}$ (in green) is divided by a factor 10.}
\label{fig:LyapExp}
\end{figure}

Decreasing $\rho$ towards the attractor crisis at $\rho_A$,
one may wonder whether the positive Lyapunov exponent becomes negative.
However, since for $\rho < \rho_A$ a chaotic invariant set persists,
the positive Lyapunov exponent must remain positive.
Indeed, the negative value of the average volume contraction rate
implies that the Lyapunov exponents different from zero must be either both negative
or, if one of them is positive, the other must be negative and larger in absolute value.
However, if both non-zero Lyapunov exponents of the invariant set $\Lambda$ were negative
for $\rho < \rho_A$, this set would not be chaotic, leading to a contradiction.
This heuristic argument is supported by numerical estimations
\footnote{Following the algorithm described in \cite{Kuptsov2012}, also yielding the CLVs.}
of the Lyapunov exponents
of $\Lambda$ for varying values of $\rho$, represented in figure \ref{fig:LyapExp}.
One can see on this figure that the Lyapunov exponents vary only slightly close to the crisis
and that none of them cross the imaginary axis.

In addition, since a break of hyperbolicity may be due to a homoclinic tangency
where the stable and unstable manifolds are tangent,
the angle between the CLVs has been suggested in \cite{Ginelli2007} and \cite{Yang2009}
as an indicator of the degree of hyperbolicity of the dynamics.
Moreover, it has been conjectured in \cite{Beims2016} that the alignment of the CLVs could
provide a criterion to predict crises, although the latter are understood there as chaotic bursts
(in other words, an extreme fluctuation in an otherwise weakly chaotic trajectory) rather than as an attractor crisis.
However, due to the singularity in the Lorenz attractor, the angle between the unstable
and the central CLVs can be arbitrarily small even away from the crisis,
so that the loss of hyperbolicity should not provide a precursor of the crisis.
This is confirmed from the computations of the CLVs for varying $\rho$,
as well as for the angle between the unstable and the stable CLVs,
and between the center and the stable CLVs (not shown here).

\section{Approximation of the Ruelle-Pollicott resonances}
\label{sec:approxSpectrum}

The Ruelle-Pollicott resonances are given by the eigenvalues
of the generator of the semigroup of transfer operators,
governing the evolution of densities.
While analytical results on the properties of this semigroup
for chaotic systems are difficult to obtain, the relationship
between the ergodic properties of dynamical systems and
the spectrum of the semigroup of transfer operators is well known.
A summary of this is given in appendix \ref{sec:transfer}.
Here, we explain how these operators can be approximated from
transition matrices estimated from time series.
While the method is well known, we make clear the difference between two
variants allowing either for the approximation of both the stable and the unstable
resonances from the transfer operators $\mathcal{P}_t^m$
with respect to the Lebesgue measure,
or only for the approximation of the unstable resonances
from the transfer operators $\mathcal{P}_t^\mu$ with respect to the invariant measure
supported by a chaotic attractor.

Calculating the Ruelle-Pollicott resonances is a difficult task.
Explicit formulas have only been obtained for low-dimensional systems
from a trace formula or from the decomposition of the eigenvectors
on basis functions \cite{Hasegawa1992b,Gaspard1992,Gaspard1995,Gaspard2001a}.
For high-dimensional systems, to our knowledge,
only qualitative results limited to uniformly hyperbolic systems
have been obtained regarding the distribution of the resonances
in the complex plane \cite{Gouezel2006,Butterley2007,Faure2014}.

Numerical methods are thus required to approximate the Ruelle-Pollicott resonances
and the associated eigenvectors.
For still relatively low-dimensional systems, but with chaotic dynamics,
one approach is to calculate the spectrum of transition matrices resulting
from the projection of the transfer operators on a finite family of basis functions.
The transition probabilities can then be estimated from time series.
This Galerkin truncation with estimation is referred to as \emph{Ulam's method} \cite{ulam1964collection,Dellnitz1999}
in the literature. The method is not limited to the use of characteristic functions
and has also been referred to as the \emph{Extended Dynamical Mode Decomposition} \cite{Klus2015a,Williams2015}. In particular, Ulam's method has been applied to the Lorenz flow
in \cite{Lucarini2016} to calculate its linear response to forcing from transition matrices
approximating the transfer operators.
See also \cite{Shutte1999}, \cite{Chekroun2014} and \cite{Chekroun2016}
for generalizations to high-dimensional systems and \cite{Tantet2015} for an application
to atmospheric regimes detection.

Following this approach, the transfer operators $\mathcal{P}_t^\eta$,
defined by \eqref{eq:transfer} with respect to some probability measure $\eta$,
are projected on a truncated family $G = \{\chi_1, ..., \chi_n\}$ of orthogonal basis functions.
That is, $\left<\chi_i, \chi_j \right>_\eta = 0$ when $i \ne j$, where the scalar product is defined as
$\left<f, g\right>_\eta = \int_X f(x) g(x) \eta(dx)$ and $\eta(f) = \int_X f(x) \eta(dx)$.
Any vector of components $\mathbf{f} = (f_1, ..., f_n)$ in $\mathbb{R}^n$,
defines an observable
\begin{align*}
	f = \sum_{i = 1}^n f_i \frac{\chi_i}{\eta(\chi_i)} \quad \mathrm{such~that}
	\quad f_i = \left<f, \chi_i\right>_\eta \quad \mathrm{and} \quad \int_X f d\eta = \sum_{i = 1}^n f_i.
\end{align*}
For any such $f$, the component of $\mathcal{P}_t^\eta f$
on the basis function $\chi_j$ is then given by
\begin{align*}
	(\mathcal{P}_t^\eta f)_j = \left<\mathcal{P}^\eta_t f, \chi_j \right>_\eta
	&= \sum_{i = 1}^n f_i \frac{\left<\mathcal{P}^\eta_t \chi_i, \chi_j\right>_\eta}{\eta(\chi_i)}
	= \left(\bv{f} \bv{P}^\eta_t\right)_j \quad t \ge 0,
\end{align*}
where we have defined the transition matrix $\bv{P}^\eta_t$
with elements the normalized correlations
\begin{align}
	[\bv{P}^\eta_t]_{ij}
	&:= \frac{\left<\mathcal{P}^\eta_t \chi_i, \chi_j\right>_\eta}{\eta(\chi_i)}.
	\label{eq:correlationMatrix}
\end{align}

In this study, we will only consider families of characteristic functions $G = \{\mathbf{1}_{B_1}, ..., \mathbf{1}_{B_n}\}$
on a grid of disjoint boxes $\{B_1, ...,B_n\}$ such that $\cup_{i = 1}^n B_i \subseteq \mathcal{B}(R_o)$,
where $\mathcal{B}(R_o)$ is the Borel $\sigma-$algebra of the ball $R_o$ defined in section
\ref{sec:crisis}
\footnote{In the uniform discretization of the ball $B_o$ in spherical coordinates, in section \ref{sec:design},
different grid boxes will have different Lebesgue volumes.
The component $f_i$ of an observable $f$ being the integral of $f$ over the box $B_i$,
this component will tend to be larger for boxes of larger volume.
This does not affect the eigenvalues,  but clearly affects the eigenvectors,
where large values of the components might be associated to large boxes, ceteris paribus.}.
In this case, the correlations are simply given by the transition probabilities
\begin{align*}
	[\bv{P}^\eta_t]_{ij}
	&= \frac{\left<\mathcal{P}^\eta_t \mathbf{1}_{B_i}, \mathbf{1}_{B_j}\right>_\eta}{\left<\mathbf{1}_{B_i}\right>_\eta}
	= \frac{\eta\left(B_i \cap \Phi_t^{-1} B_j\right)}{\eta(B_i)}.
\end{align*}

Discrete approximations of the eigenvalues, eigenvectors
and adjoint eigenvectors of $\mathcal{P}_\tau$
can then be obtained from those of a transition matrix $\bv{P}^\eta_\tau$
at some time $t = \tau$.
Assuming that the family of transition matrices $\bv{P}^\eta_t, t \ge 0,$
preserves the semigroup property of $\mathcal{P}^\eta_t, t\ge 0,$
the \emph{spectral mapping theorem} \cite[Chap.~I.3,~IV]{Engel2001}
ensures that the eigenfunctions of the generator of the semigroup
are given by those of $\bv{P}^\eta_\tau$ and that the generator eigenvalues
$\lambda_k, k \ge 0,$ can be calculated from the eigenvalues
$\zeta_k(\tau)$ of $\bv{P}^\eta_\tau$ according to
\begin{align}
	\lambda_k = \frac{1}{\tau} \log \zeta_k(\tau) \label{eq:SMF}.
\end{align}
The eigenvalues of the generator, as poles of the resolvent of the semigroup,
correspond to the Ruelle-Pollicott resonances.
If the semigroup property is preserved, they should not depend on the transition
time $\tau$ and are thus more amenable to analysis
\footnote{The dependence of the $\lambda_k$'s on $\tau$ thus constitute
an important test for the quality of the approximations. This is discussed in appendix \ref{sec:robust}.}.
Before that, however, it is necessary to estimate the transition probabilities in $\bv{P}^\eta_\tau$.

\subsection{Estimation of $\mathcal{P}^m_t$ from short time series sampling the phase space}
\label{sec:estimPm}
In the original version of Ulam's method (see e.g.~\cite{Dellnitz1999}),
the transition probabilities are estimated from an ensemble of $N_s$ short time series
$x^{(i)}_t, 0 \le t \le \tau$, $1 \le i \le N_s$, integrated numerically, with initial states
sampling a given volume in phase space ($R_o$, for the Lorenz flow).
The transition probabilities are then estimated by
\begin{align}
	[\bv{P}^m_\tau]_{ij}
	= \frac{\#\{x^{(i)}_0 \in B_i, x^{(i)}_{\tau} \in B_j\}}{\#\{x^{(i)}_0 \in B_i\}},
	\label{eq:MLEm}
\end{align}
Because the initial states uniformly sample the Lebesgue measure $m$ (restricted to $R_o$),
the transition matrix $\bv{P}^m_\tau$ gives an approximation
of the transfer operator $\mathcal{P}^m_\tau$, defined in \eqref{eq:transfer}
as the dual of the Koopman operator $\mathcal{U}^m_\tau$
with respect to the scalar product induced by $m$.

\subsection{Estimation of $\mathcal{P}_t^\mu$ from a long time series on the attractor}
\label{sec:estimPmu}
While the semigroup of transfer operator $\mathcal{P}_t^m, t \ge 0$
encapsulates all the information necessary to propagate densities in phase space,
we now take the point of view of an observer of a system in a statistical steady-state.
The latter has then only access to time series of trajectories on the attractor and sampling
the physical measure $\mu$ (the statistical steady-state) associated to it
(see remarks~\ref{rmk:physical} and \ref{rmk:physicalCorr}).
In particular, according to the formula \eqref{eq:corrTime},
this is all one needs to calculate the correlation function \eqref{eq:defCorr}.

It turns out that this information is encapsulated 
in the semigroup $\mathcal{P}_t^\mu, t \ge 0$,
defined in \eqref{eq:transfer} as the dual of the Koopman semigroup $\mathcal{U}^\mu_t$
with respect to the scalar product induced by $\mu$.
It follows that the transition matrix $\bv{P}^\mu_\tau$
approximating the transfer operator $\mathcal{P}_\tau^\mu$
can be estimated from a single long time series $x_{t_s}, 0 \le t_s \le T_\mathrm{samp}$,
of length $T_\mathrm{samp}$ and converged to the attractor $\Lambda$, according to
\begin{align}
	[\bv{P}^\mu_\tau]_{ij}
	= \frac{\frac{1}{T_\mathrm{samp}} \sum_{s = 0}^{T_\mathrm{samp}} \chi_i(x_{t_s}) \chi_j (x_{t_s + \tau})}{\frac{1}{T_\mathrm{samp}} \sum_0^{T_\mathrm{samp}} \chi_i (x_{t_s})}
	= \frac{\#\{x_{t_s} \in B_i, x_{t_s + \tau} \in B_j\}}{\#\{x_{t_s} \in B_i\}}.
	\label{eq:MLEmu}
\end{align}
The second equality in \eqref{eq:MLEmu} is valid only for the case
of characteristic basis functions so that the transition probabilities
are given by the number of times the time series transits
from one box to another in a time $\tau$ normalized by the number
of trajectories in the initial box.
In practice, the time series is sampled at a finite rate for a finite time,
so that the time means yield the Maximum Likelihood Estimator (MLE, see e.g~\cite{Billingsley1961})
of the transition probabilities.

In this study, the invariant measure $\mu$ of interest is that supported
by the chaotic set $\Lambda$, so that it may only be physical when $\Lambda$ is attracting,
i.e. after the crisis, for $\rho > \rho_A$.
%

\subsection{Experiment design}
\label{sec:design}

In order to discuss the changes in the evolution of statistics during the crisis,
in the next section \ref{sec:results},
the spectrum of both semigroups $\mathcal{P}_t^m$ and $\mathcal{P}_t^\mu$ are approximated
via the methods presented in the previous sections \ref{sec:estimPm} and \ref{sec:estimPmu}.
Before that, let us summarize the choice of the numerical parameters used to obtain these results.
This choice is based on careful robustness tests given in appendix \ref{sec:robust}
and can be divided into five items:
\begin{enumerate}
\item \emph{Numerical integration}: we use a Runge-Kutta sheme of order four
with a time step of $10^{-4}$ time units.
\item \emph{Grid}: the ball $R_o$ (see Sect.~\ref{sec:crisis}) is discretized into $3.2 \cdot 10^{7}$ boxes
by dividing each spherical coordinate $(r, \theta, \phi)$ in 400, 200 and 400 intervals
of the same length, respectively.
\item \emph{Transition time}: the generator eigenvalues were calculated from transition matrices estimated for
a transition time $\tau = 0.05$ time units.
\item \emph{Number and length of trajectories}: $6.4 \cdot 10^{9}$ trajectories of length
the transition time $\tau$
and uniformly sampling $R_o$ were integrated in order to estimate $\bv{P}^m_\tau$,
following section \eqref{sec:estimPm}.
To estimate $\bv{P}^\mu_\tau$, following section \ref{sec:estimPmu},
24 trajectories of length $T_\mathrm{samp} = 1 \cdot 10^5$ where integrated,
with a spinup of $10^4$ time units removed.
\item \emph{Eigenproblem solver}: to solve the eigenproblem for the transition matrices,
the block Krylov Schur algorithm implemented in the Anasazi package \cite{Baker2009}
of the Trilinos library \cite{Heroux2003} was used.
The generator eigenvalues are then calculated according to \eqref{eq:SMF}
from those of the transition matrix.
\end{enumerate}

\section{Results}
\label{sec:results}

In this main section, the generator eigenvalues,
calculated according to \eqref{eq:SMF}
and following both methodologies presented in section \ref{sec:approxSpectrum},
are analyzed for varying control parameter values $\rho$ around the attractor crisis.
The difference between the behavior of the eigenvalues approximating
the stable and the unstable resonances
will be emphasized and the consequences regarding critical slowing down discussed.

\subsection{Stable and unstable resonances from $\bv{P}^m_\tau$}
\label{sec:resultsM}

\begin{figure}[ht]
\centering
\begin{subfigure}{0.47\textwidth}
\includegraphics[width=\textwidth]{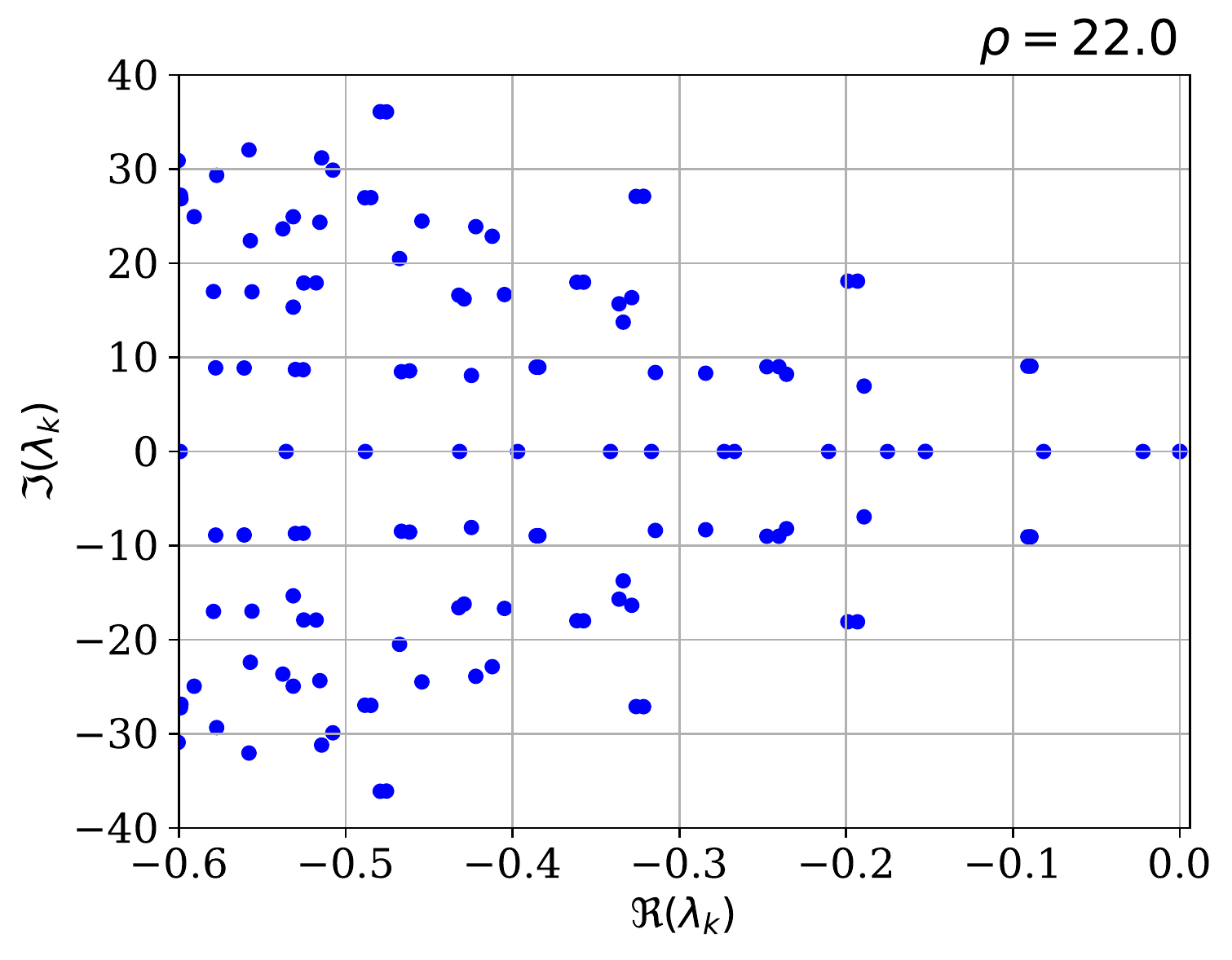}
\end{subfigure}
\begin{subfigure}{0.47\textwidth}
\includegraphics[width=\textwidth]{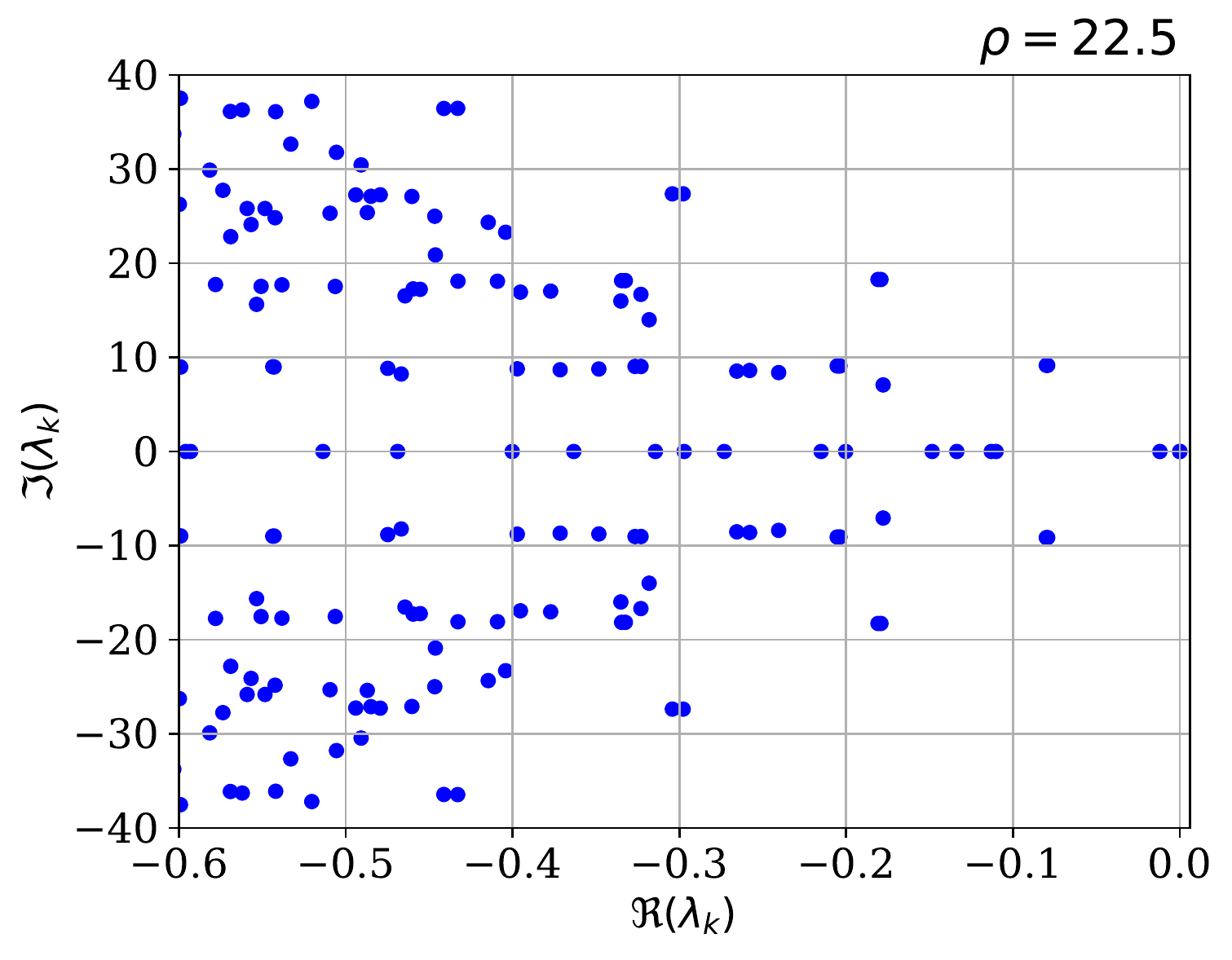}
\end{subfigure}
\begin{subfigure}{0.47\textwidth}
\includegraphics[width=\textwidth]{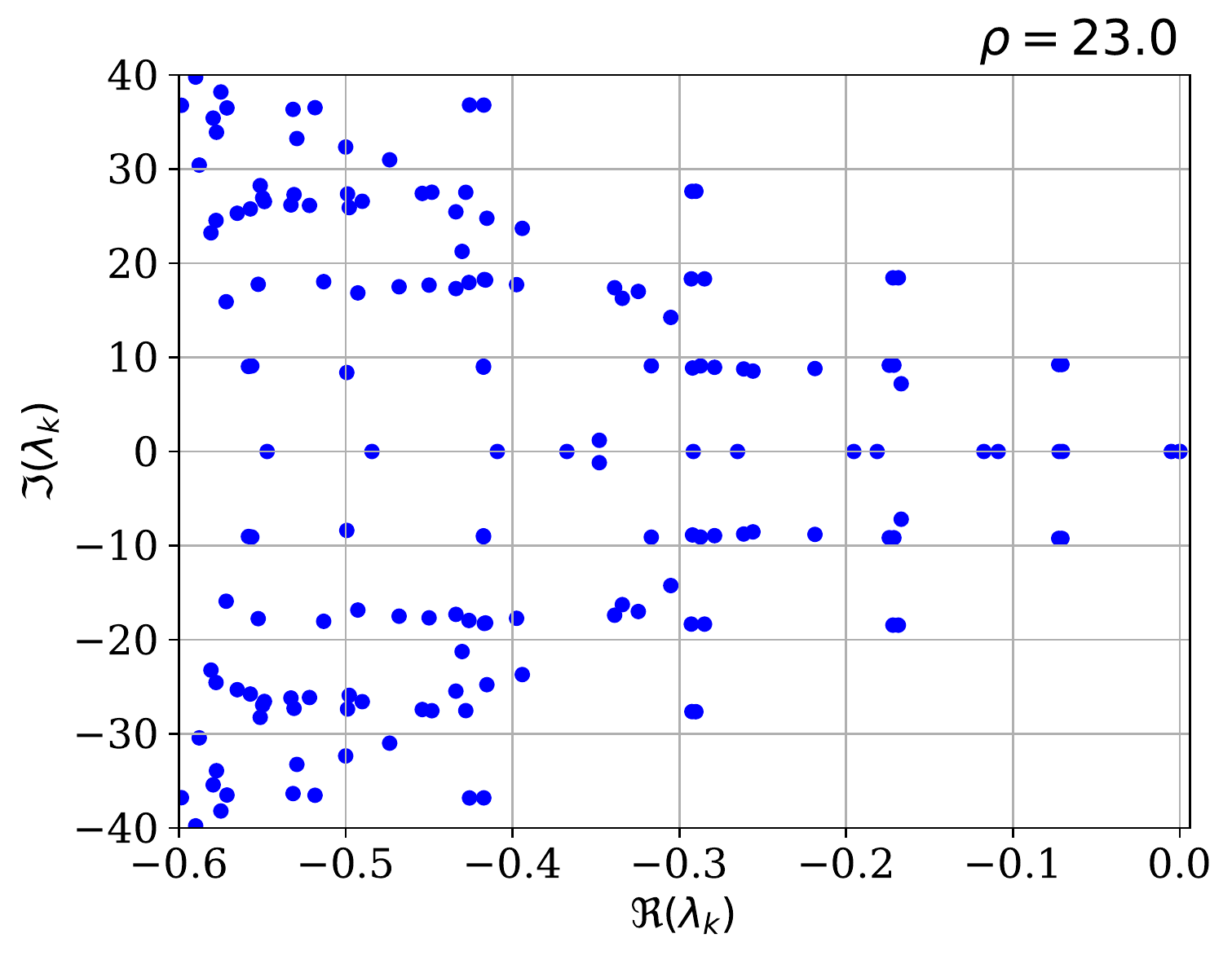}
\end{subfigure}
\begin{subfigure}{0.47\textwidth}
\includegraphics[width=\textwidth]{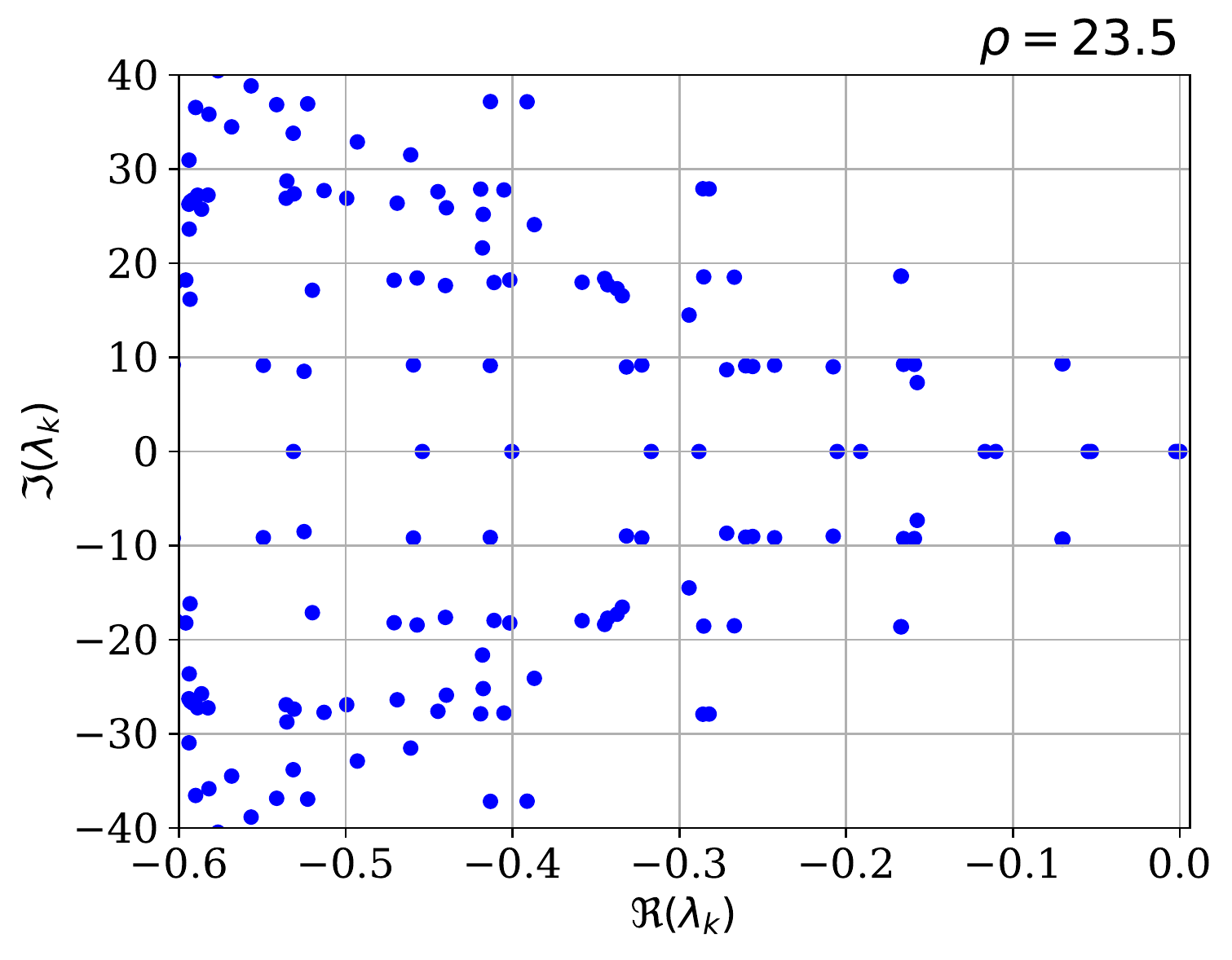}
\end{subfigure}
\begin{subfigure}{0.47\textwidth}
\includegraphics[width=\textwidth]{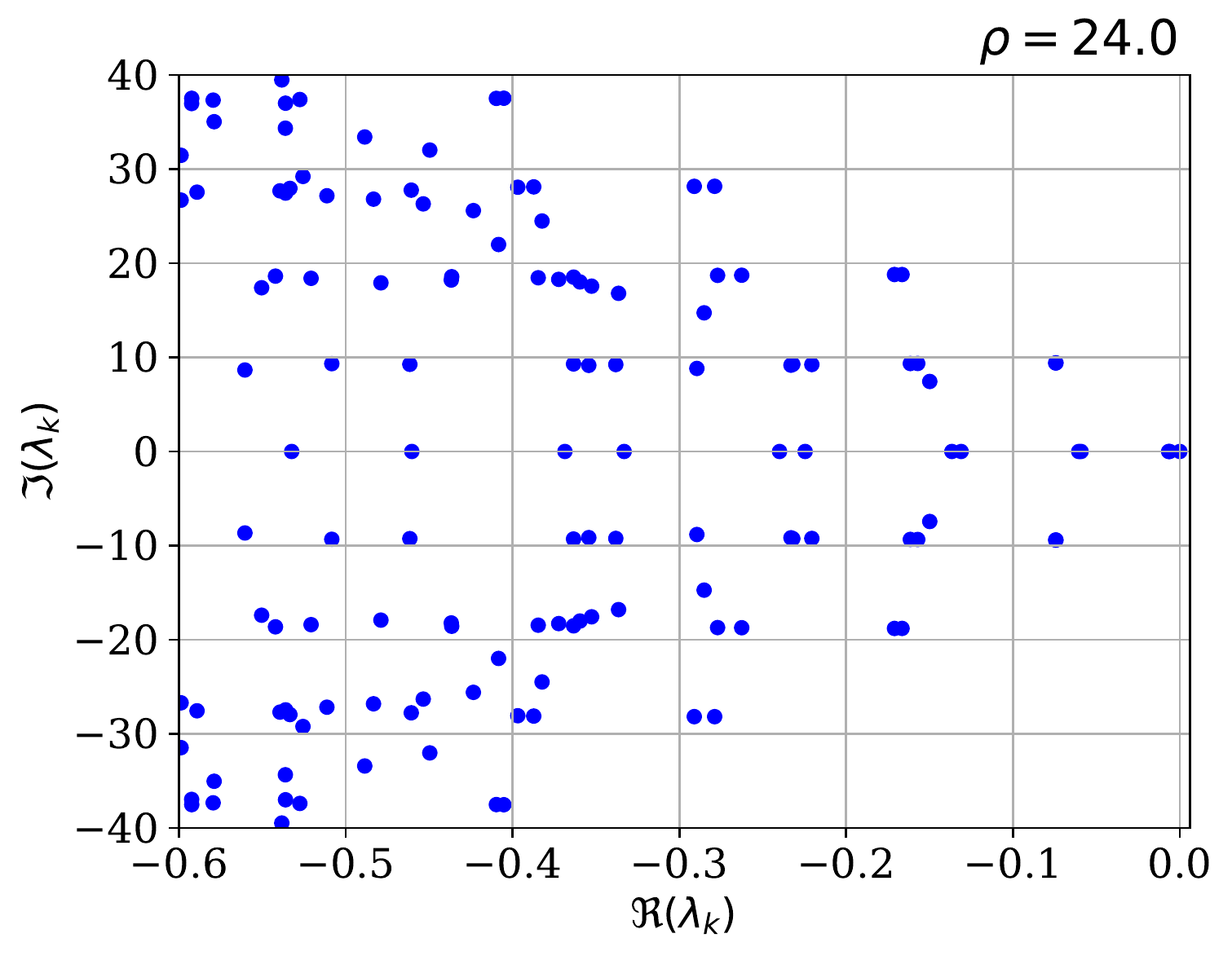}
\end{subfigure}
\begin{subfigure}{0.47\textwidth}
\includegraphics[width=\textwidth]{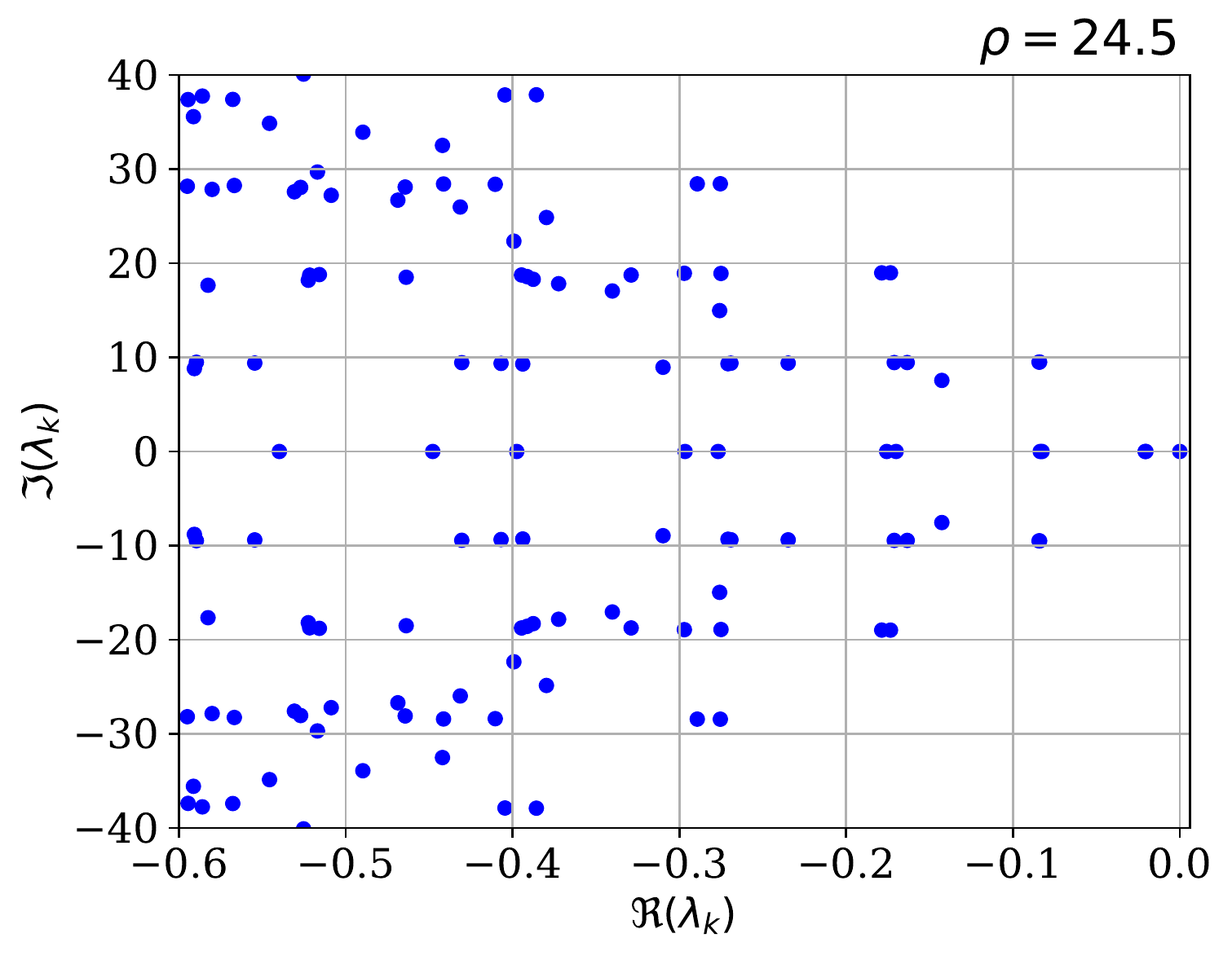}
\end{subfigure}
\caption{Leading generator eigenvalues calculated from the transition matrices $\bv{P}^m_\tau$
for different values of $\rho$.}
\label{fig:RPPhase}
\end{figure}

The approximations of both the stable and unstable resonances
from the eigenvalues of the transition matrix $\bv{P}^m_\tau$ (Sect.~\ref{sec:estimPm}),
are represented in figure \ref{fig:RPPhase}, for values of $\rho$ 
ranging from 22 (between the homoclinic bifurcation and the boundary crisis)
and 25 (after the boundary crisis).
Overall, a nontrivial arrangement of eigenvalues is found,
as expected from the presence of the chaotic set $\Lambda$.
%

\begin{figure}[ht]
\centering
\begin{subfigure}{\textwidth}
\includegraphics[width=\textwidth]{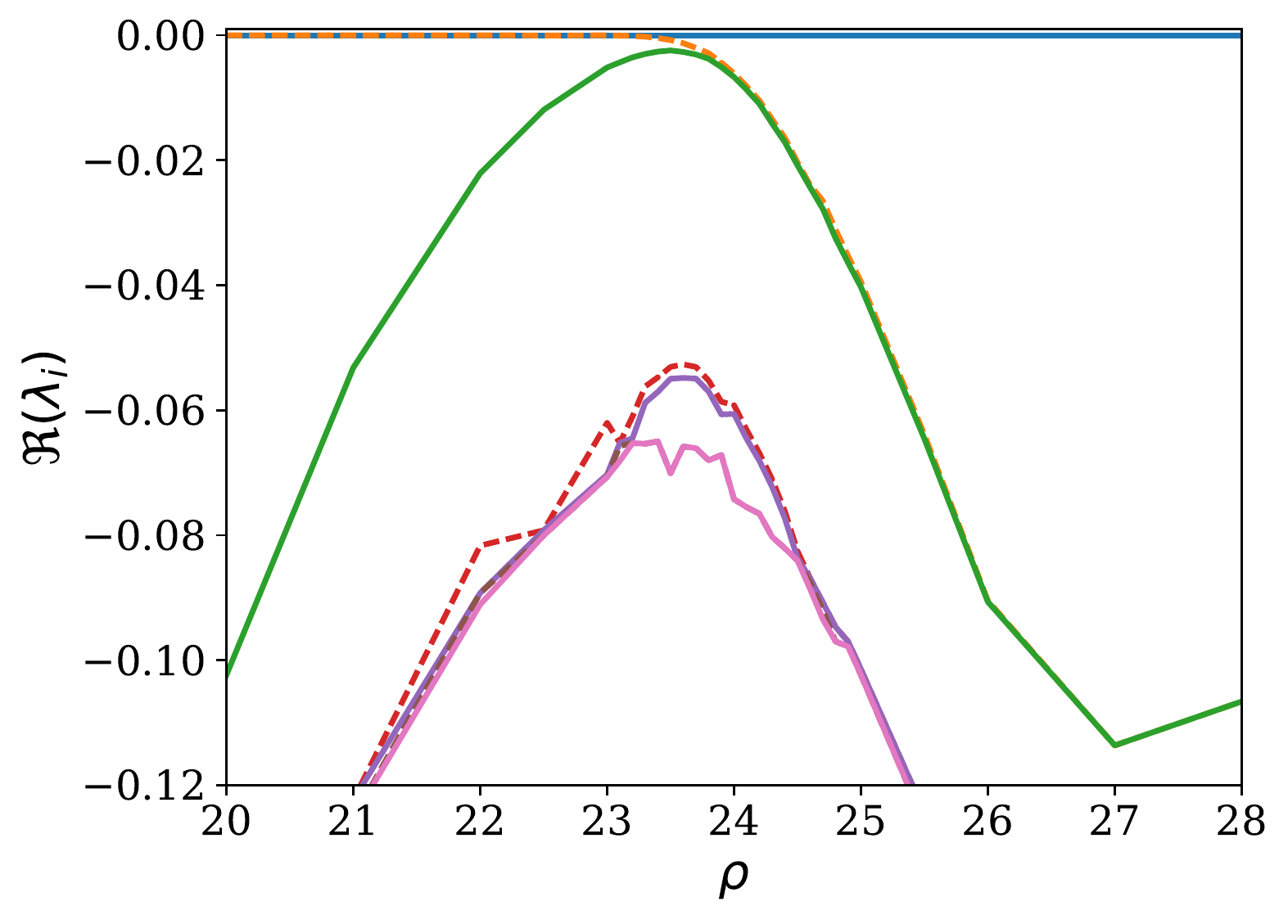}
\end{subfigure}
\caption{Real parts of the leading eigenvalues calculated from the transition matrices $\bv{P}^m_\tau$
versus $\rho$.}
\label{fig:gap}
\end{figure}

In order to focus on the generator eigenvalues closest to the imaginary axis,
expected to be affected by the crisis,
the real parts of the leading eigenvalues are represented in figure \ref{fig:gap}
versus $\rho$.
One can see that for $\rho$  smaller than $\rho_A \sim 24.06$, there are two eigenvalues zero.
As $\rho$ is increased to $\rho_A$, a third eigenvalue approaches the imaginary axis.
As $\rho$ is increased further that $\rho_A$, there is an exchange of eigenvalues,
so that only one eigenvalue remains zero and two eigenvalues get further from the imaginary axis.
There are also other eigenvalues approaching and then escaping from the imaginary axis,
although with a gap remaining finite.

Zero eigenvalues are associated with physical measures corresponding to attractors.
For $\rho < \rho_A$ there are two attractors, namely the two stationary points, so that it
does not come as a surprise that two zero eigenvalues are found.
For $\rho_A < \rho < \rho_{\mathrm{Hopf}}$, however, there are three coexisting attractors,
namely the two stationary points and the strange attractor $\Lambda$,
so that one would expect three zero eigenvalues instead of only one.
It thus appears that the discretization does not allow to resolve
the boundary of the different basin of attractions.
One can interpret the errors induced by the discretization
as numerical diffusion \cite{Froyland2011a}
hindering the distinction of the three different basins of attraction.
The three attractors are, at our level of description, merged as a result of diffusion.
\begin{figure}[ht]
\centering
\begin{subfigure}{0.48\textwidth}
\includegraphics[width=\textwidth]{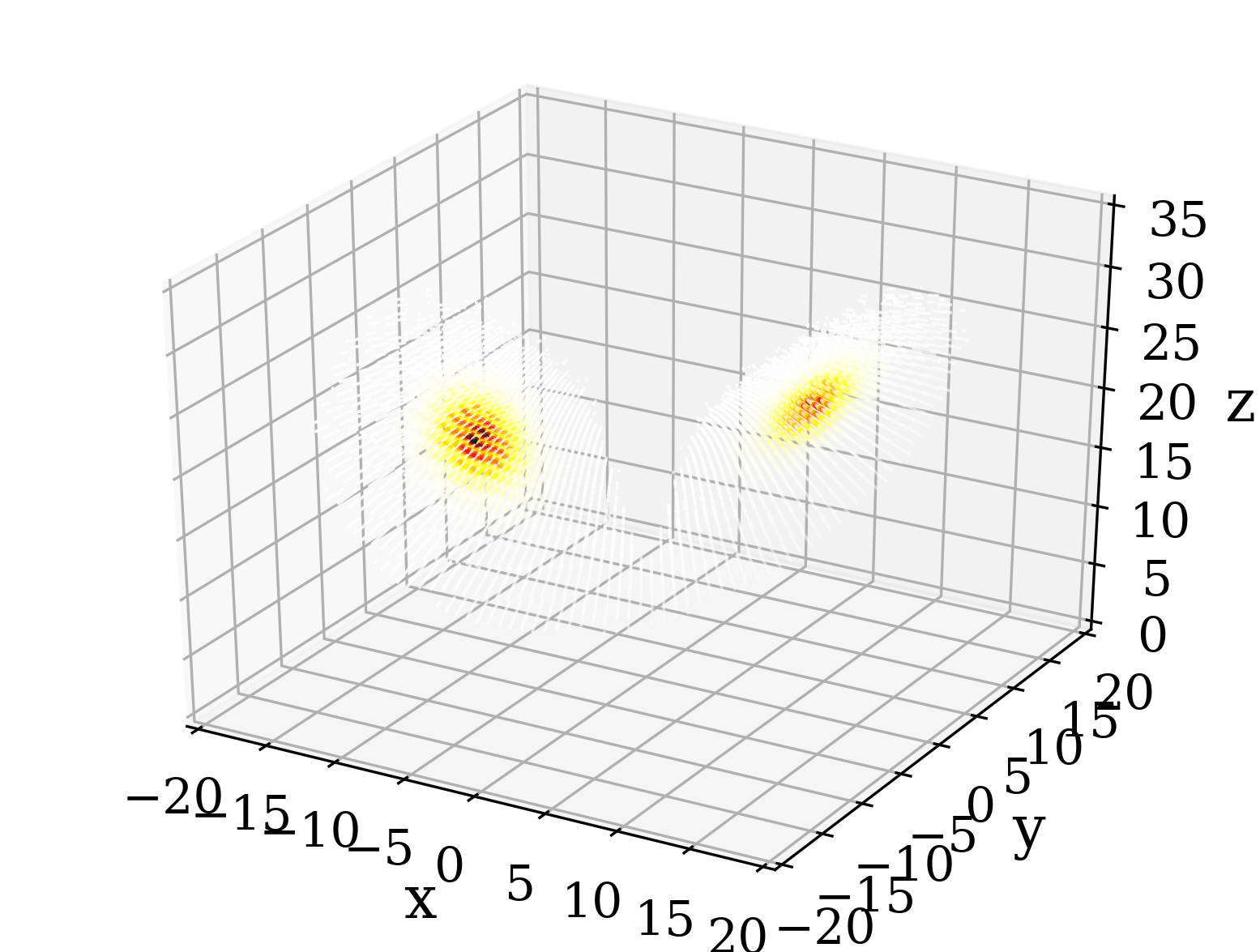}
\end{subfigure}
\begin{subfigure}{0.48\textwidth}
\includegraphics[width=\textwidth]{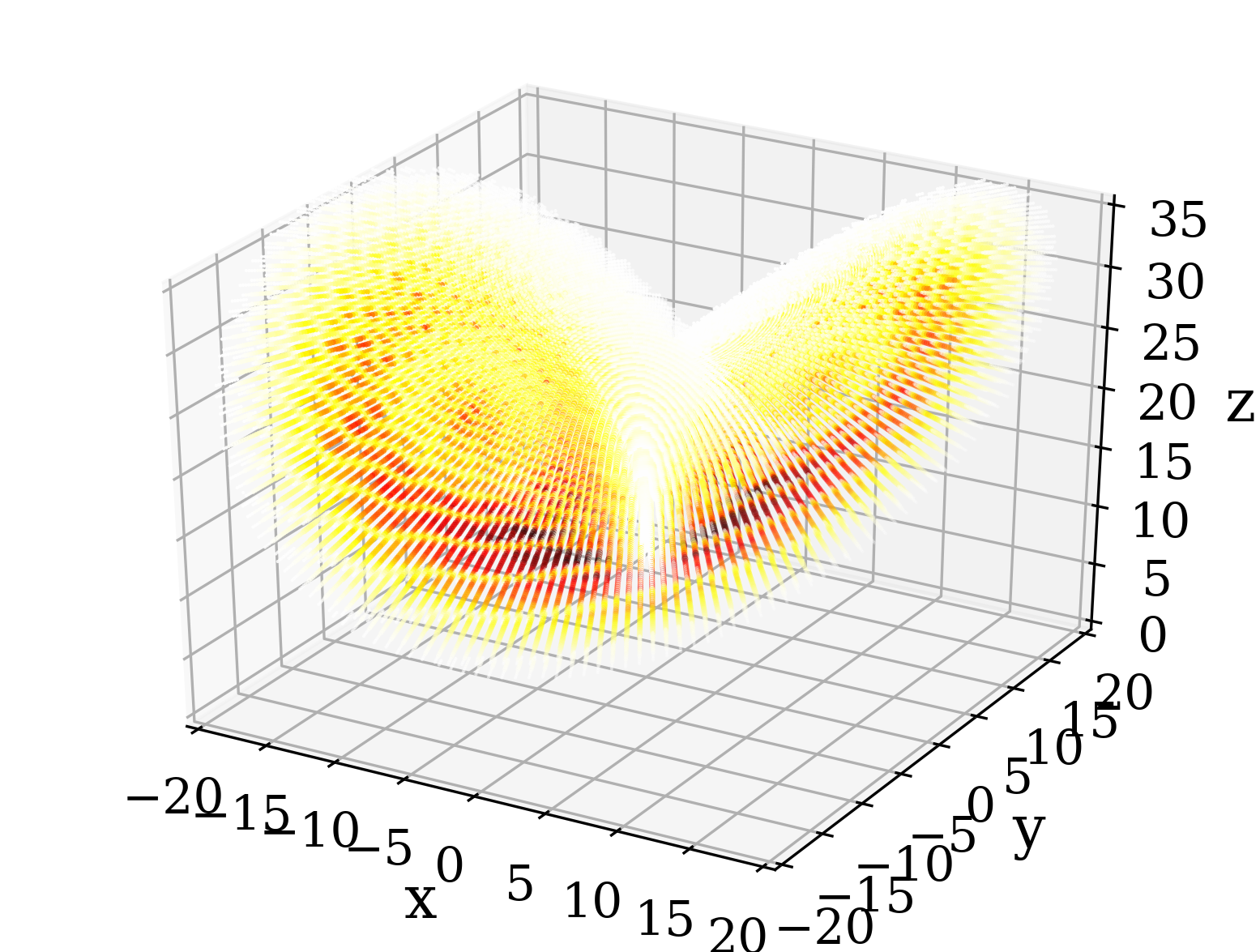}
\end{subfigure}\\
\begin{subfigure}{0.48\textwidth}
\includegraphics[width=\textwidth]{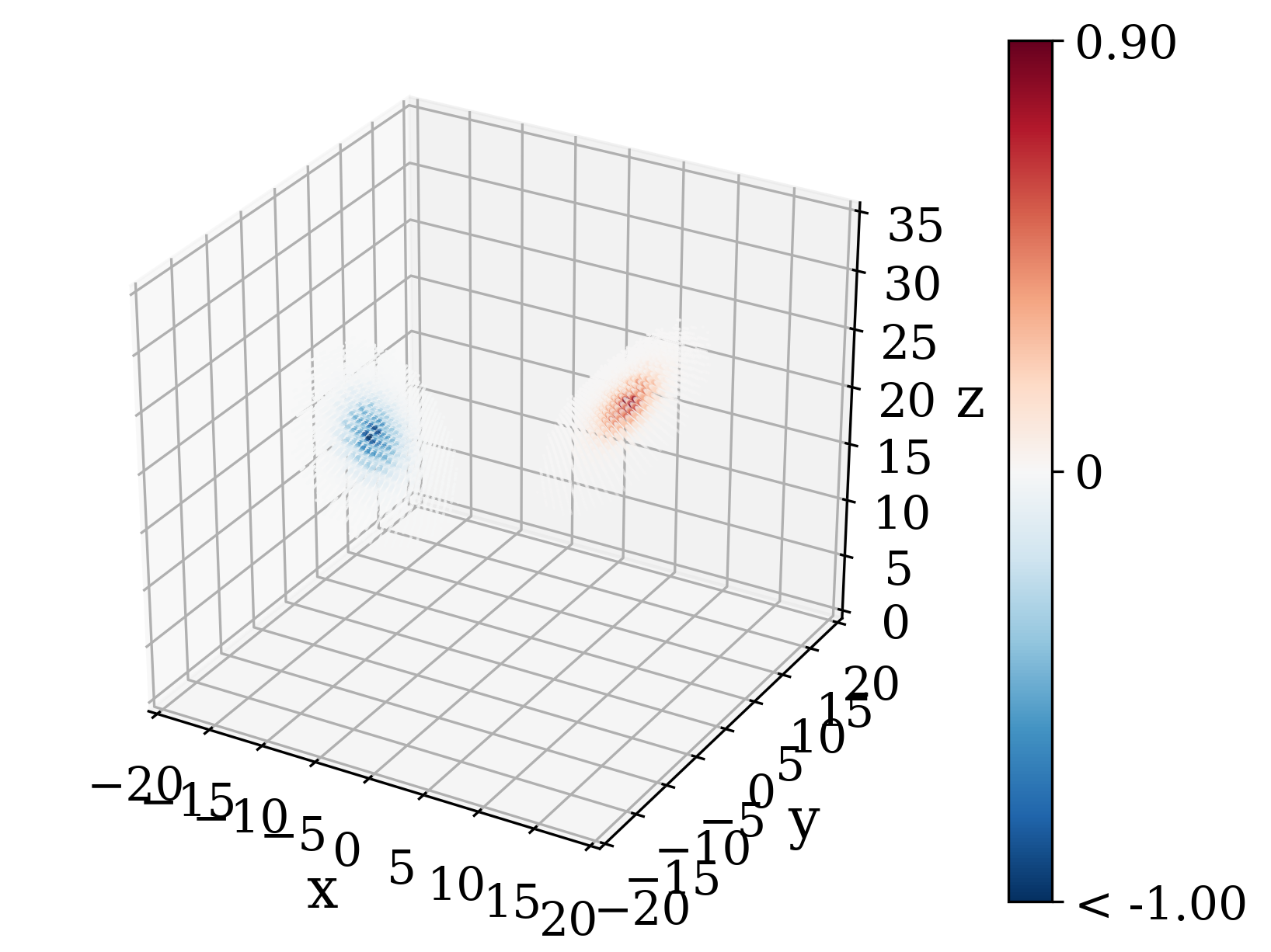}
\end{subfigure}
\begin{subfigure}{0.48\textwidth}
\includegraphics[width=\textwidth]{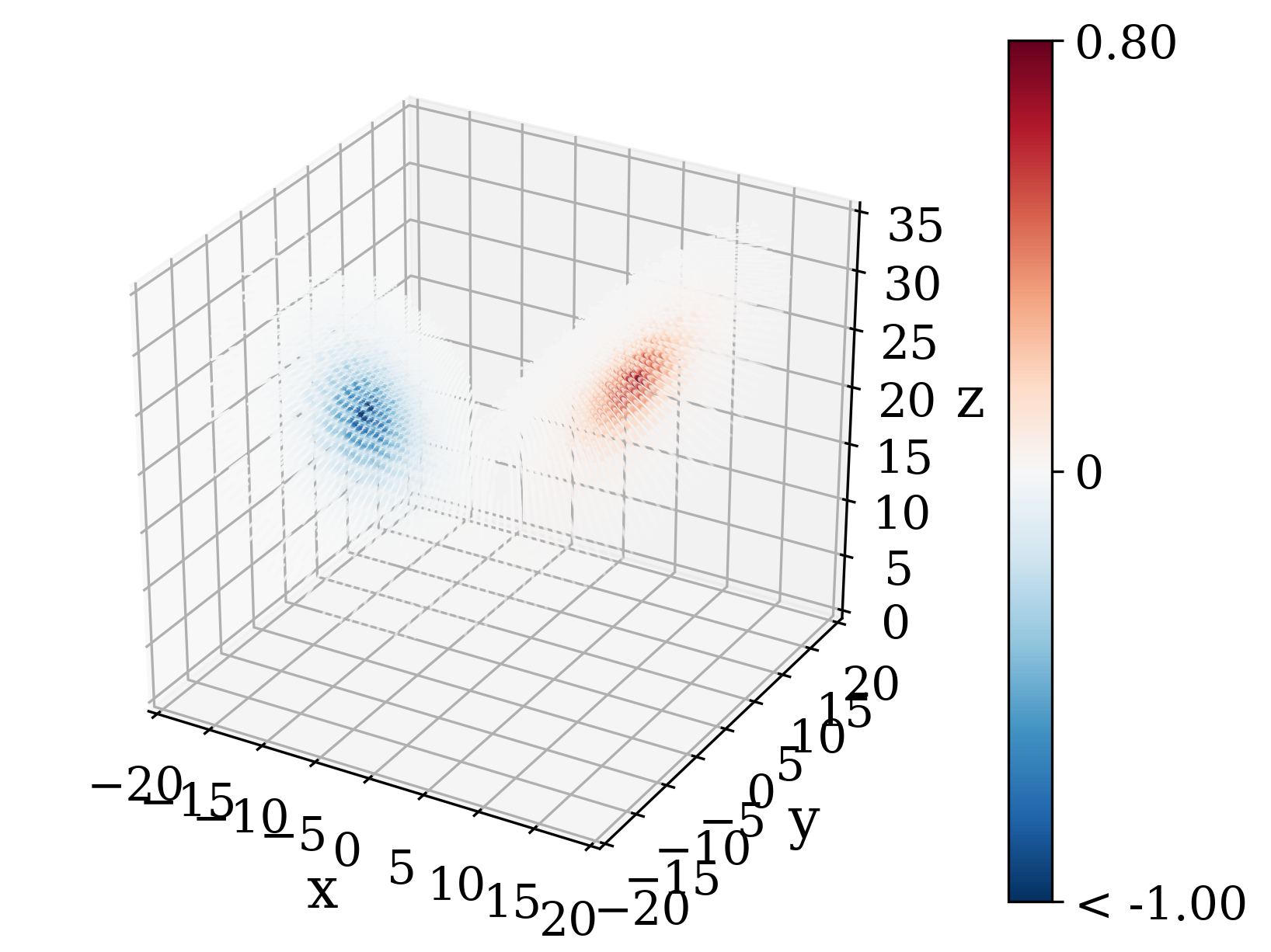}
\end{subfigure}\\
\begin{subfigure}{0.48\textwidth}
\includegraphics[width=\textwidth]{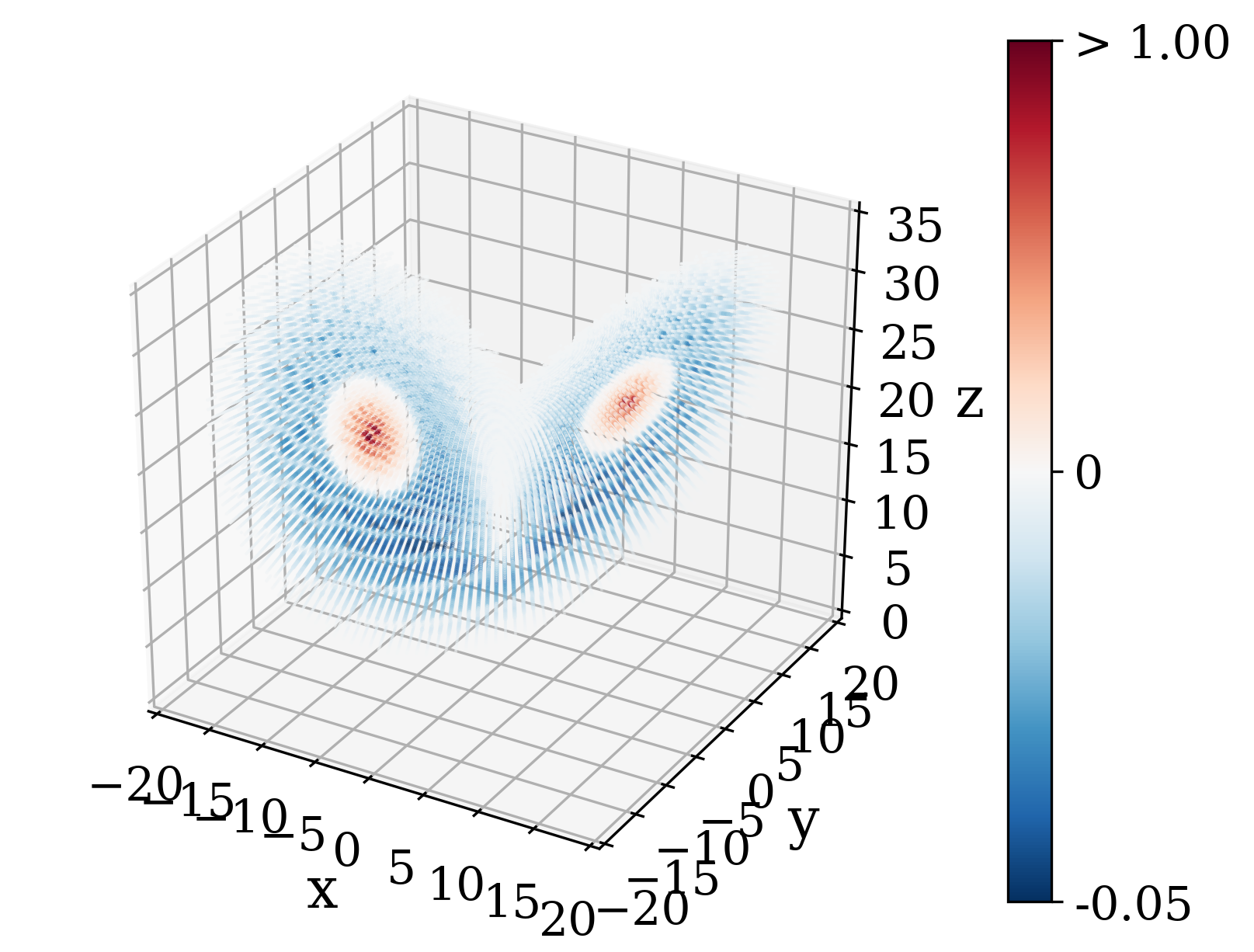}
\end{subfigure}
\begin{subfigure}{0.48\textwidth}
\includegraphics[width=\textwidth]{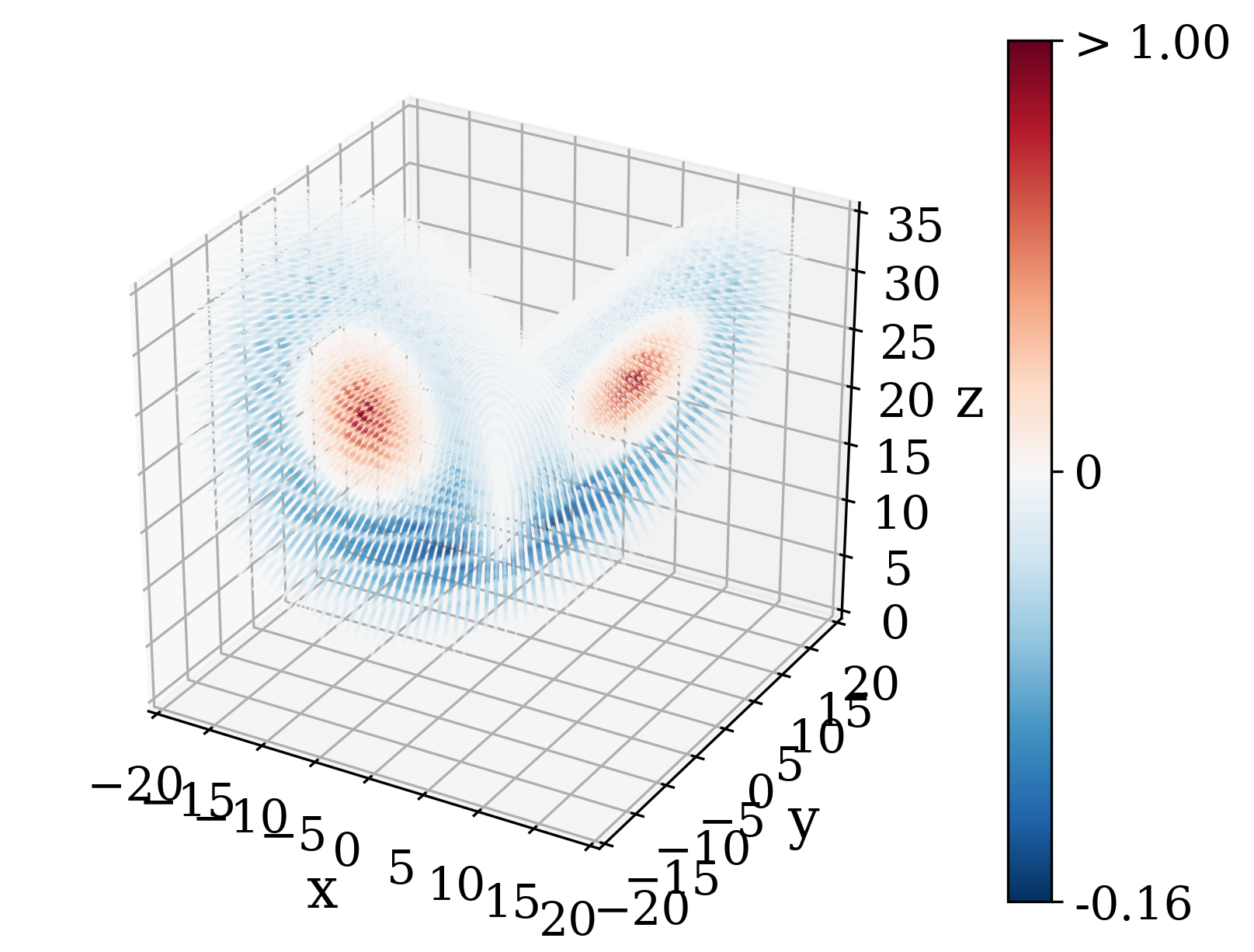}
\end{subfigure}
\caption{Eigenvectors associated with the first (top), second (middle) and third (bottom)
generator eigenvalues calculated from the transition matrices $\bv{P}^m_\tau$ for
$\rho = 22$ (left) and $\rho = 24.5$ (right).
For practical purposes, the eigenvectors have been calculated from transition matrices
on a grid of 200-by-100-by-200, instead of 400-by-200-by-400. 
Care has been taken for the values of $\rho$ of 22 and 24 to be chosen
before and after the minimum in the spectral gap, respectively.}
\label{fig:eigvec}
\end{figure}

A more detailed understanding is obtained from the eigenvectors, represented in figure \ref{fig:eigvec},
associated with the three leading generator eigenvalues (from top to bottom),
calculated from the transition matrices $\bv{P}^m_\tau$
for $\rho = 22$ (left) and $\rho = 24.5$ (right).
The first eigenvectors (top) give an approximation of the physical measure of the system
by a Lebesgue-density which is stationary, as far as the numerical approximations are concerned.
One can see that, for $\rho < \rho_A$ (left), this stationary density has two modes
over the two stable stationary points $p^\pm$.
There is, however, some spread around these points, due to the numerical approximation.
The second eigenvector is also nonzero only about $p^\pm$, but with opposite values
over each stationary point
\footnote{Note that the secondary eigenvectors being orthogonal
to the leading adjoint eigenvectors, which is constant (see Appendix~\ref{sec:basin}),
they integrate to zero with respect to the Lebesgue measure.}.
It thus accounts for the fact that transitions between these points are very rare
(and only possible due to the discretization).
The third eigenvector, however, is positive about the stationary points
but negative about the chaotic set $\Lambda$.
For $\rho$ close, but smaller than $\rho_A$,
the set $\Lambda$ is invariant but weakly unstable,
so that the measure $\mu$ supported by $\Lambda$ is not yet physical.
Transitions to $p^\pm$ from points around $\Lambda$, but not exactly on it, are thus
possible although very slow, since the crisis is near.
This explains the dichotomy between $p^\pm$ and $\Lambda$
visible in the third eigenvector.

For $\rho_A < \rho < \rho_\mathrm{Hopf}$ (right panels) the second and third eigenvectors
are similar to those for $\rho < \rho_A$.
However, the stationary density (top right panel) now spreads about
$\Lambda$, in agreement with the fact that $\Lambda$ is now an attractor.
For $\rho < \rho_A$, one can thus interpret
the real part of the third eigenvalue as the escape rate 
from the chaotic set $\Lambda$ to each stationary point,
while for $\rho > \rho_\mathrm{Hopf}$, the second and third eigenvalues
can be interpreted as the escape rate from the stationary points to $\Lambda$.
However, while this interpretation is valid for $\rho > \rho_\mathrm{Hopf}$,
for the value of $\rho$ between $\rho_A$ and $\rho_\mathrm{Hopf}$ represented in the left panels,
the stationary points $p^\pm$ are still attracting.
One would thus expect three eigenvalues to be zero (see appendix \ref{sec:basin})
instead of one (Fig.~\ref{fig:gap}).
It thus appears that the discretization is still too coarse to get around
numerical diffusion \cite{Froyland2011a} preventing the
resolution of the three distinct basins of attraction.
This difficulty is likely to arise from the small size of the basins of attraction of $p^\pm$
close to the Hopf bifurcation and from the possibly convoluted geometry
of their boundaries.

As a conclusion, the first key result of this study
is the clear indication of the crisis in the shrinkage of the spectral gap
in the approximation of the generator eigenvalues for $\mathcal{P}_t^m, t \ge 0$.
However, as discussed in section \ref{sec:estimPmu} (see also appendix~\ref{sec:spectrumDecay}),
for this change in the spectrum to be detectable from observations on the attractor alone,
the eigenvalues of the generator of $\mathcal{P}_t^\mu, t \ge 0$
with respect to the physical measure $\mu$, for $\rho > \rho_A$, should also be affected.
This is tested in the next section \ref{sec:resultsMu}.

\subsection{Unstable resonances from $\bv{P}^\mu_\tau$ and decay of correlations}
\label{sec:resultsMu}
To better understand which of the changes in the evolution of
statistics are available from observations on the attractor,
the unstable resonances alone are calculated from the generator eigenvalues
of $\bv{P}^\mu_\tau$ (Sect.~\ref{sec:estimPmu}).
The resulting generator eigenvalues are represented in the left panels of figure \ref{fig:RPAttractor},
for values of $\rho$ ranging from about $\rho_A$ to $\rho = 28$.
As recalled in appendix \ref{sec:spectrumDecay},
the spectrum of the transfer semigroup $\mathcal{P}_t^\mu, t \ge 0,$ governs
the decay of correlations between any pair of observables in appropriate functional spaces.
As an example, the right panels of Fig.~\ref{fig:RPAttractor} represent sample estimates
\footnote{The sample correlation function is estimated from a long time series
initialized in the basin of attraction of $\Lambda$
as a discrete approximation of the time average \eqref{eq:corrTime}
(see e.g.~\cite{VonStorch1999b}).}
of the correlation function $C_{z, z}(t)$ for the observable $z: (x, y, z) \to z$.
This is but one example of correlation function, which will be sufficient for the present discussion.
To facilitate the comparison with the results of the previous section \ref{sec:resultsM},
the corresponding generator eigenvalues from $\bv{P}^m_\tau$
have also been represented by black crosses
in the top left panel in figure \ref{fig:RPAttractor}, for $\rho = 24.1$.
\begin{figure}[ht]
\centering
\begin{subfigure}{0.47\textwidth}
\includegraphics[width=\textwidth]{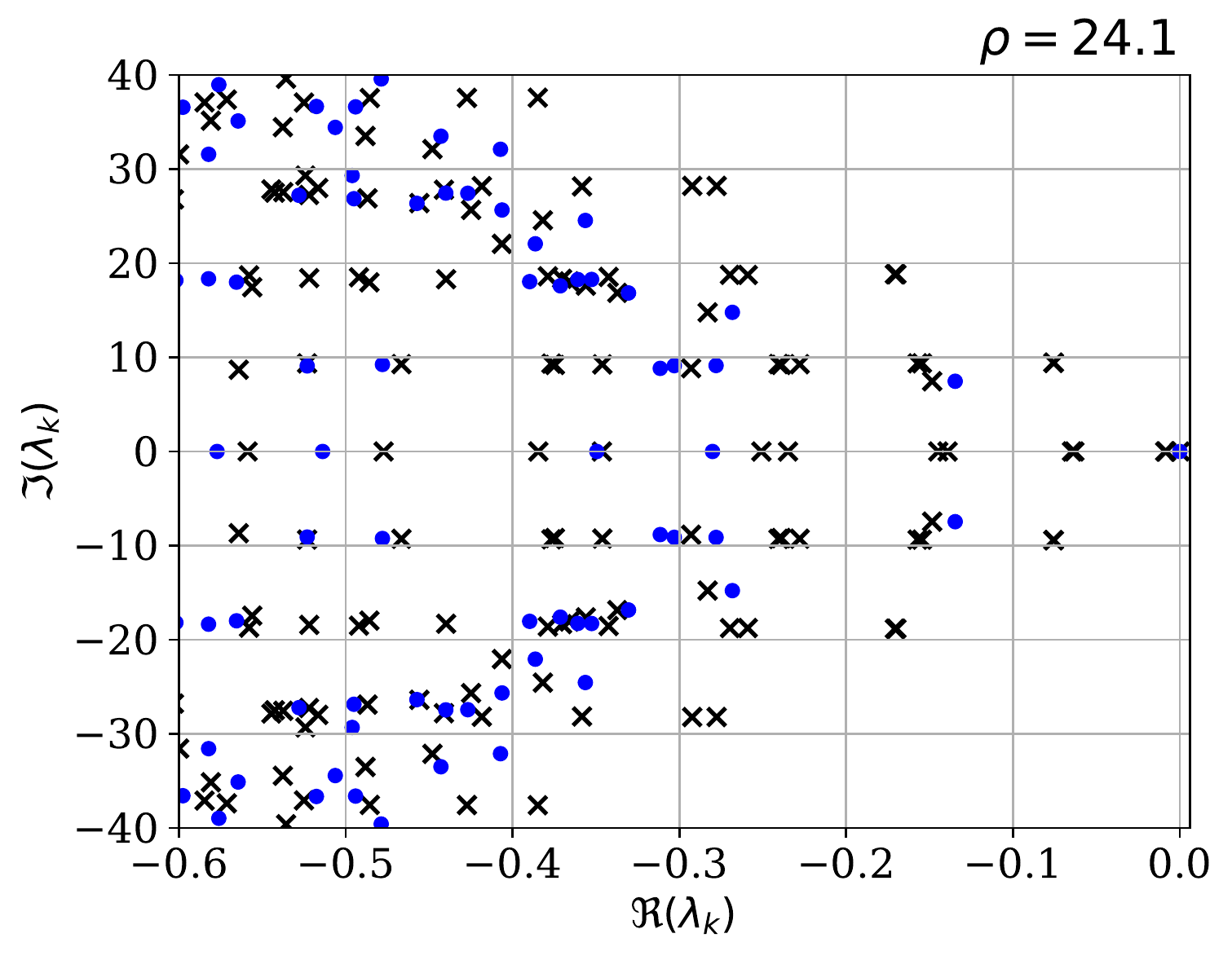}
\end{subfigure}
\begin{subfigure}{0.47\textwidth}
\includegraphics[width=\textwidth]{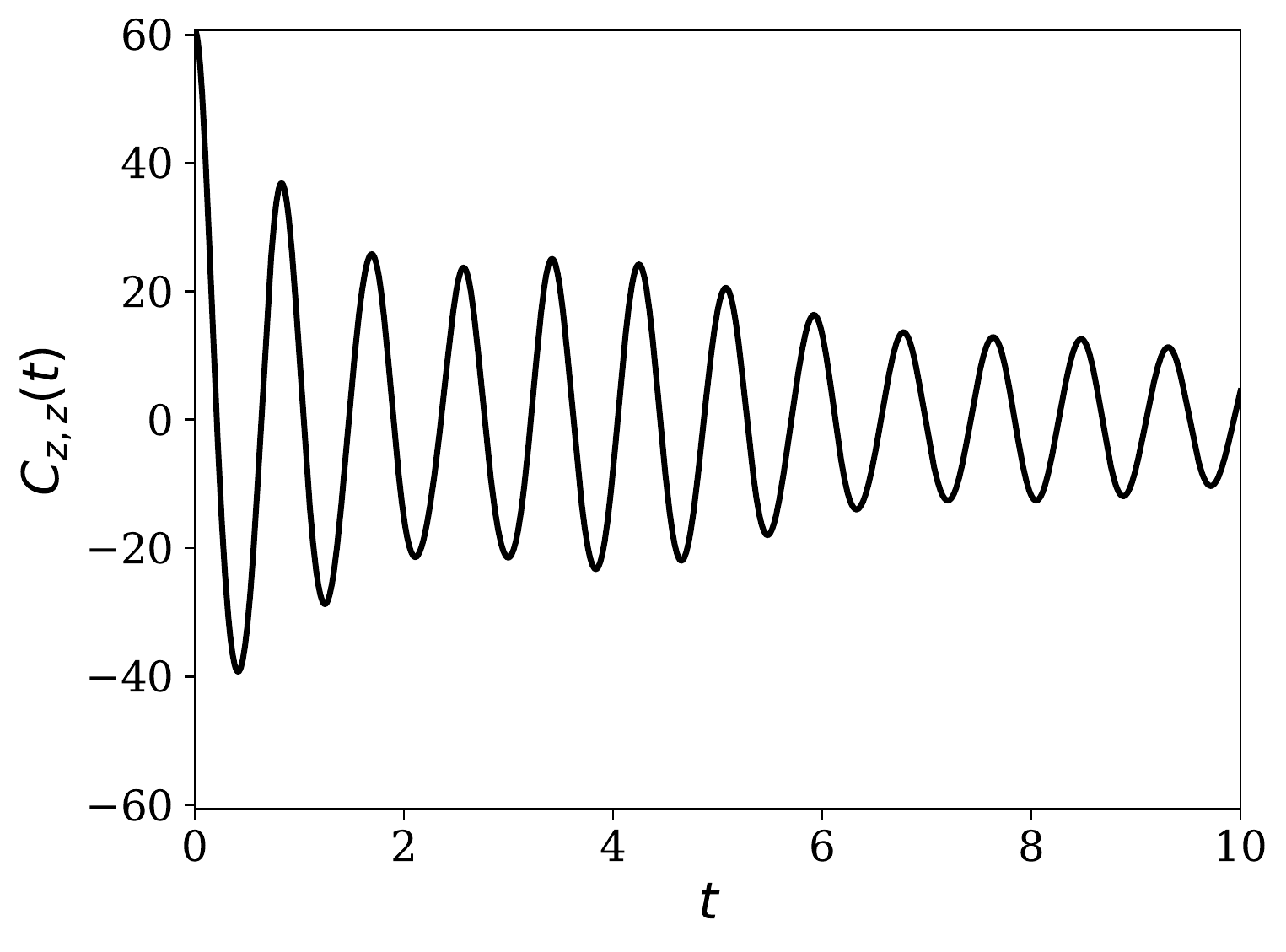}
\end{subfigure}\\
\begin{subfigure}{0.47\textwidth}
\includegraphics[width=\textwidth]{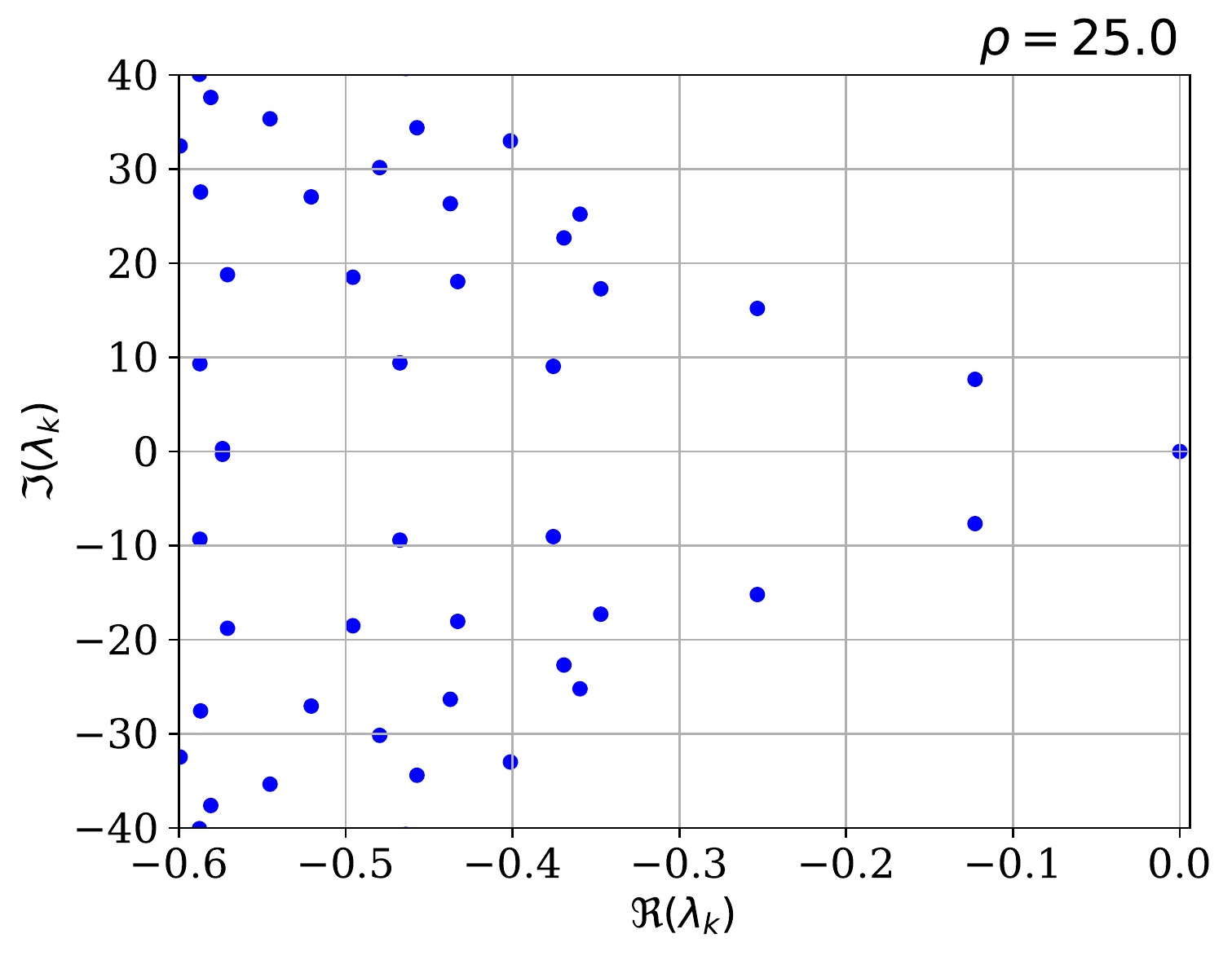}
\end{subfigure}
\begin{subfigure}{0.47\textwidth}
\includegraphics[width=\textwidth]{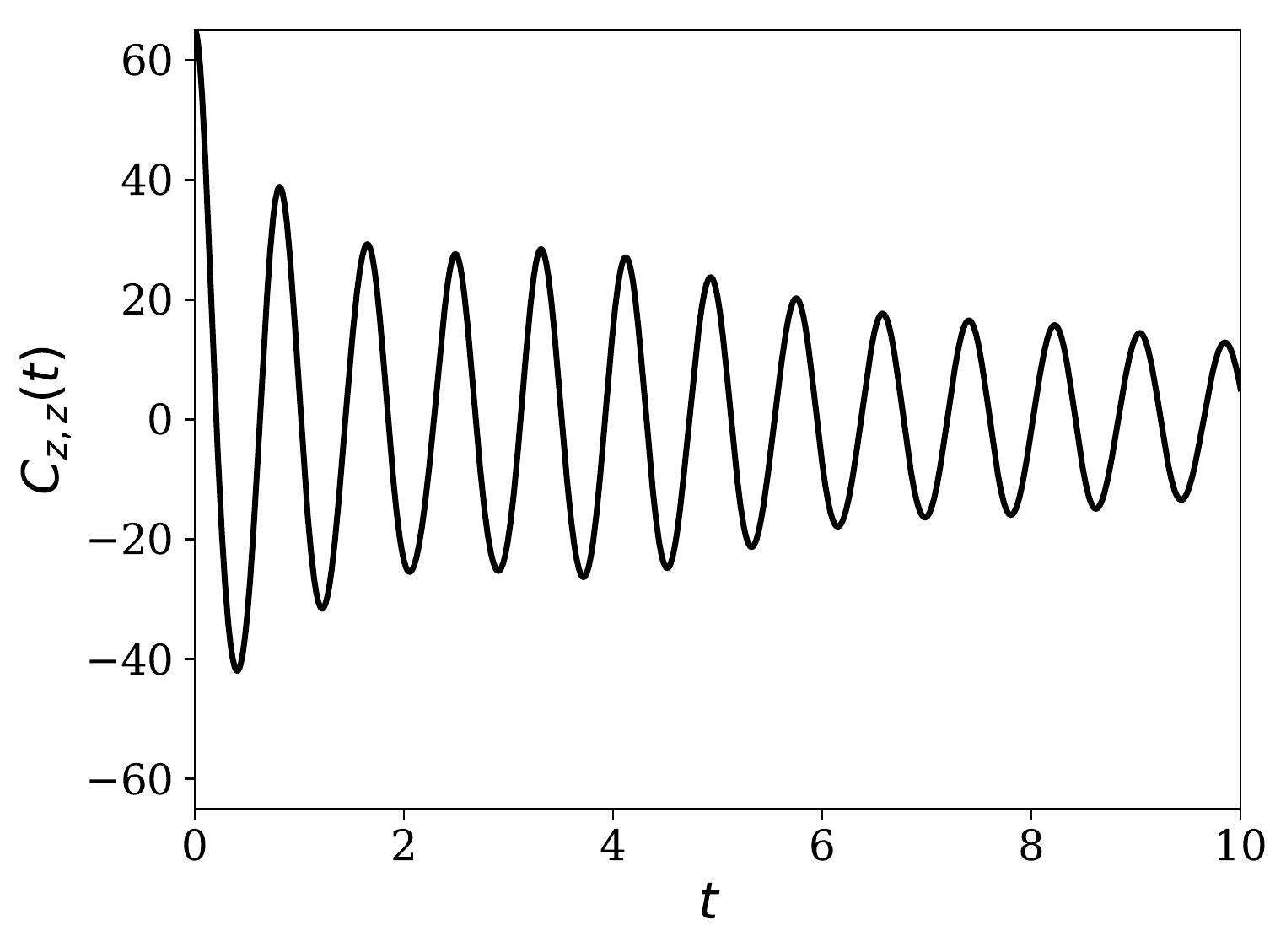}
\end{subfigure}\\
\begin{subfigure}{0.47\textwidth}
\includegraphics[width=\textwidth]{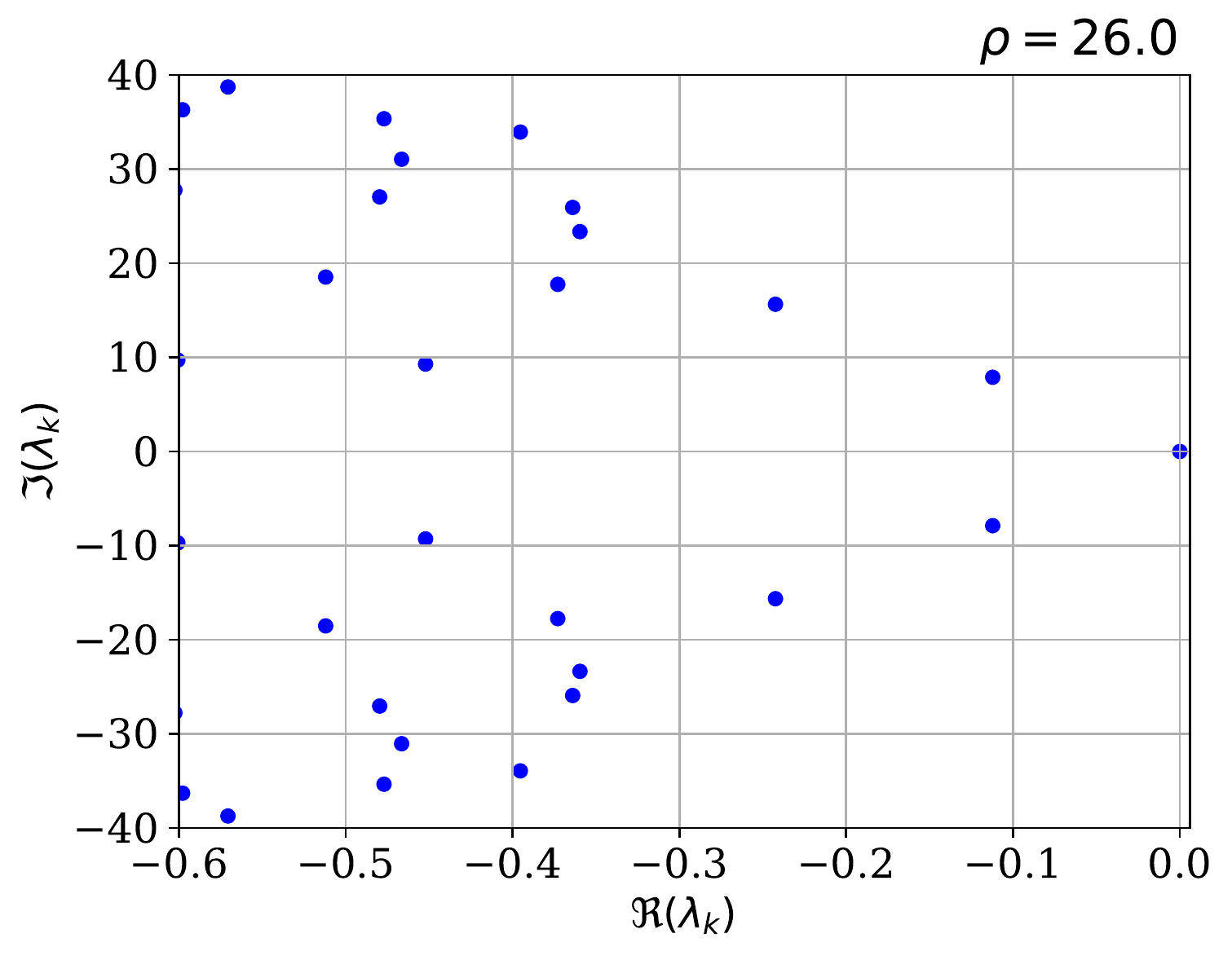}
\end{subfigure}
\begin{subfigure}{0.47\textwidth}
\includegraphics[width=\textwidth]{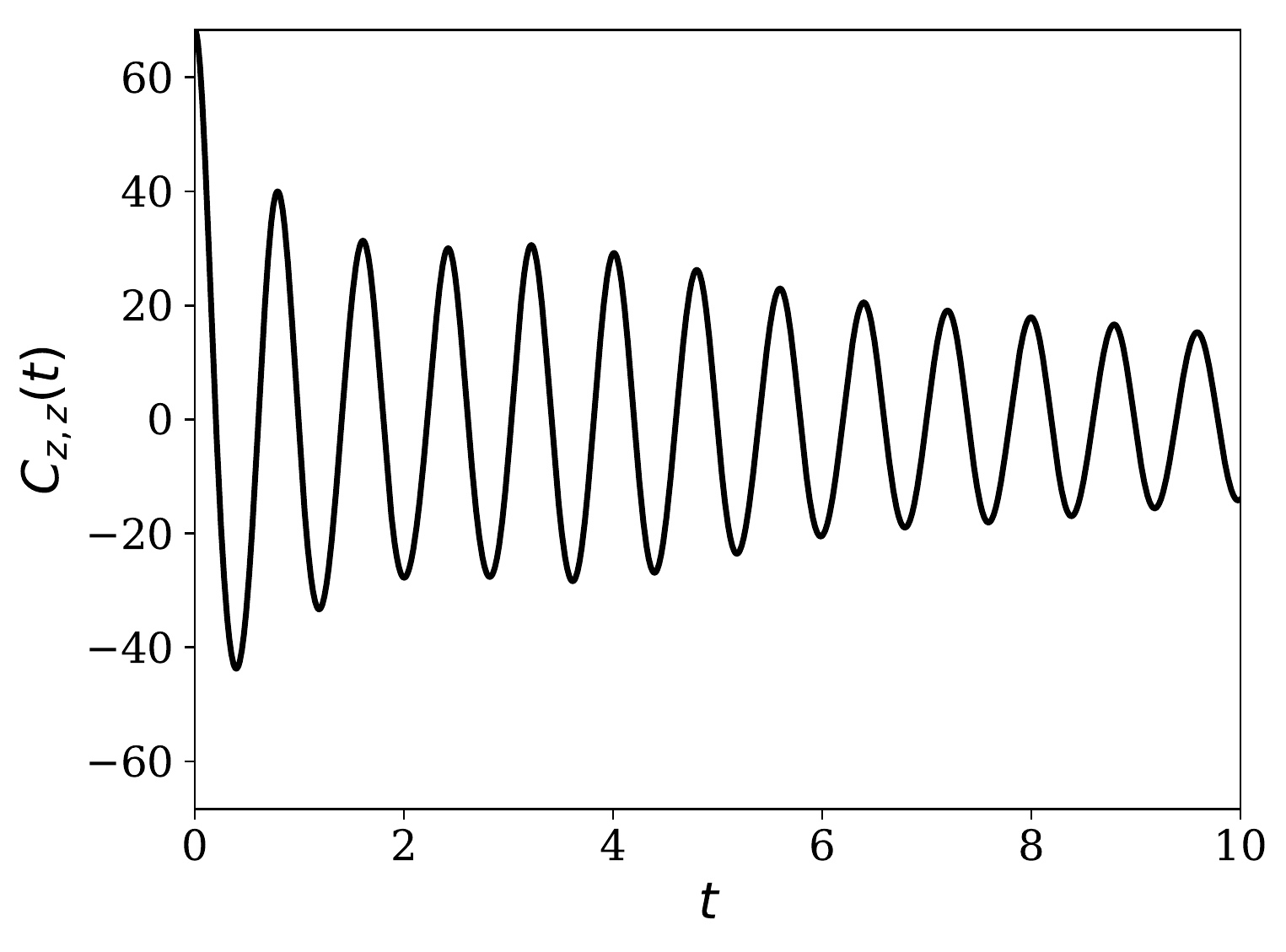}
\end{subfigure}\\
\begin{subfigure}{0.47\textwidth}
\includegraphics[width=\textwidth]{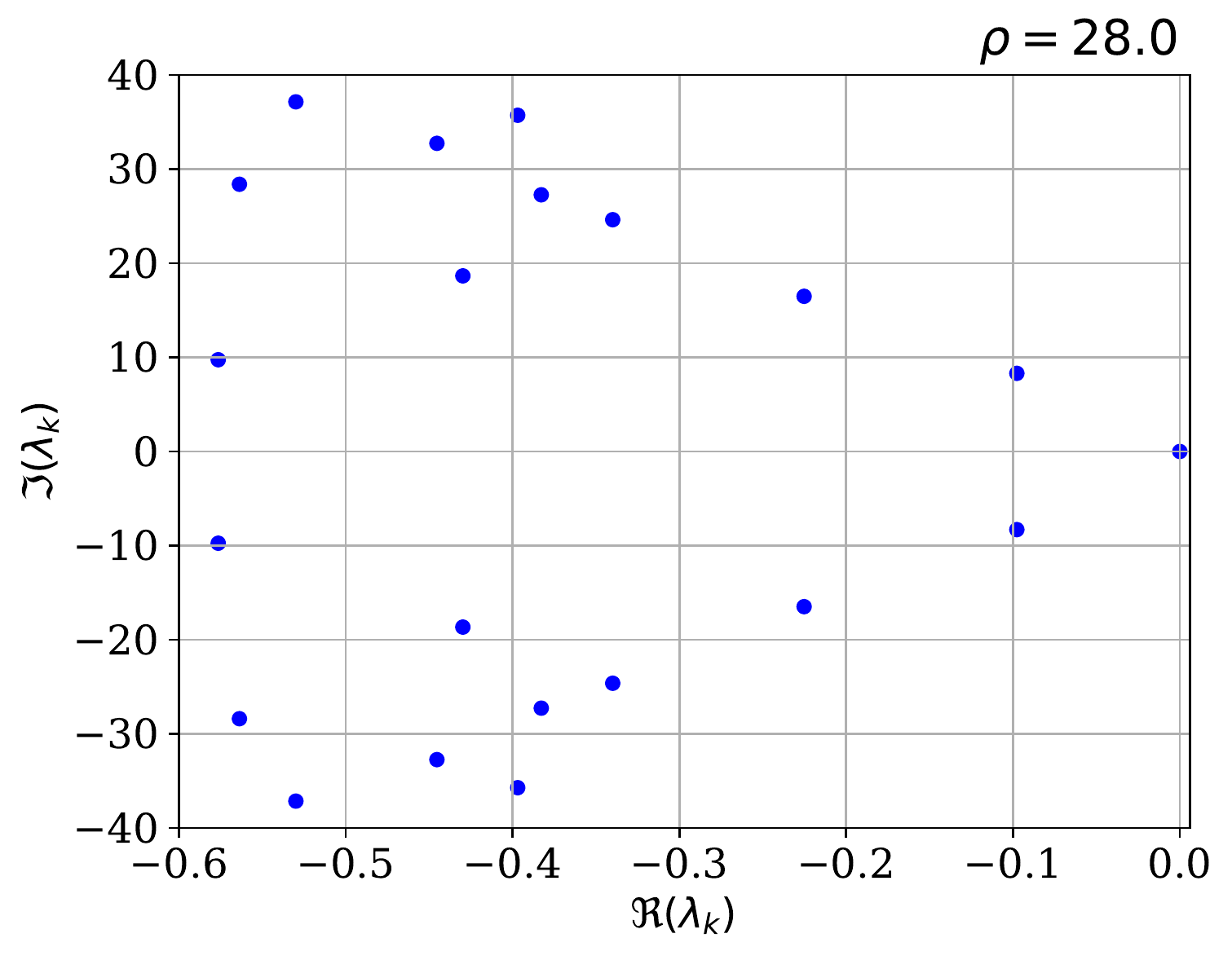}
\end{subfigure}
\begin{subfigure}{0.47\textwidth}
\includegraphics[width=\textwidth]{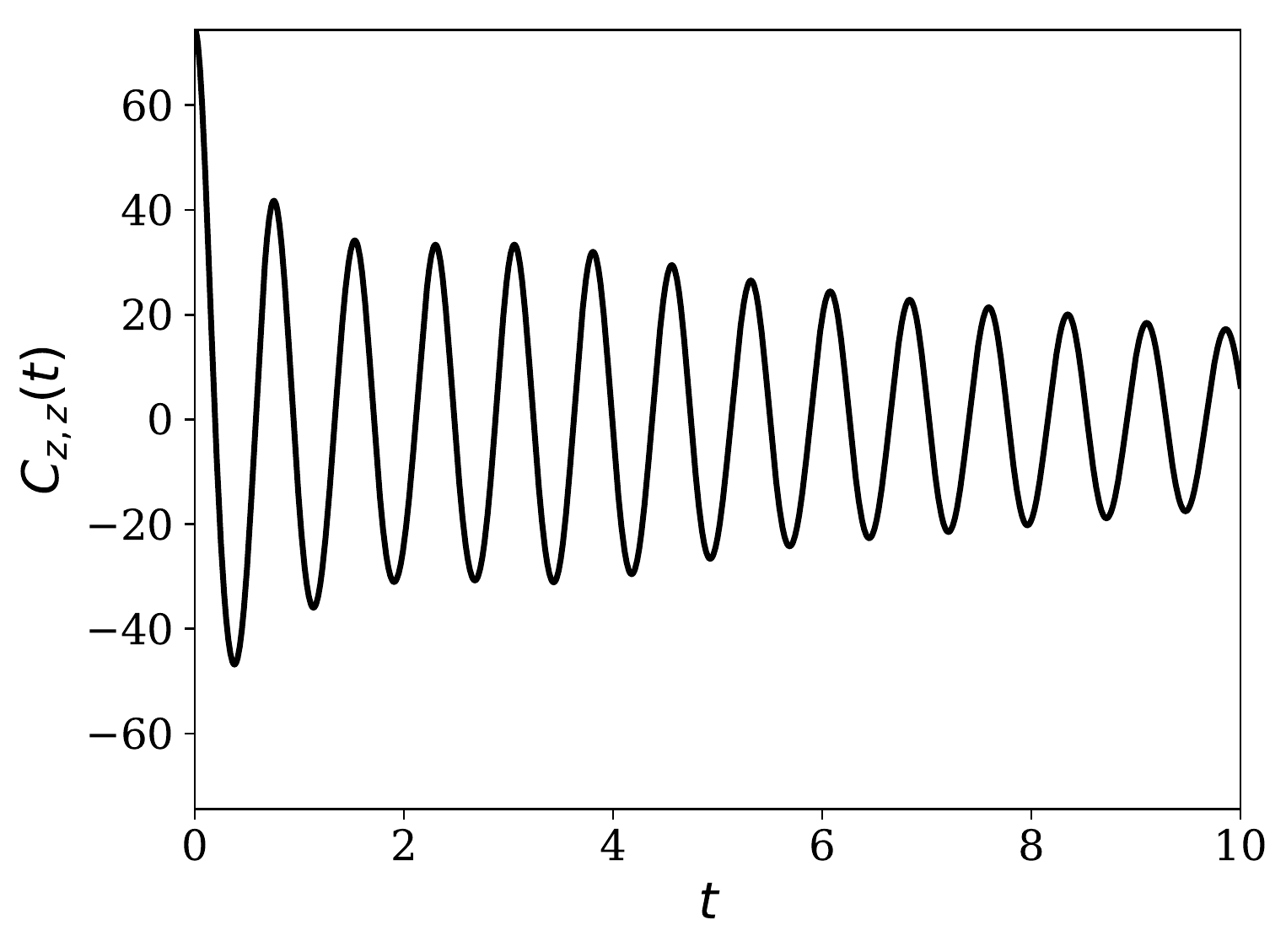}
\end{subfigure}
\caption{Leading generator eigenvalues calculated from the transition matrices $\bv{P}^\mu_\tau$ (left)
and sample correlation functions $C_{zz}(t)$ (right), for different values of $\rho$.
Here, according to the formula \eqref{eq:corrTime}, the correlation function is not normalized
by the covariance at $t = 0$. Thus, $C_{zz}(0)$ yields the variance of $z$.}
\label{fig:RPAttractor}
\end{figure}

As discussed in appendix \ref{sec:basin},
we expect the eigenvalues of $\bv{P}_\tau^\mu$
to constitute a subset of the eigenvalues of $\bv{P}_\tau^m$,
within numerical errors.
This seems to be indeed the case as the generator eigenvalues calculated from $\bv{P}^\mu_\tau$
(blue dots in the top left panel)
roughly correspond to some of the generator eigenvalues calculated from $\bv{P}^m_\tau$ (black crosses).
However, the complex plane is less densely populated by generator eigenvalues 
of $\bv{P}^\mu_\tau$ than of $\bv{P}^m_\tau$.
This is not surprising, since the transfer operators $\mathcal{P}_t^\mu$
only gives access to the unstable resonances and not to the stable resonances
describing the relaxation of densities transversally to $\Lambda$.

The second key result of this study is that the generator eigenvalues of 
$\bv{P}^m_\tau$, which where found in section \ref{sec:resultsM}
to approach/escape from the imaginary axis during the crisis,
do not belong to the set of generator eigenvalues of $\bv{P}^\mu_\tau$.
In fact, the spectral gap between the leading generator eigenvalues
of $\bv{P}^\mu_\tau$, in figure \ref{fig:RPAttractor}, is not significantly
affected by the attractor crisis.
As a result, neither is the rate of decay of the correlation function $C_{z, z}(t)$,
whose sample estimates are represented in the right panels of the same figure \ref{fig:RPAttractor}.
This applies to the correlation function between any pair of observables (not shown here).
In conclusion, for this attractor crisis in the Lorenz flow,
critical slowing down is not observable from time series converged to $\Lambda$.
This can be interpreted by the fact that the evolution of densities is affected
by the weaker stability of the attractor close to the crisis, but that this change in stability
is not felt along the attractor.
The stability of the attractor should here be understood as the rate of convergence
of densities to it, as measured by the spectral gap between the zero and the
secondary generator eigenvalues of $\mathcal{P}_t^m$,
as opposed to the rate of divergence of nearby trajectories measured
by the Lyapunov exponents.
In conclusion, only the stable resonances are affected by the crisis.

\section{Summary and discussion}
\label{sec:summary}
`
The global stability of the chaotic attractor in the Lorenz flow
as it undergoes a boundary crisis is investigated numerically in terms of Ruelle-Pollicott resonances.
The results are now summarized and their implications
regarding early-warning of chaotic attractor crises
and response theory discussed.

For the boundary crisis of the Lorenz flow,
neither the Lyapunov exponents nor the covariant Lyapunov vectors provide a precursor.
For that purpose, global information appears to be necessary.
While the Lyapunov spectrum describes the divergence of nearby trajectories,
the semigroup of transfer operators governs the
convergence/escape of densities to/from an invariant measure.
The rate at which this occurs is characterized by the eigenvalues
of the generator of the semigroup with negative real part, the Ruelle-Pollicott resonances.
These resonances are divided into a stable and an unstable family describing
the evolution of densities about and on the support of the invariant measure, respectively.
The gaps between the stable (unstable) resonances and the imaginary axis thus
provides an indicator of the global stability (mixing rate)
of a possibly chaotic invariant set.

To monitor the changes in the spectrum during the crisis,
a discretization of the transfer operators is estimated
either from many short time series
sampling the phase space or from a few long trajectories converged to the attractor.
The difference between the two approaches is essential,
as the first one yields an approximation of transfer operators
acting on functions of the full phase space, thus yielding information on global stability,
while the second one gives a restriction of the transfer operators to
the attractor supporting a physical measure.
The second approach thus only allows to approximate the unstable resonances
associated with mixing dynamics on the attractor, while the first
also gives an approximation of the stable resonances corresponding to contraction
towards the attractor, thus providing global information on the properties of the system
outside the attractor.

The main result of this study is twofold.
First, as the crisis is approached, some stable resonances
approach the imaginary axis, as indicative of the weaker stability of the attractor.
After the crisis, these stable resonances get further from the imaginary axis,
their distance to the imaginary axis giving a measure of the escape rate from the chaotic saddle.

Second, only the stable resonances are affected by the crisis.
This implies that, in the boundary crisis considered here,
the approach of the crisis cannot be inferred from time series
evolving along the attractor, i.e. from long observations of the unperturbed system.
This explains why the correlation functions between different observables
are not showing any sign of slowing down of their decay before the crisis.

As a consequence, early-warning indicators \cite{Scheffer2009}
based on correlation functions \cite{Held2004a} or power spectra \cite{Kleinen2003a}
are in general unable to give a precursor of attractor crises showing the same behavior as the one studied here.
On the other hand, indicators based on the recovery from perturbations \cite{Nes2007}
may be useful.
The perturbation should then be transverse to the attractor,
in the sense that it should project on the stable manifold of the attractor.
In fact, in the case of a stationary point,
the recovery rates as given by the eigenvalues of the Jacobian of this point \cite{Nes2007}
coincide with the leading Ruelle-Pollicott resonances \cite{Gaspard1995,Lan2013}.
A conclusion of this study is that, in the chaotic case,
the recovery rates are not given by the Lyapunov exponents,
but rather by the stable resonances.

In fact, that some resonances correspond to motions on the attractor,
which can be observed from long time series,
while other correspond to motions transverse to the attractor,
is the central difference between the response theory
for systems in thermodynamic equilibrium \cite{Kubo1957a}
and Ruelle's response theory for dissipative systems \cite{Ruelle2009}.
Indeed, invariant measures of equilibrium systems have a density with respect
to the Lebesgue measure, allowing for the expression of the linear term in the response to forcing 
in terms of correlation functions. This is the statement of the celebrated \emph{flucutation-dissipation theorem}.
On the other hand, for dissipative systems such as the Lorenz flow \cite{Lucarini2009b},
the invariant measure supported by the attractor is singular and
an additional term accounting for perturbations along the stable manifold of the attractor
must be taken into account.
In any case, the susceptibility of the system to forcing depends
on the presence of resonances close to the imaginary axis \cite{Ruelle2009,Cessac2007}.
Eventually, when resonances touch the imaginary axis, response theory breaks down,
which is what is observed in this study during the attractor crisis, where
response theory cannot be expected to work.

The story can, however, be very different in the presence of noise.
Noisy perturbations may indeed push the system away from the attractor,
in the direction of its stable manifold,
allowing for the system to explore the phase space.
This argument can be made rigorous by verifying the Lie bracket H\"ormander
condition \cite{Hormander1968a,Hairer2011} to ensure,
from the hypoellipticity of the generator, that, thanks to the noise,
the transfer semigroup has a smoothing effect.
If this condition is verified, the invariant measure may admit a density,
allowing for the fluctuation-dissipation theorem to hold \cite{Hairer2010}.
In this case, the transfer semigroup in phase space and the one restricted to
the support of the invariant measure $\mu$ can be identified with
the same semigroup on $L^2_\mu$.
As a consequence, recovery rates and decorrelation rates can both
be identified with the real part of the leading generator eigenvalues \cite{Chekroun2016}.
Moreover, when the unperturbed deterministic system is hyperbolic,
singular perturbation theory may be applied to relate the eigenvalues
of the perturbed system to those of the unperturbed system \cite{Gaspard2002a,Tantet2016}.

A final issue remains regarding the genericity of the results obtained here for the Lorenz flow. The crisis scenario described here might be of general relevance for the study of three-dimensional flows, since the robust chaotic sets in three dimensions are singular hyperbolic. It is not a priori clear whether the basic difference between the behavior of stable vs. unstable resonances near the crisis persists in higher dimension. 
In particular, indications were given in \cite{Tantet2015a} of the shrinkage of the spectral gap of transfer operators and of the slowing down of the decay of correlations during a boundary crisis in a deterministic and autonomous general circulation climate model, featuring exclusively internally generated fluctuations. Nonetheless, those results are not conclusive enough, because we looked at the transfer operator in a severely projected space and we could not follow the statistical properties mentioned above up to the exact value of the parameter determining the crisis. 

A next step in our analysis is to try to understand what determines whether critical transitions due to boundary crises in a deterministic chaotic system are flagged by the properties of the unstable resonances or whether looking at the stable resonances is instead required. These two scenarios pertain to cases where the attractor contains all the needed information, or, instead, a neighborhood of the attractor needs to be considered. In the latter case, a natural way to probe the system is by adding a small stochastic perturbation.

A possible way ahead is to look at whether a role in this sense is played by different scenarios pertaining to the geometrical properties of the basin boundary, and in particular at looking at the details of how the edge state, i.e. the unstable periodic orbits in the case studied here, and the attractor collide at the bifurcation. It is important to note that the edge state separating two co-existing attractors is the gate for the noise-induced transitions between the attractors. Therefore, near the crisis, adding noise does not only allow for probing a small region near the attractor, but also leads to sampling the properties of edge state.

\appendix

\section{Transfer operators, resonances and decay of correlations}
\label{sec:transfer}

Let us shortly review the spectral theory of transfer operators relevant for this study.
A more detailed exposition can be found in \cite{Tantet2015a} and references therein.

\subsection{Semigroup of transfer operators and ensemble}
\label{sec:specTransfer}

We consider a dynamical system on a phase space $X$ in  $\mathbb{R}^d$
and with global invertible flow $\Phi_t : x_0 \to x(t), t \ge 0$ on $X$
generated by a sufficiently smooth autonomous vector field $F:X \to X$, i.e.
\begin{align}
	\dot{x}(t) &= F(x(t)),
	\quad x(0) = x_0 \in X.
	\label{eq:ODE}
\end{align}
In other words, the transformation $\Phi_t$ associates to any initial condition $x_0$ in $X$
the corresponding solution to (\ref{eq:ODE}) at time $t$.

When the system is chaotic, it is fruitful to follow the evolution of 
probability densities and observables rather than that of solutions of \eqref{eq:ODE}.
For that purpose, we endow the phase space $X$ with its Borel $\sigma$-algebra
$\mathcal{B}$ and some probability measure $\eta$ (to be specified) on $(X, \mathcal{B})$.
The flow $\Phi_t, t \ge 0$ 
\footnote{
The flow will be assumed to be nonsingular for the Lebesgue measure,
so that it maps sets of null measure into sets of null measure.}
induces a family of linear operators, the \emph{Koopman operators},
\begin{align}
	\mathcal{U}^\eta_t: g \to g \circ \Phi_t, \quad t \ge 0,
\end{align}
acting on observables $g$
in the space bounded measurable functions $L^q_\eta(X), 1 \le q < \infty$.
On the other hand, there exists a family of linear operators $\mathcal{P}^\eta_t, t \ge 0$,
the \emph{transfer operators} or \emph{Perron-Frobenius operators}
on $L^p_\eta(X)$, with $1/p + 1/q = 1$, and such that the duality relation
\begin{align}
\int_X g(x)~\mathcal{P}^\eta_t f(x)~\eta(dx) = \int_X \mathcal{U}^\eta_t g(x)~f(x)~\eta(dx)
\label{eq:transfer}
\end{align}
holds for any observable $f$ in $L^p_\eta(X)$ and $g$ in $L^q_\eta(X)$.
It follows that $\mathcal{U}^\eta_t$ is the adjoint of $\mathcal{P}^\eta_t$.
%
Moreover, taking $g$ in \eqref{eq:transfer} as the indicator $\mathbf{1}_A$
of some set $A \in \mathcal{B}$, one has that
\footnote{For a probability density $f$, i.e.~$f \ge 0$ and $\int_X f(x)  \eta(dx)= 1$,
\eqref{eq:transferProba} expresses the fact that the probability to find a member $x$
sampled from an initial ensemble $f$ in a set $A$ after some time $t$
is none-other than the probability of this member to be initially in the preimage of this set by the flow.}
%
\begin{align}
	\int_A \mathcal{P}^\eta_t f(x) ~ \eta(dx)
	= \int_{\Phi_t^{-1}A} f(x) ~ \eta(dx), \quad t \ge 0.
	\label{eq:transferProba}
\end{align}
The families of linear operators $\mathcal{P}^\eta_t$ and $\mathcal{U}^\eta_t, t \ge 0$
inherit from the \emph{semigroup} property of the flow, thus imposing
a strong constrain on the operators for different times.

The relationship between the nonlinear flow $\Phi_t$ and these semigroups
provides the connection between the ergodic properties of dynamical systems
and the functional analysis of linear operators \cite{Yosida1980,Engel2001,Davies2007},
as was first recognized in \cite{Koopman1931} and \cite{Neumann1932}.

\subsection{Ergodicity, mixing and correlations}

A key concept in ergodic theory is that of an \emph{invariant measure} for the flow $\Phi_t$,
that is, a probability measure $\mu$ such that
\begin{align*}
	\mu(\Phi_t^{-1} A) = \mu(A), \quad \mathrm{for~any~} A \in \mathcal{B}.
\end{align*}
In other words, a measure is invariant if, according to this measure, the probability
of a state to be in some set does not change as this state is propagated by the flow.
It follows then that the average with respect to the invariant measure $\mu$
of any integrable observable $g$ is also invariant with time, i.e.
\begin{align}
	\int_X g(\Phi_t x) \mu(d x) = \int_X g(x) \mu(d x) :=  \left<g\right>_\mu,
	\label{eq:invariantAverage}
\end{align}
so that the invariant measure gives a statistical steady-state.

A flow $\Phi_t$ with an invariant measure $\mu$ has interesting statistical properties
when $\mu$ is \emph{ergodic}, that is, when the sets $A$ which are invariant,
i.e.~$\Phi_t^{-1} A = A$, are either of measure 0 or 1
\footnote{In particular, the ergodicity of the invariant measure implies that each set of positive measure
is visited infinitely often by orbits starting from $\mu$-almost every point \cite[Lemma~6.15]{Eisner2015}.}.
Then, by the celebrated individual ergodic theorem of Birkhoff \cite[Chap.~7.3]{Lasota1994},
the average of any $\mu$-integrable observable $g$ is such that
\begin{align}
	 \left<g\right>_\mu = \lim_{T \to \infty} \frac{1}{T} \int_0^T g(\Phi_t x) dt := \bar{g},
	 \quad \mathrm{for ~} \mu\mathrm{-almost~every~} x.
	 \label{eq:Birkhoff}
\end{align}
Thus, when $\mu$ is ergodic, the time mean is independent of the initial state $x$
except for a set of null measure.
\begin{remark}
\label{rmk:physical}
There may exist many ergodic measures.
This is actually the case for the Lorenz flow for $\rho_{\mathrm{Homo}} < \rho < \rho_A$,
where, as seen in section \ref{sec:crisis}, three attractors coexist.
Each attractor then supports at least one invariant measure \cite[Chap.~4]{Katok1996}.
In this case, the equality (\ref{eq:Birkhoff}) between ensemble averages and time averages
may hold only for initial states belonging to the attractor supporting the measure,
while time averages for two initial conditions in different basins of attraction will not coincide in general.
More useful for experiments is then the eventual \emph{physical} property of the measure.
The latter ensures that the equality (\ref{eq:Birkhoff}) between ensemble averages and time averages
holds not only for initial states in a set of positive measure $\mu$,
but also for states in any set of positive Lebesgue measure $m$ in the basin of attraction
of a given attractor \cite{Young2002}.
\end{remark}

A particularly important quantity in ergodic theory is the correlation function,
\begin{align} 
C_{f,g}(t) :=  \int_X  f(x) g(\Phi_t x)  \mu(dx) - \left<f\right>_\mu \left<g\right>_\mu, \quad t \ge 0,
\label{eq:defCorr} 
\end{align}
between any observables $f, g \in L^2_\mu(X)$.
It gives a measure of the relationship between the two observables as one evolves with time.
The decay of correlation functions with time is a macroscopic manifestation
of chaos. Indeed, this decay is equivalent to the mixing property
(see e.g.~\cite[Chap.~4]{Lasota1994} and \cite[Chap.~4]{Katok1996}),
\begin{align}
	\lim_{t\to\infty} \mu(A \cap \Phi_t^{-1} B) = \mu(A) \mu(B),
	\quad \mathrm{for~any~} A, B \in \mathcal{B}, \label{eq:defMixing}
\end{align}
of the invariant measure $\mu$, which is stronger than ergodicity.
In other words, the probability for some state to be in any set $B$ after some time $t$
is independent of the probability of this state to be in any set $A$ initially.
Any information about the initial state of an ensemble is thus gradually forgotten.

\begin{remark}
\label{rmk:physicalCorr}
For a physical measure $\mu$ supported by an attractor $\Lambda$,
the correlation function can be estimated by the time mean
\begin{align} 
	C_{f,g}(t) = \lim_{T \rightarrow \infty} \frac{1}{T} \int_0^T (f(x) - \bar{f})(g(\Phi_t  x) - \bar{g}) ~ dt,
	\label{eq:corrTime} 
\end{align}
from a single time initialized in the basin of attraction of $\Lambda$.
For this reason, we will see in section \ref{sec:estimPmu} that
complete information on $\mathcal{P}^\mu_t, t \ge 0$
can be obtained from from observations on the corresponding attractor alone.
Unfortunately, for the forced and dissipative systems considered here,
volumes contract on average \cite[Chap.~2.8]{Gallavotti2014a} so that the invariant measure $\mu$
is supported by an attractor with zero Lebesgue measure $m$.
It follows that, as opposed to the semigroup $\mathcal{P}^m_t, t \ge 0$,
no information on the dynamics away from the attractor
is carried by $\mathcal{P}^\mu_t$.
%
\end{remark}

\subsection{Spectral theory of ergodicity and mixing}
\label{sec:spectrumDecay}

One can see from the definition \eqref{eq:defCorr}
that the correlation function is fully determined by the family of
transfer (Koopman) operators $\mathcal{P}_t^\mu$ ($\mathcal{U}_t^\mu$) on $L^2_\mu(X)$,
i.e.~\emph{with respect to $\mu$}.
In fact, classical results relate the ergodic properties of
measure-preserving dynamical systems to the behaviour of the semigroups
\cite{Halmos1956,Arnold1968,Lasota1994}.
Note first, that the invariance of the measure $\mu$ together with
the invertibility of the flow ensure that the semigroups are isometries
and constitute a unitary group on $L^2_\mu(X)$.
As a consequence, the spectrum of the operators is contained in the unit circle
$|z| = 1, z \in \mathbb{C}$ and $\mathcal{P}_t^\mu$ and $\mathcal{U}_t^\mu$
have the same eigenvalues and eigenfunctions, for any real $t$.
Moreover, the following results hold.
\begin{theorem}
If the measure-preserving dynamical system
$(X, \mathcal{B}, (\Phi_t)_{t\in\mathbb{R}}, \mu)$ is
\begin{enumerate}[label=(\roman*)]
\item ergodic, then $1$ is a simple eigenvalue of $\mathcal{P}_t^\mu, t\in\mathbb{R}$, and conversely.
\item mixing, then the only one eigenvalue is one.
\end{enumerate}
\end{theorem}
The first item of this theorem can be understood from the fact that
ergodic systems have only one stationary density with respect to the invariant measure.
The second is due to the fact that eigenfunctions associated with eigenvalues
on the unit circle do not decay and thus prevent mixing for general observables.

The rate of decay of correlations, or mixing, characterizes
the (weak) convergence of ensembles to a statistical steady-state
and has been the subject of intense research these past decades.
Once again, the latter can be studied from the spectral properties
of the semigroups \cite{Liverani1995a}.
However, the eigenvalues responsible for such decay,
the \emph{Ruelle-Pollicott resonances} \cite{Pollicott1985,Ruelle1986},
lie inside the unit disk and are thus not accessible from operators on regular functional spaces.
Anisotropic Banach spaces of distributions capturing
the dynamics of contraction and expansion of the system and on which the semigroups
are contracting should instead be considered
\cite{Liverani1995a,Blank2001a,Gouezel2006,Butterley2007,Faure2014,Baladi2017}
\footnote{Contraction by semigroups can be interpreted in physical systems as determining entropy production. Recognizing this has been essential
to understand how reversible microscopic evolutions can lead
to irreversible macroscopic properties of systems out of thermal equilibrium
(see the pioneering work \cite{Misra1979} and \cite{Gaspard1998,Garbaczewski2002}
for reviews).}.

A certain degree of hyperbolicity is necessary to adapt these spaces to the dynamics,
and the robustness to perturbations of the spectrum or the convergence of numerical
algorithms is not guarantied in general
(see \cite{Keller1998,Gouezel2006} and \cite{Baladi1999,Froyland2007a},
respectively, for results in this direction).
In the present study, it is found, however, that the discretization scheme
used in section \ref{sec:approxSpectrum} yields reasonably good approximations
of the resonances.

The spectral properties of the semigroups with respect to an invariant measure $\mu$
thus allows for the study of the ergodic properties of this measure.
On the other hand, for the study of the global stability of this set,
operators acting on a neighborhood of positive Lebesgue measure
of a chaotic set should be considered.

\subsection{Spectral theory of global stability}
\label{sec:basin}

In the introduction, we have advocated on heuristic grounds
that the global stability of invariant sets could be studied from the
evolution of densities and observables.
In this case, however, the transfer (Koopman) operators
$\mathcal{P}_t^m$ ($\mathcal{U}_t^m$) \emph{with respect to
the Lebesgue measure $m$} should be considered.
Indeed, this should allow one not only to study the mixing dynamics
of an invariant measure, but also the contraction to/escape from such a measure.
This has allows for new developments in the theory 
of the global stability of stationary points and periodic orbits
from the behavior of $\mathcal{P}_t^m$ \cite{Vaidya2008}
and $\mathcal{U}_t^m$ \cite{Mauroy2016}.
This approach should also be amenable to chaotic invariant sets.
However, the same difficulties as mentioned in the previous section \ref{sec:spectrumDecay},
regarding the functional analytic framework appropriate to capture
the resonances inside the unit disk, are encountered.
In order to better interpret the numerical results obtained in section \ref{sec:results},
let us however give a few comments regarding the eigenfunctions
of the Koopman operators $\mathcal{U}^X_t$ and $\mathcal{U}^\Lambda_t$
on the space $C(X)$ of continuous functions on the compact metric space $(X, d)$
(e.g.~$X=\mathbb{R}^n$ with $d$ the distance induced by the Euclidean norm)
and its restriction to the support $\Lambda$
of some invariant measure $\mu$, respectively.
Since $(X, d)$ is a normal space and $\Lambda$ is a closed subset of $X$,
the Tietze-Urysohn Extension Theorem (e.g.~\cite{Lang1993}) ensures that
for any continuous function $g^\Lambda$ on $\Lambda$, there exists
a continuous function $g^X$ on $X$ whose restriction to $\Lambda$ is $g^\Lambda$.
The restriction $\mathcal{U}^\Lambda_t$ of $\mathcal{U}^X_t$ on $C(\Lambda)$
is thus defined such that
$\mathcal{U}^\Lambda_t g^\Lambda = \mathcal{U}^X_t g^X$.

\subsubsection{Multiple attractors}

We first discuss the coexistence of multiple attractors with
disjoint basins of attraction,
(such as in the Lorenz flow for $\rho_{\mathrm{Homo}} < \rho < \rho_A$),
in terms of multiplicity of the eigenvalue $1$.
When the phase space $X$ is forward invariant, i.e. such that $X \subset \Phi_t^{-1} X$
then one has that
\begin{align*}
\mathcal{U}^X_t \mathbf{1}_X = \mathbf{1}_{\phi_t^{-1} X \cap X} = \mathbf{1}_{X}, \quad t \ge 0.
\end{align*}
Thus, the forward invariance of $X$ implies that the function $\mathbf{1}_X$, constant on $X$,
is an eigenfunction of the Koopman semigroup associated with the eigenvalue $1$.
This is for example the case when the flow has a single globally asymptotically stable attractor on $X$.
If instead, there is a single repeller and $\Phi_t^{-1} X \subset X$, then
$\mathcal{U}^X_t \mathbf{1}_X = \mathbf{1}_{\phi_t^{-1} X \cap X} = \mathbf{1}_{\phi_t^{-1}  X}$
so that $\mathbf{1}_X$ is no longer an eigenfunction associated with $1$.

When several attractors $\Lambda_1, ..., \Lambda_l$ coexist,
each basin of attraction $B(\Lambda_i)$ is forward invariant
and $\mathcal{U}^X_t \mathbf{1}_{B(\Lambda_i)} = \mathbf{1}_{B(\Lambda_i)}, t \ge 0$.
Thus, to each attractor $\Lambda_i$ corresponds a Koopman eigenfunction $\mathbf{1}_{B(\Lambda_i)}$
associated with a generator eigenvalue zero.
However, only $l$ of the $l+1$ functions
$\mathbf{1}_X, \mathbf{1}_{B(\Lambda_1)}, ..., \mathbf{1}_{B(\Lambda_l)}$
are independent, so that there are only $l$ eigenvalues 1.
For example, when two attractors coexist,
a possibility is to have as independent eigenfunctions
$\mathbf{1}_X$ and $\mathbf{1}_{B(\Lambda_1)} - \mathbf{1}_{B(\Lambda_2)}$.
Adding suitably defined (e.g. Gaussian) noise leads to a unique invariant measure; reducing the intensity of the noise to zero leads in the limit to selecting a special invariant measure constructed as linear combination of all the invariant measure of the deterministic case. 
Note, however, that the numerical approximations of the basins of attraction
may be difficult when the geometry of the boundary is convoluted. The fact that the numerical discretization of the phase space leads effectively to introducing some noise in the system might explain why also in the deterministic case one can miss the presence of various coexisting and independent invariant measures.

\subsubsection{Correspondence between the eigenvalues of $\mathcal{U}_t^\Lambda$ and $\mathcal{U}_t^X$}
Let us discuss the correspondence between the eigenvalues of
$\mathcal{U}_t^X$ and $\mathcal{U}_t^\Lambda$.
Following \cite{Cessac2007}, we refer to the eigenvalues of $\mathcal{U}_t^X$
as the \emph{unstable resonances} and those of the $\mathcal{U}_t^\Lambda$
which do not correspond to the latter as the stable resonances.

What we show below is strictly applicable only for eigenvalues on the unit circle, as discussed in Remark \ref{pitfall}. Nonetheless, we believe that it may be practically relevant in the region near the unit circle, and, using more advanced mathematical tools, could be extended for the unit disk, away form the essential spectrum.

We consider the particular case where the invariant measure $\mu$
is supported by a uniformly hyperbolic (compact invariant) set $\Lambda \subset X$ for the continuously
differentiable flow $\Phi_t, t \in \mathbb{R}$ on $(X,d)$.
For convenience, we will assume that $X$ is the basin of attraction of $\Lambda$.
The global stable manifold $W_s$ through the point $x \in X$
can then be characterized topologically by
\begin{align*}
	W_s(x) = \{y \in X: \quad d(\Phi_t(x), \Phi_t(y)) \to 0, \quad t \to \infty\}.
\end{align*}
It follows directly that $W_s$ is invariant, i.e.~$\Phi_t W_s(x) \subset W_s(x)$
and $\cup_{x \in \Lambda} W_s(x) = X$, since $X$ is the basin of attraction of $\Lambda$.
Moreover, as a consequence of the stable manifold theorem \cite[Chap.~6]{Katok1996},
$x \mapsto W_s(x)$ is continuous.
\begin{prop}
	Let $\Phi_t, t \in \mathbb{R}$ be a continuously differentiable flow on the compact space $X$.
	Let $\Lambda$ be a uniformly hyperbolic compact invariant set
	and denote by $\mathcal{U}_t^\Lambda : C(\Lambda) \to C(\Lambda), t\ge 0,$
	the restriction to $C(\Lambda)$ of the Koopman operator
	$\mathcal{U}^X_t: C(X) \to C(X), g \mapsto g \circ \Phi_t$.
	Assume that $\psi^\Lambda$ in $C(\Lambda)$ is an eigenfunction
	of $\mathcal{U}_t^\Lambda$ associated with the eigenvalue $\zeta \in \mathbb{C}$
	for some $t > 0$, i.e.
	$\mathcal{U}_t^\Lambda \psi^\Lambda = \zeta \psi^\Lambda$.
	Then the function $\psi^X$ such that
	\begin{align*}
		\psi^X(y) = \psi^\Lambda(x), \quad \mathrm{whenever~} y \in W_s(x)
	\end{align*}
	is in $C(X)$ and is an eigenfunction for $\mathcal{U}_t^X$ associated with the eigenvalue $\zeta$.
\end{prop}
In other words, $\psi^X$ takes on a leaf of the global stable manifold
a constant value given by that of $\psi^\Lambda$ on this leaf.
\begin{proof}
	Let us first verify that $\psi^X$ is indeed in $C(X)$.
	Note first that $\psi^X$ is defined on $X$, since $W_s = X$.
	The continuity of $\psi^X$ follows from that of
	$\psi^\Lambda$ and $x \mapsto W_s(x)$.
	To see this, let $\{y_n\}$ be a sequence in $X$ converging to $y$.
	For each $y_n$ there is an $x_n$ in $\Lambda$ such that $y_n \in W_s(x_n)$.
	Thus
	\begin{align*}
		\psi^X(y_n) = \psi^\Lambda(x_n) 
	\end{align*}
	From the continuity of $x \mapsto W_s(x)$,
	the limit $x$ of $x_n$ exists and is such that $y \in W_s(x)$
	and $\psi^X(y) = \psi^\Lambda(x)$.
	From the continuity of $\psi^\Lambda$, it follows that
	\begin{align*}
		\lim_{n \to \infty} \psi^X(y_n) = \lim_{n \to \infty} \psi^\Lambda(x_n) = \psi^\Lambda(x) = \psi^\Lambda(y),
	\end{align*}
	so that $\psi^X$ is continuous on the metric space $(X, d)$.
	
	That $\psi^X$ is an eigenfunction follows directly from the invariance
	of the global stable manifold:
	\begin{align*}
		\mathcal{U}_t^X \psi^X(y) = \psi^X(\Phi_t y)
		&= \psi^\Lambda(x) \quad\mathrm{whenever~} \Phi_t y \in W_s(x) \\
		&= \psi^\Lambda(x) \quad\mathrm{whenever~} y \in W_s(\Phi_t^{-1} x) \\
		&= \psi^\Lambda(\Phi_t z) \quad\mathrm{whenever~} y \in W_s(z) \\
		&= \zeta \psi^\Lambda(z) \quad\mathrm{whenever~} y \in W_s(z) = \zeta \psi^X(y).
	\end{align*}
\end{proof}

Thus, to each eigenfunction $\mathcal{U}_t^\Lambda$ on $C(\Lambda)$
corresponds an eigenfunction of $\mathcal{U}_t^X$ on $C(X)$ associated with the same eigenvalue,
i.e.~the spectrum of $\mathcal{U}_t^\Lambda$ is a subset of the spectrum of $\mathcal{U}_t^X$.
\begin{remark}\label{pitfall}
	There is, however, a major caveat to the applicability of this results.
	Indeed, it cannot be applied to eigenfunctions associated with eigenvalues
	inside the unit disk, since $\mathcal{U}_t^\Lambda$
	has a roughening effect due to the contraction on the unstable manifold
	manifold of $\Lambda$, backward in time.
	For that purpose, spaces a distributions should be considered for $\psi^\Lambda$.
\end{remark}
\begin{remark}
	A similar result holds for the nonuniformly hyperbolic case \cite{Barreira2002}.
	However, the stable foliation is then only measurable
	so that eigenfunctions in spaces of measurable functions should be considered.
\end{remark}

\section{Robustness of the resonances}
\label{sec:robust}

In this appendix, we explain and give support to the choice of parameters
used to obtain the results of section \ref{sec:results} and given in section \ref{sec:design}.

\subsection{Numerical integration}

In this study, the time step of integration is particularly important for the position of the attractor
crisis. It appears that for larger time steps, the attractor crisis occurs for values of $\rho$ smaller
than $\rho_A$. A time step of $10^{-4}$ allows for the crisis to occur within $1\%$ of $\rho_A$,
while keeping the numerical integration tractable.

\subsection{Discretization}

The choice of the grid used to approximate the transfer operators by transition matrices
is the most critical step of the numerical application in this study.
First, in order for probabilities to be conserved,
the grid should cover a bounded set within which any trajectory remains.
This condition if fulfilled by a grid covering the ball $R_o$ (see Sect.~\ref{sec:crisis}).
A discretization of $R_o$ is then easily implemented when working in spherical coordinates $(r, \theta, \phi)$.

\begin{figure}[ht]
	\centering
	\includegraphics[width=0.48\textwidth]{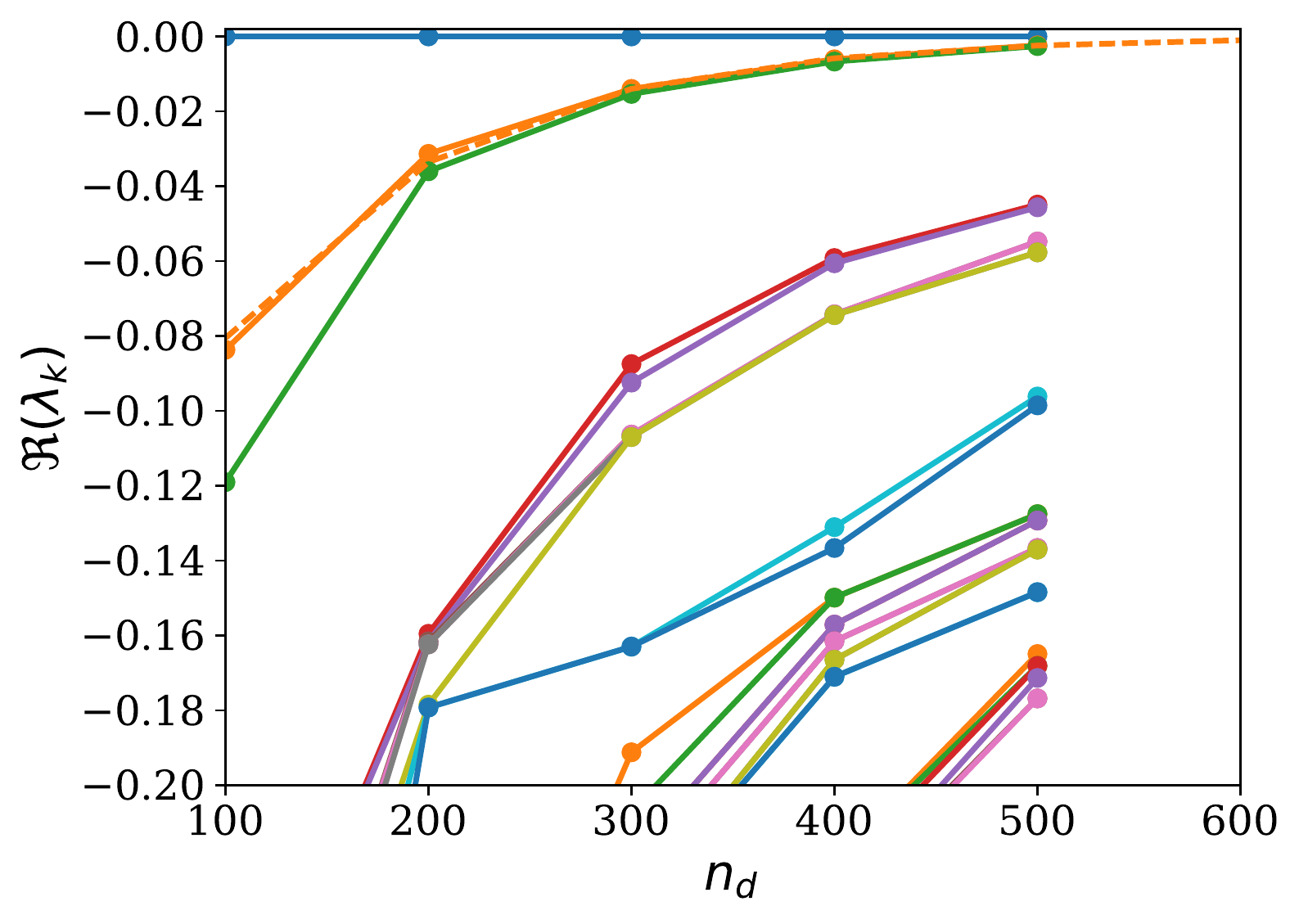}
	\caption{Convergence test of the real parts of the five leading generator eigenvalues
	calculated from $\bv{P}^m_\tau$ with respect to the grid resolution $n_d$-by-$n_d/2$-by-$n_d$,
	for $\rho = 24$, $\tau = 0.05$ and from $3.2 \cdot 10^{10}$ trajectories.}
	\label{fig:testGrid}
\end{figure}

Second, due to the fine-grained geometry of the eigenvectors associated with chaotic sets,
the generator eigenvalues are slow to converge with the grid resolution
\footnote{There is no general result regarding the convergence with the grid resolution
of eigenvalues of transition matrices to the Ruelle-Pollicott resonances.
This is, however, the case for uniformly hyperbolic systems \cite{Gouezel2006,Butterley2007}
for which Ulam's method may converge (see~\cite{Froyland2007a}, for the case of hyperbolic maps).}.
In this study, the focus is, however, on the generator eigenvalues close to the imaginary axis,
which are expected to be more robust to perturbations that eigenvalues further from the imaginary axis.
The results of a test of convergence with respect to the grid resolution $n_d$-by-$n_d/2$-by-$n_d$
is given in figure \ref{fig:testGrid}.
One can see that for $n_d \ge 400$, the real part of the first nonzero eigenvalue is close to convergence.
To get an idea of the value to which this real part would converge for higher resolutions,
the dashed orange line represents a least square fit of an exponential $a e^{b n_d},$ with $a, b \in \mathbb{R}$,
 to it.
The quality of this fit and the fact that the fitted curve is an exponential
converging to zero suggests that this eigenvalue, for $\rho \approx 24$, would
converge to the imaginary axis if the resolution were to be further increased.
Eventually, a grid resolution of 400-by-200-by-400 is chosen, allowing for the real part of the first nonzero
generator eigenvalue to remain within $2\%$ of the corresponding value obtained for $n_d = 500$.
The same grid is used to estimate $\bv{P}^\mu_\tau$, for which similar numerical convergence is
also observed (not shown here).

\subsection{Transition time}

The transition time $\tau$ for which the transition matrices $P_\tau^m$ and $P_\tau^\mu$
are estimated is a key parameter.
In theory, the spectral mapping formula \eqref{eq:SMF} allows to calculate the
generator eigenvalues $\lambda_k$ from the eigenvalues $\zeta_k(\tau), k \ge 0$
for the transfer operator $\mathcal{P}_\tau^\eta$, for any transition time $\tau$.
A first issue is, however, that, in taking the complex logarithm divided by $\tau$ to get
the $\lambda_k$ from the $\zeta_k(\tau)$, the imaginary part of the $\lambda_k$
is only known modulo $2\pi / \tau$.
An arbitrary choice of the principal part would then be valid only for true $\lambda_k$
such that $|\Im(\lambda_k)| \le \pi / \tau$, so that this window shortens as $\tau$ is increased
(see~\cite[Sect.~2.4]{Crommelin2009a}).

Second, a compromise should be found in order to estimate correctly as many eigenvalues
as possible, while only approximating those for which the eigenvectors can be resolved for a given grid.
Indeed, the longer the transition time $\tau$, the smaller $|\zeta_k(\tau)| = e^{\Re(\lambda_k) \tau}$
for $\lambda_k$ with a small real part.
Thus, in order to be able to estimate the eigenvalue $\zeta_k(\tau)$ numerically,
$\tau$ should be sufficiently small for $e^{\Re(\lambda_k) \tau}$ to be larger than
a threshold under which numerical errors become important.
This threshold depends on several factors such as the sampling, the nonnormality
of the transfer operators and roundoff errors \cite{Crommelin2009a}.

On the other hand, it is not always a good strategy to take $\tau$ short to approximate as many $\lambda_k$
far from the imaginary axis as possible.
Indeed, eigenvalues further from the imaginary tend to be associated with eigenvectors
with more and more changes of sign.
For the latter to be resolved, the grid resolution should be higher and higher.
However, if for a fixed grid, eigenvalues $\zeta_k(\tau)$ associated
with eigenvectors which cannot be appropriately resolved have not decayed,
their imprecise approximation will also have an impact on eigenvalues closer to the imaginary axis.
As a rule of thumb, the lower the grid resolution, the larger should $\tau$ be,
so as to approximate only the eigenvalues for which the eigenvectors can be properly resolved
at this resolution.
For a grid of 400-by-200-by-400, a transition time $\tau$ of 0.05 time units
is found to give of a good compromise.

\subsection{Number and length of trajectories}

\begin{figure}[ht]
	\centering
	\begin{subfigure}{0.48\textwidth}
	\includegraphics[width=\textwidth]{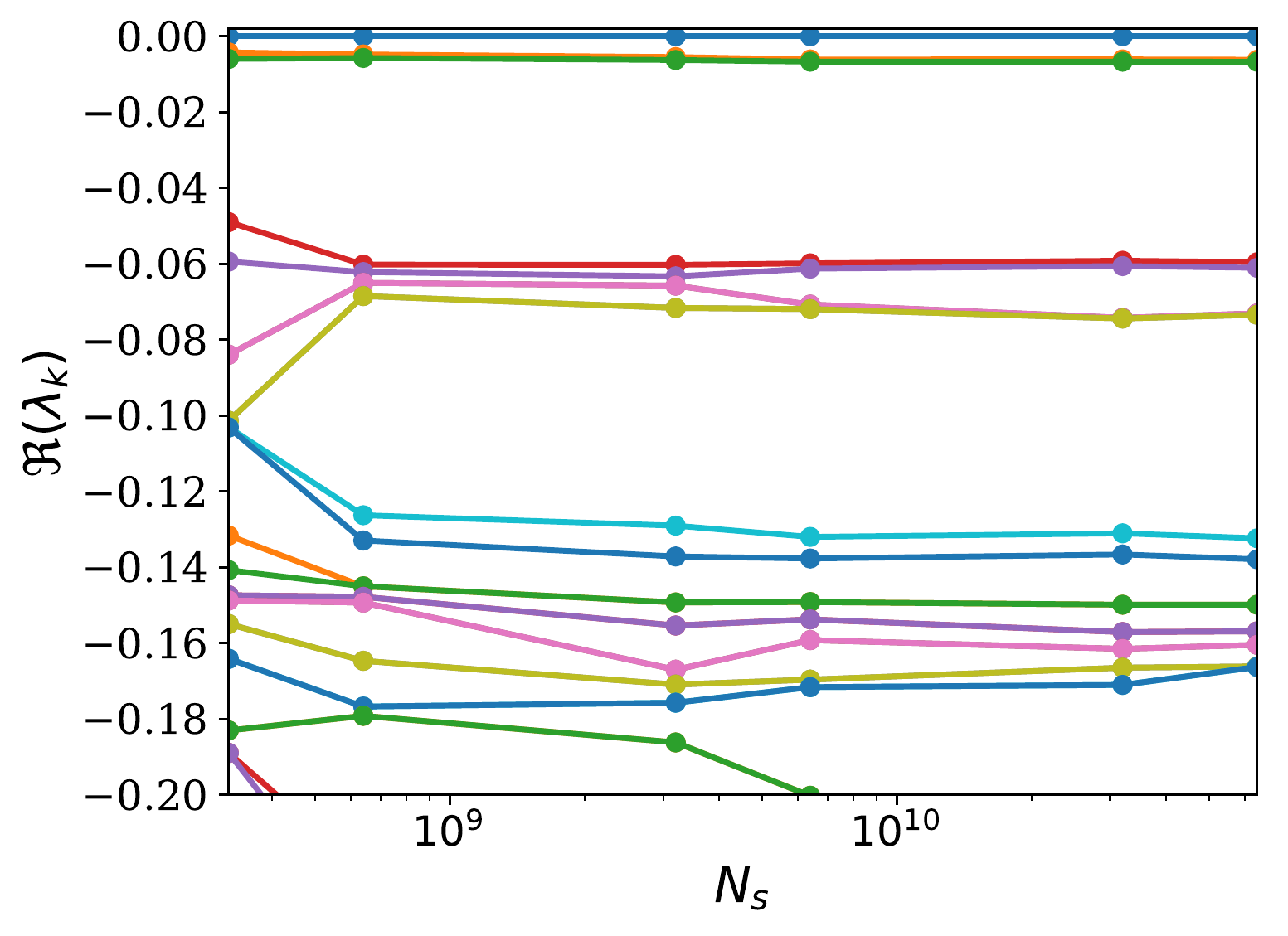}
	\end{subfigure}
	\begin{subfigure}{0.48\textwidth}
	\includegraphics[width=\textwidth]{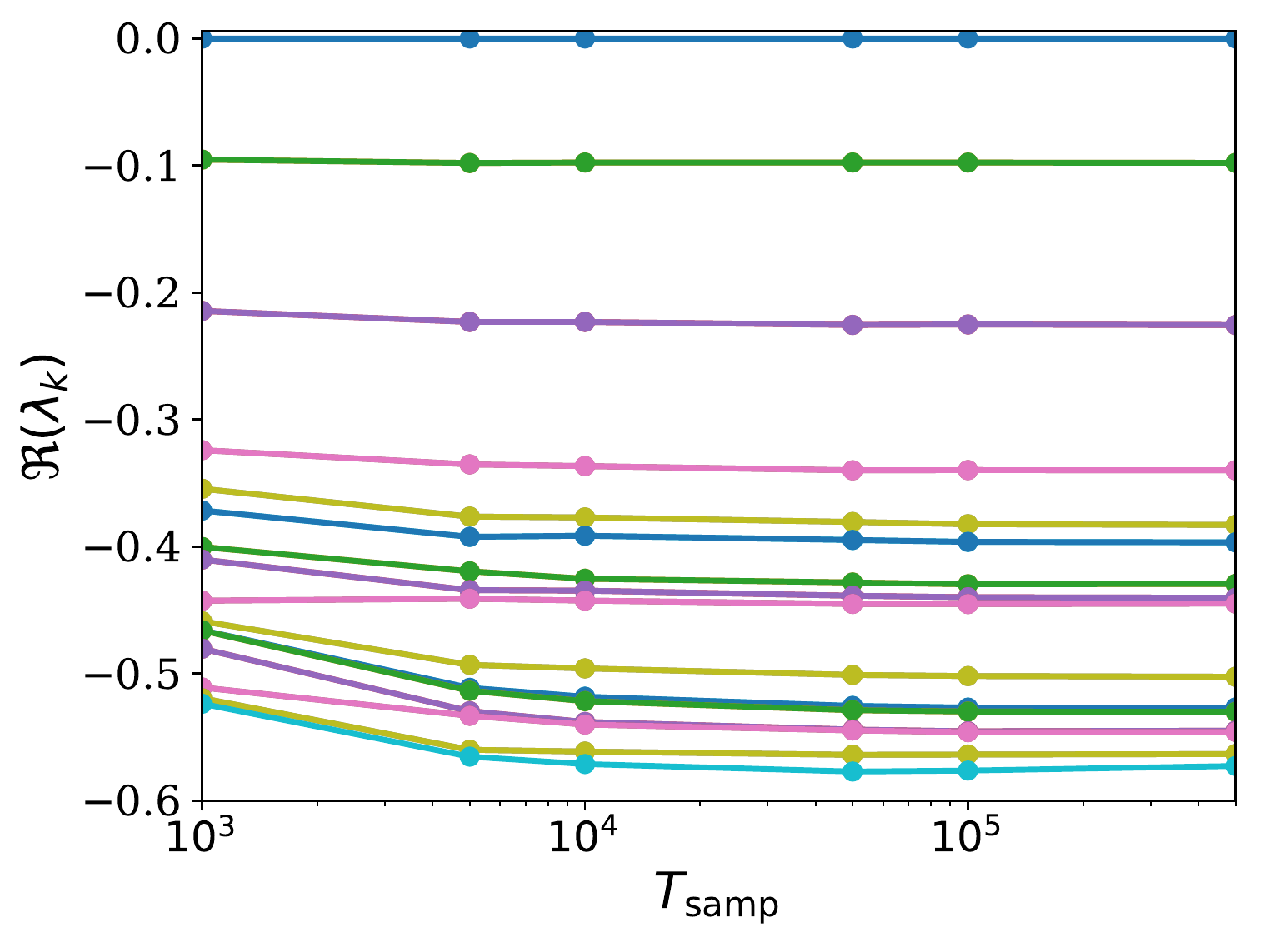}
	\end{subfigure}
	\caption{Convergence test of the real parts of the leading generator eigenvalues
	of $\bv{P}_\tau^m$ (left) and $\bv{P}_\tau^\mu$ (right)
	with respect to the number of samples $N_s$, for $\rho = 24$ and $\rho = 28$, respectively,
	and a grid of $400-by-200-by-400$.}
	\label{fig:testSamp}
\end{figure}

Ulam's method relies on the estimation of transition probabilities from time series.
The quality of these estimations depend on the sampling.
To test the sampling, one strategy is to estimate confidence intervals
(see e.g.~\cite[SI]{Chekroun2014} and \cite{Tantet2015}).
Another approach, followed here, is to directly test the convergence
of the generator eigenvalues with respect to the number of samples $N_s$.
This convergence of the transition probabilities and the eigenvalues with $N_s$
is known to occur at a rate of $\mathcal{O}(N_s^{1/2})$ \cite{Billingsley1961,Crommelin2009a}.
When estimating $\bv{P}_\tau^m$ from many short time series,
the number of samples is given by the number of trajectories.
In this case, the robustness of the eigenvalues to $N_s$ is shown
in the left panel in figure \ref{fig:testSamp}.
One can see that a number $6.4 \cdot 10^{9}$ of trajectories
is more than enough for an estimation on the chosen grid of 400-by-200-400.
When estimating $\bv{P}_\tau^\mu$ from a few long time series,
the number of samples is given by the number of trajectories
by their lengths $T_\mathrm{samp}$ divided by their sampling rate.
Here, a sampling rate of $100$ samples per time unit is used
and 24 long trajectories are used (in order to distribute each on a computer thread).
The robustness of the eigenvalues to $T_\mathrm{samp}$ is shown
in the right panel in figure \ref{fig:testSamp}.
One can see that a length $T_\mathrm{samp}$ of $1 \cdot 10^{5}$ time units
is more than enough for an estimation on the chosen grid of 400-by-200-400.

\subsection{Numerical eigenvalue problem}

We have seen that  the fine-grained geometry of the eigenvectors
of the Lorenz flow require a large grid resolution in order to achieve numerical convergence,
if only for the leading eigenvalues.
While, the estimation of the transition matrices from time series is,
an embarrassingly parallel problem, which can be easily distributed on several nodes of a calculator,
solving the eigenproblem for such large transition matrices is more challenging,
both in terms of computations and memory.
On the other hand, the sparse structure of the transition matrices allows to use iterative eigensolvers.
Here, we have chosen the block Krylov Schur algorithm implemented in the Anasazi package \cite{Baker2009}
of the Trilinos library.
This algorithm is a common and robust choice for such problems and the Anasazi implementation
makes it straightforward to distribute it on several nodes of a calculator.

\begin{acknowledgements}
AT would like to thank Sebastian Schubert for very helpful discussions on Lyapunov exponents and vectors and for providing his numerical code to estimate them. AT would also like to thank Andrey Gritsun for valuable discussions on the role of periodic orbits in chaotic systems as well as their numerical continuation.
AT and VL acknowledge support from the DFG Project MERCI.\\
VL would like to thank C. Liverani, T. Kuna, and V. Baladi for various inspiring conversations. 
VL acknowledges the support of the Horizon2020 Project CRESCENDO and of the DFG SFB/Transregio project TRR181.  \\
HD likes to acknowledge the support of the Netherlands Center
for Earth System Science (NESSC) funded by  the Netherlands
Foundation for Scientific Research (NWO). Part of this work was carried 
out on the Dutch national e-infrastructure (Cartesius) with the support of 
SURF Cooperative under the project SH284.
\end{acknowledgements}

\bibliographystyle{spphys}       
\bibliography{atantet}

\begin{thebibliography}{100}
\providecommand{\url}[1]{{#1}}
\providecommand{\urlprefix}{URL }
\expandafter\ifx\csname urlstyle\endcsname\relax
  \providecommand{\doi}[1]{DOI \discretionary{}{}{}#1}\else
  \providecommand{\doi}{DOI \discretionary{}{}{}\begingroup
  \urlstyle{rm}\Url}\fi

\bibitem{Guckenheimer1983}
J.M. Guckenheimer, P.~Holmes, \emph{{Nonlinear Oscillations, Dynamical Systems,
  and Bifurcation of Vector Fields}} (Springer, New York, 1983)

\bibitem{Ruelle1989a}
D.~Ruelle, \emph{{Elements of Differentiable Dynamics and Bifurcation Theory}}
  (Academic Press, San Diego, 1989).
\newblock \doi{10.1016/B978-0-12-601710-6.50001-8}

\bibitem{Kuznetsov1998}
Y.A. Kuznetsov, \emph{{Elements of Applied Bifurcation Theory, Second Edition}}
  (Springer-Verlag, New York, 1998)

\bibitem{Kloeden2011}
P.E. Kloeden, M.~Rasmussen, \emph{{Nonautonomous dynamical systems}} (American
  Mathematical Society, Providence, 2011)

\bibitem{Arnold2003}
L.~Arnold, \emph{{Random Dynamical Systems}} (Springer, Berlin, 2003)

\bibitem{Held2004a}
H.~Held, T.~Kleinen, Geophys. Res. Lett. \textbf{31}(December), 1 (2004).
\newblock \doi{10.1029/2004GL020972}

\bibitem{Kleinen2003a}
T.~Kleinen, H.~Held, G.~Petschel-Held, Ocean Dyn. \textbf{53}, 53 (2003).
\newblock \doi{10.1007/s10236-002-0023-6}

\bibitem{Nes2007}
E.H. van Nes, M.~Scheffer, Am. Nat. \textbf{169}(6), 738 (2007)

\bibitem{Scheffer2009}
M.~Scheffer, J.~Bascompte, W.a. Brock, V.~Brovkin, S.R. Carpenter, V.~Dakos,
  H.~Held, E.H. van Nes, M.~Rietkerk, G.~Sugihara, Nature \textbf{461}(7260),
  53 (2009).
\newblock \doi{10.1038/nature08227}

\bibitem{Lenton2011a}
T.M. Lenton, Nat. Clim. Chang. \textbf{1}(4), 201 (2011).
\newblock \doi{10.1038/nclimate1143}

\bibitem{Eckmann1985}
J.P. Eckmann, D.~Ruelle, Rev. Mod. Phys. \textbf{57}(July), 617 (1985).
\newblock \doi{10.1103/RevModPhys.57.617}

\bibitem{Young2002}
L.S. Young, J. Stat. Phys. \textbf{108}(5), 733 (2002)

\bibitem{Grebogi1983}
C.~Grebogi, E.~Ott, J.A. Yorke, Phys. D Nonlinear Phenom. \textbf{7}(1-3), 181
  (1983).
\newblock \doi{10.1016/0167-2789(83)90126-4}

\bibitem{Ashwin1996}
P.~Ashwin, J.~Buescu, I.~Stewart, Nonlinearity \textbf{9}, 703 (1996).
\newblock \doi{10.1088/0951-7715/9/3/006}

\bibitem{Skufca2006}
J.D. Skufca, J.a. Yorke, B.~Eckhardt, Phys. Rev. Lett. \textbf{96}(17), 5
  (2006).
\newblock \doi{10.1103/PhysRevLett.96.174101}

\bibitem{Schneider2007}
T.M. Schneider, B.~Eckhardt, J.a. Yorke, Phys. Rev. Lett. \textbf{99}(3), 1
  (2007).
\newblock \doi{10.1103/PhysRevLett.99.034502}

\bibitem{Eckhardt2008}
B.~Eckhardt, Nonlinearity \textbf{21}, T1 (2008).
\newblock \doi{10.1088/0951-7715/21/1/T01}

\bibitem{Bodai2015}
T.~B{\'{o}}dai, V.~Lucarini, F.~Lunkeit, R.~Boschi, Clim. Dyn. \textbf{44},
  3361 (2015).
\newblock \doi{10.1007/s00382-014-2206-5}

\bibitem{Lucarini2017}
V.~Lucarini, T.~Bodai, Nonlinearity \textbf{In Press} (2017)

\bibitem{Oseledets1968}
V.I. Oseledets, Tr. Mosk. Mat. Obs. \textbf{19}, 179 (1968)

\bibitem{LaSalle1976}
J.P. {La Salle}, \emph{{The Stability of Dynamical Systems}} (Society for
  Industrial and Applied Mathematics, Philadelphia, 1976)

\bibitem{Faranda2014e}
D.~Faranda, V.~Lucarini, P.~Manneville, J.~Wouters, Chaos, Solitons and
  Fractals \textbf{64}(1), 26 (2014).
\newblock \doi{10.1016/j.chaos.2014.01.008}

\bibitem{Faranda2014b}
D.~Faranda, B.~Dubrulle, F.M.E. Pons, J. Phys. A Math. Theor. \textbf{47}(25),
  252001 (2014).
\newblock \doi{10.1088/1751-8113/47/25/252001}

\bibitem{Lasota1994}
A.~Lasota, M.C. Mackey, \emph{{Chaos, Fractals and Noise}} (Springer, Berlin,
  1994)

\bibitem{Halmos1956}
P.R. Halmos, \emph{{Lectures on Ergodic Theory}} (Chelsea Publishing Company,
  New York, 1956)

\bibitem{Arnold1968}
V.I. Arnold, A.~Avez, \emph{{Ergodic Problems of Classical Mechanics}}
  (Advanced Book Classics, New York, 1968).
\newblock \doi{10.1002/zamm.19700500721}

\bibitem{Baladi2001}
V.~Baladi, in \emph{Eur. Congr. Math.}, ed. by C.~Casacuberta, R.M.
  Mir{\'{o}}-Roig, J.~Verdera, S.~Xamb{\'{o}}-Descamps (Birkh{\"{a}}user,
  Basel, 2001), pp. 203--223.
\newblock \doi{10.1007/978-3-0348-8268-2_11}

\bibitem{Young2013}
L.S. Young, Commun. Pure Appl. Math. \textbf{66}(9), 1439 (2013).
\newblock \doi{10.1002/cpa.21468}

\bibitem{Pollicott1985}
M.~Pollicott, Invent. Math. \textbf{81}(3), 413 (1985).
\newblock \doi{10.1007/BF01388579}

\bibitem{Ruelle1986}
D.~Ruelle, Phys. Rev. Lett. \textbf{56}(5), 405 (1986)

\bibitem{Liverani1995a}
C.~Liverani, Ann. Math. \textbf{142}, 239 (1995)

\bibitem{Blank2001a}
M.~Blank, G.~Keller, C.~Liverani, Nonlinearity \textbf{15}(6), 58 (2001).
\newblock \doi{10.1088/0951-7715/15/6/309}

\bibitem{Gouezel2006}
S.~Gou{\"{e}}zel, C.~Liverani, Ergod. Theory Dyn. Syst. \textbf{26}(01), 26
  (2006).
\newblock \doi{10.1017/S0143385705000374}

\bibitem{Butterley2007}
O.~Butterley, C.~Liverani, J. Mod. Dyn. \textbf{1}(2), 301 (2007)

\bibitem{Faure2014}
F.~Faur{\'{e}}, M.~Tsujii, in \emph{Anal. Probabilistic Approaches to Dyn.
  Negat. Curvature}, ed. by F.~Dal'Bo, M.~Peign{\'{e}}, A.~Sambusetti
  (Springer, Cham, 2014), chap.~2, pp. 65--138

\bibitem{Baladi2017}
V.~Baladi, J. Stat. Phys. \textbf{166}(3), 525 (2017)

\bibitem{Collet2004}
P.~Collet, J.P. Eckmann, J. Stat. Phys. \textbf{115}(April), 217 (2004)

\bibitem{Alves2004}
F.~Alves, S.~Luzzatto, V.~Pinheiro, Ergod. Theory Dyn. Syst. \textbf{24}(3),
  637 (2004)

\bibitem{Pires2011}
C.J.A. Pires, A.~Saa, R.~Venegeroles, Phys. Rev. E \textbf{84}, 066210 (2011).
\newblock \doi{10.1103/PhysRevE.84.066210}

\bibitem{Slipantschuk2013}
J.~Slipantschuk, O.F. Bandtlow, W.~Just, J. Phys. A \textbf{46}, 1 (2013).
\newblock \doi{10.1088/1751-8113/46/7/075101}

\bibitem{Vaidya2008}
U.~Vaidya, P.G. Mehta, IEEE Trans. Automat. Contr. \textbf{53}(1), 307 (2008).
\newblock \doi{10.1109/TAC.2007.914955}

\bibitem{Mauroy2016}
A.~Mauroy, I.~Mezi{\'{c}}, IEEE Trans. Automat. Contr. \textbf{61}(11), 3356
  (2016)

\bibitem{Tantet2015a}
A.~Tantet, V.~Lucarini, F.~Lunkeit, H.A. Dijkstra, arXiv pp. 1--28 (2015)

\bibitem{Ruelle2009}
D.~Ruelle, Nonlinearity \textbf{22}, 855 (2009).
\newblock \doi{10.1088/0951-7715/22/4/009}

\bibitem{Cessac2007}
B.~Cessac, J.A. Sepulchre, Phys. D Nonlinear Phenom. \textbf{225}, 13 (2007).
\newblock \doi{10.1016/j.physd.2006.09.034}

\bibitem{Gritsun2017}
A.S. Gritsun, V.~Lucarini, Phys. D Nonlinear Phenom. pp. 1--15 (2017).
\newblock \doi{http://dx.doi.org/10.1016/j.physd.2017.02.015}

\bibitem{Hairer2010}
M.~Hairer, A.J. Majda, Nonlinearity \textbf{23}(4), 909 (2010).
\newblock \doi{10.1088/0951-7715/23/4/008}

\bibitem{Lucarini2012c}
V.~Lucarini, J. Stat. Phys. \textbf{146}(4), 774 (2012).
\newblock \doi{10.1007/s10955-012-0422-0}

\bibitem{Gaspard2002a}
P.~Gaspard, J. Stat. Phys. \textbf{106}(1-2), 57 (2002).
\newblock \doi{10.1023/A:1013167928166}

\bibitem{Tantet2016}
A.~Tantet, M.D. Chekroun, J.D. Neelin, H.A. Dijkstra, Phys. D Nonlinear Phenom.
   (2017)

\bibitem{Lorenz1963a}
E.N. Lorenz, J. Atmos. Sci. \textbf{20}, 130 (1963)

\bibitem{Sparrow1982}
C.~Sparrow, \emph{{The Lorenz Equations: Bifurcations, Chaos and Strange
  Attractors}} (Springer, New York, 1982)

\bibitem{Guckenheimer1979}
J.M. Guckenheimer, R.F. Williams, Publ. math{\'{e}}matiques l'I.H.{\'{E}}.S.
  \textbf{50}, 59 (1979)

\bibitem{Araujo}
V.~Ara{\'{u}}jo, M.J. Pacifico, \emph{{Three-Dimensional Flows}} (Springer,
  Heidelberg, 2010).
\newblock \doi{10.1007/978-3-642-11414-4}

\bibitem{Tucker1999}
W.~Tucker, C. R. Acad. Sci. Paris \textbf{328}, 1197 (1999)

\bibitem{Reick2002a}
C.H. Reick, Phys. Rev. E - Stat. Nonlinear, Soft Matter Phys.
  \textbf{66}(October 2001), 1 (2002).
\newblock \doi{10.1103/PhysRevE.66.036103}

\bibitem{Lucarini2009b}
V.~Lucarini, J. Stat. Phys. \textbf{134}, 381 (2009).
\newblock \doi{10.1007/s10955-008-9675-z}

\bibitem{Rollins1984}
R.W. Rollins, E.R. Hunt, Phys. Rev. A \textbf{29}(6), 3327 (1984)

\bibitem{Pompe1988}
B.~Pompe, R.W. Leven, Phys. Scr. \textbf{38}(5), 651 (1988).
\newblock \doi{10.1088/0031-8949/38/5/003}

\bibitem{Mehra1996}
V.~Mehra, R.~Ramaswamy, Phys. Rev. E \textbf{53}(4), 3420 (1996)

\bibitem{Beims2016}
M.W. Beims, J.A.C. Gallas, Sci. Rep. \textbf{6}(November), 37102 (2016).
\newblock \doi{10.1038/srep37102}

\bibitem{Lorenz1979b}
E.N. Lorenz, in \emph{Glob. Anal.}, ed. by M.~Grmela, J.E. Marsden (Springer,
  Berlin, 1979), pp. 53--75

\bibitem{Hartman1964}
P.~Hartman, \emph{{Ordinary Differential Equations}}, vol.~53 (John Wiley {\&}
  Sons, New York, 1964)

\bibitem{Kaplan1979a}
J.L. Kaplan, J.a. Yorke, Commun. Math. Phys. \textbf{67}(2), 93 (1979).
\newblock \doi{10.1007/BF01221359}

\bibitem{Katok1996}
A.~Katok, B.~Hasselblatt, \emph{{Introduction to the Modern Theory of Dynamical
  Systems}} (Cambridge University Press, Cambridge, 1996)

\bibitem{Barreira2002}
L.~Barreira, Y.B. Pesin, O.~Sarig, in \emph{Handb. Dyn. Syst. 1B}, ed. by
  B.~Hasselblatt, A.~Katok (Elsevier, 2006), chap.~2, pp. 57--263

\bibitem{Kuptsov2012}
P.V. Kuptsov, U.~Parlitz, J. Nonlinear Sci. \textbf{22}(5), 727 (2012).
\newblock \doi{10.1007/s00332-012-9126-5}

\bibitem{Pesin2004}
Y.B. Pesin, \emph{{Lectures on partial hyperbolocity and stable ergodicity}},
  zurich lec edn. (European Mathematical Society, Zurich, 2004).
\newblock \doi{10.4171/003}

\bibitem{Ginelli2007}
F.~Ginelli, P.~Poggi, A.~Turchi, H.~Chat{\'{e}}, R.~Livi, A.~Politi, Phys. Rev.
  Lett. \textbf{99}(13), 1 (2007).
\newblock \doi{10.1103/PhysRevLett.99.130601}

\bibitem{Yang2009}
H.L. Yang, K.A. Takeuchi, F.~Ginelli, H.~Chat{\'{e}}, G.~Radons, Phys. Rev.
  Lett. \textbf{102}(7), 1 (2009).
\newblock \doi{10.1103/PhysRevLett.102.074102}

\bibitem{Hasegawa1992b}
H.~Hasegawa, W.~Saphir, Phys. Rev. A \textbf{46}(12), 7401 (1992).
\newblock \doi{10.1103/PhysRevA.46.7401}

\bibitem{Gaspard1992}
P.~Gaspard, D.A. Ramirez, Phys. Rev. A \textbf{45}(12), 8383 (1992).
\newblock \doi{10.1103/PhysRevA.45.8383}

\bibitem{Gaspard1995}
P.~Gaspard, G.~Nicolis, A.~Provata, S.~Tasaki, Phys. Rev. E \textbf{51}(1), 74
  (1995)

\bibitem{Gaspard2001a}
P.~Gaspard, S.~Tasaki, Phys. Rev. E \textbf{64}(5), 056232 (2001).
\newblock \doi{10.1103/PhysRevE.64.056232}

\bibitem{ulam1964collection}
S.M. Ulam, \emph{{Problems in Modern Mathematics}}, science edn. (Wiley, New
  York, 1964)

\bibitem{Dellnitz1999}
M.~Dellnitz, O.~Junge, SIAM J. Numer. Anal. \textbf{36}(2), 491 (1999).
\newblock \doi{10.1137/S0036142996313002}

\bibitem{Klus2015a}
S.~Klus, P.~Koltai, C.~Sch{\"{u}}tte, arXiv pp. 1--19 (2015)

\bibitem{Williams2015}
M.O. Williams, I.G. Kevrekidis, C.W. Rowley, J. Nonlinear Sci. \textbf{25}(6),
  1307 (2015).
\newblock \doi{10.1007/s00332-015-9258-5}

\bibitem{Lucarini2016}
V.~Lucarini, J. Stat. Phys. \textbf{162}(2), 312 (2016).
\newblock \doi{10.1007/s10955-015-1409-4}

\bibitem{Shutte1999}
C.~Sch{\"{u}}tte, {Conformational Dynamics: Modelling, Theory, Algorithm and
  Application to Biomolecules}.
\newblock Tech. Rep. July, Konrad-Zuse-Zentrum f{\"{u}}r Informationstechnik,
  Berlin (1999)

\bibitem{Chekroun2014}
M.D. Chekroun, J.D. Neelin, D.~Kondrashov, J.C. McWilliams, M.~Ghil, Proc.
  Natl. Acad. Sci. U. S. A. \textbf{111}(5), 1684 (2014).
\newblock \doi{10.1073/pnas.1321816111}

\bibitem{Chekroun2016}
M.D. Chekroun, A.~Tantet, J.D. Neelin, H.A. Dijkstra, Phys. D Nonlinear Phenom.
   (2017)

\bibitem{Tantet2015}
A.~Tantet, F.R. van~der Burgt, H.A. Dijkstra, Chaos An Interdiscip. J.
  Nonlinear Sci. \textbf{25}(3), 036406 (2015).
\newblock \doi{10.1063/1.4908174}

\bibitem{Engel2001}
K.J. Engel, R.~Nagel, \emph{{One-parameter semigroups for linear evolution
  equations}} (Springer, New York, 2001)

\bibitem{Billingsley1961}
P.~Billingsley, \emph{{Statistical Inference for Markov process}} (University
  of Chicago Press, Chicago, 1961)

\bibitem{Baker2009}
C.G. Baker, U.L. Hetmaniuk, R.B. Lehoucq, H.K. Thornquist, ACM Trans. Math.
  Softw. \textbf{36}(3), 13 (2009).
\newblock \doi{10.1145/1527286.1527287}

\bibitem{Heroux2003}
M.~Heroux, R.~Bartlett, V.~Howle, R.~Hoekstra, J.~Hu, T.~Kolda, R.B. Lehoucq,
  K.~Long, R.~Pawlowski, E.~Phipps, A.~Salinger, H.~Thornquist, R.~Tuminaro,
  J.~Willenbring, A.~Williams, {An Overview of Trilinos}.
\newblock Tech. rep., Sandia National Laboratories, Albuquerque (2003)

\bibitem{Froyland2011a}
G.~Froyland, O.~Junge, P.~Koltai, SIAM J. Numer. Anal. \textbf{51}(1), 223
  (2011).
\newblock \doi{10.1137/110819986}

\bibitem{VonStorch1999b}
H.~von Storch, F.~Zwiers, \emph{{Stastistical Analysis in Climate Research}}
  (Cambridge University Press, Cambridge, 1999)

\bibitem{Lan2013}
Y.~Lan, I.~Mezi{\'{c}}, Phys. D Nonlinear Phenom. \textbf{242}(1), 42 (2013).
\newblock \doi{10.1016/j.physd.2012.08.017}

\bibitem{Kubo1957a}
R.~Kubo.
\newblock {Statistical-mechanical theory of irreversible processes. I. general
  theory and simple applications to magnetic and conduction problems} (1957).
\newblock \doi{10.1143/JPSJ.12.570}

\bibitem{Hormander1968a}
L.R. H{\"{o}}rmander, Acta Math. \textbf{119}(1), 147 (1968).
\newblock \doi{10.1007/BF02392081}

\bibitem{Hairer2011}
M.~Hairer, Bull. des Sci. Math. \textbf{135}(6-7), 650 (2011).
\newblock \doi{10.1016/j.bulsci.2011.07.007}

\bibitem{Yosida1980}
K.~Yosida, \emph{{Functional Analysis}}, vol. 123 (Springer-Verlag, Berlin
  Heidelberg New York, 1980)

\bibitem{Davies2007}
E.B. Davies, \emph{{Linear Operators and Their Spectra}} (Cambridge University
  Press, Cambridge, 2007)

\bibitem{Koopman1931}
B.O. Koopman, Proc. Natl. Acad. Sci. U. S. A. \textbf{17}(5), 315 (1931).
\newblock \doi{10.1073/pnas.17.5.315}

\bibitem{Neumann1932}
J.~von Neumann, Proc. Natl. Acad. Sci. U. S. A. \textbf{18}(2), 70 (1932).
\newblock \doi{10.1073/pnas.18.1.70}

\bibitem{Eisner2015}
T.~Eisner, B.~Farkas, M.~Haase, R.~Nagel, \emph{{Operator Theoretic Aspects of
  Ergodic Theory}} (Springer International Publishing, 2015).
\newblock \doi{10.1007/978-3-319-16898-2}

\bibitem{Gallavotti2014a}
G.~Gallavotti, \emph{{Nonequilibrium and irreversibility}} (Springer, Cham,
  2014).
\newblock \doi{10.1007/978-3-319-06758-2}

\bibitem{Misra1979}
B.~Misra, I.~Prigogine, M.~Courbage, Phys. A Stat. Mech. its Appl.
  \textbf{98}(1-2), 1 (1979).
\newblock \doi{10.1016/0378-4371(79)90163-8}

\bibitem{Gaspard1998}
P.~Gaspard, \emph{{Chaos, Scattering and Statistical Mechanics}} (Cambridge
  University Press, Cambridge, 1998)

\bibitem{Garbaczewski2002}
P.~Garbaczewski, R.~Olkiewicz (eds.), \emph{{Dynamics of Dissipation}}
  (Springer, Berlin, 2002)

\bibitem{Keller1998}
G.~Keller, C.~Liverani, T.U.D. Roma, {Stability of the spectrum for transfer
  operators}.
\newblock Tech. rep., Scuola Norm. Sup. Pisa, Pisa (1998)

\bibitem{Baladi1999}
V.~Baladi, M.~Holschneider, Nonlinearity \textbf{12}(December), 525 (1999).
\newblock \doi{10.1088/0951-7715/12/3/006}

\bibitem{Froyland2007a}
G.~Froyland, Discret. Contin. Dyn. Syst. \textbf{17}(3), 671 (2007).
\newblock \doi{10.3934/dcds.2007.17.671}

\bibitem{Lang1993}
S.~Lang, \emph{{Real and Functional Analysis}} (Springer, New York, 1993)

\bibitem{Crommelin2009a}
D.~Crommelin, E.~Vanden-Eijnden, Multiscale Model. Simul. \textbf{7}(4), 1
  (2009)

\end{thebibliography}

\end{document}